\documentclass[preprint, fleqn]{acmart}
\usepackage{booktabs} 
\usepackage{subcaption}
\usepackage{algorithm}
\usepackage{algpseudocode}

\setcopyright{none}
\begin{document}
\title{Force vs Nudge : Comparing Users Pattern Choices on SysPal and TinPal}

\author{Harshal Tupsamudre}
\orcid{1234-5678-9012-3456}
\affiliation{%
  \institution{TCS Research}
  \city{Pune}
  \country{India}}
\email{harshal.tupsamudre@tcs.com}
\author{Sukanya Vaddepalli}
\affiliation{
  \institution{TCS Research}
  \city{Pune}
  \country{India}}
\email{sukanya.vaddepalli@tcs.com}
\author{Vijayanand Banahatti}
\affiliation{%
 \institution{TCS Research}
  \city{Pune}
  \country{India}}
\email{vijayanand.banahatti@tcs.com}
\author{Sachin Lodha}
\affiliation{%
\institution{TCS Research}
  \city{Pune}
  \country{India}}
\email{sachin.lodha@tcs.com}

\begin{abstract}
Android's $3 \times 3$ graphical pattern lock scheme is one of the widely used authentication method on smartphone devices. However, users choose $3 \times 3$ patterns from a small subspace of all possible 389,112 patterns. The two recently proposed interfaces, {\em SysPal} by Cho {\em et al.} \cite{cho2017syspal} and {\em TinPal} by the authors \cite{TinPal}, demonstrate that it is possible to influence users $3 \times 3$ pattern choices by making small modifications in the existing interface. While SysPal {\em forces} users to include one, two or three system-assigned random dots in their pattern, TinPal employs a highlighting mechanism to {\em inform} users about the set of reachable dots from the current selected dot. Both interfaces improved the security of $3 \times 3$ patterns without affecting usability, but no comparison between SysPal and TinPal was presented. 

To address this gap, we conduct a new user study with 147 participants and collect patterns on three SysPal interfaces, 1-dot, 2-dot and 3-dot. We also consider original and TinPal patterns collected in our previous user study involving 99 participants \cite{TinPal}. We compare patterns created on five different interfaces, original, TinPal, 1-dot, 2-dot and 3-dot using a range of security and usability metrics including pattern length, stroke length, guessability, recall time and login attempts. Our study results show that participants in the TinPal group created significantly longer and complex patterns than participants in the other four groups. Consequently, the guessing resistance of TinPal patterns was the highest among all groups. Further, we did not find any significant difference in memorability of patterns created in the TinPal group and the other groups.
\end{abstract}

%
%

\ccsdesc[500]{Security and privacy~Graphical / visual passwords}
\ccsdesc[500]{Security and privacy~Usability in security and privacy}
\ccsdesc[500]{Security and privacy~Mobile platform security}

\keywords{Authentication; Graphical Passwords; Human Factors; Security}

\maketitle

\section{Introduction}
Android's pattern lock scheme in which users connect at least four dots in $3\times 3$ grid is one of the most popular authentication mechanisms to protect sensitive information on mobile devices. According to research studies \cite{mahfouz2016android,van2014studying}, nearly 40\% of the Android users employ $3 \times 3$ patterns to secure their smartphones. Other built-in authentication methods such as PINs and passwords are available, however pattern lock is perceived to be more usable than PINs and passwords \cite{vonZezschwitz:usability}. Moreover, psychological studies \cite{shepard1967recognition,Paivio1968,yuille1983imagery} show that human brain is better at remembering and recalling graphical information as compared to numbers and letters. Biometric alternatives are also usable, but they are available only in high-end Android devices and are considered to be less secure than $3\times 3$ patterns \cite{biometric}. Further, in order to use biometrics, the user is also required to choose a PIN or pattern as a fallback method which is activated whenever biometrics fail. Nowadays, Android phones (e.g., Xiaomi) come with built-in applock feature that enables users to protect their critical personal applications such as banking app, Facebook app or gallery app using $3\times 3$ patterns. Users can also choose from various $3 \times 3$ pattern lock apps available in app store (e.g., AppLock by DoMobileLab has over 350 million users). To enhance security further, some vendors ({\em e.g.}, CyanLockScreen) allow users to select patterns on larger grid sizes ranging from $4 \times 4$ up to $12 \times 12$.

Although $3\times 3$ patterns are considered to be usable, they are prone to a wide range of attacks including {\em guessing attacks}, {\em shoulder-surfing attacks} and {\em smudge attacks}. Many research studies \cite{Uellenbeck:guessing, Aviv:guessing, harshal:guessing} show that user-selected patterns are highly biased. The theoretical space of $3 \times 3$ patterns is large (389,112), however users choose simple patterns resembling English letters such as `Z', `S', `M', `N', `L', `R'  and  `G'. The characteristics such as knight moves, overlaps, direction changes, intersections (crosses) which enhance the visual complexity of patterns are almost never used \cite{Aviv:shouldersurfing, vonZezschwitz:shouldersurfing, Sun:shouldersurfing, Song:shouldersurfing, Andriotis:guessing}. Consequently, the resulting patterns could be easily memorized by an observer. The pattern lock scheme is also susceptible to smudge attack \cite{Aviv:sidechannel}, a type of side-channel attack in which the attacker infers user's pattern using physical traces left by fingers on the screen. Further related work \cite{Andriotis:sidechannel} even suggests that it is possible to recover the entire pattern from the partial traces by exploiting users' biased choices. To counter these attacks, Android enforces a lock-out policy that allows a maximum of 20 consecutive failed attempts \cite{cho2017syspal}. Even with this policy, attackers could still recover a significant portion of user-selected patterns (about 18.5\%) \cite{harshal:guessing}. Increasing the grid size to $4 \times 4$ does not help either, as users choose $4 \times 4$ patterns that are simple embeddings of $3 \times 3$ patterns \cite{Aviv:guessing}. Users' pattern choices remain biased even in the presence of pattern strength meters \cite{Song:shouldersurfing}. 

In our earlier work \cite{TinPal}, we proposed {\em TinPal}, an enhanced $3 \times 3$ interface that employs a highlighting mechanism to inform users about the set of reachable dots from the currently connected dot. The highlighting works in real-time during pattern creation as well as during pattern recall. We evaluated the efficacy of TinPal in a lab study with 99 participants. Participants were randomly split into two groups. The first group of 49 participants created patterns using the original interface and the second group of 50 participants created patterns using TinPal. We compared patterns in two groups with respect to the pattern length, stroke length, knight moves, overlaps, direction changes, intersections, and start point and end point distributions (Table \ref{tab:cmp}). We found that participants who used TinPal created significantly longer patterns containing visually complex features such as knight moves and overlaps than those who used the original interface. Consequently, TinPal patterns offered more resistance against guessing attacks. Further, we did not find any significant difference in memorability and efficiency of patterns between the two groups.
\begin{table}[h]
  \centering
 \scriptsize
  \begin{tabular}{l c c c}
\toprule
  \textit{Characteristic}
     & \textit{TinPal \cite{TinPal}}
     & \textit{SysPal \cite{cho2017syspal}}
     & \textit{Current Submission}\\
    \midrule
	pattern length & $\surd$ & $\surd$ & $\surd$ \\
	stroke length  & $\surd$ & $\times$ & $\surd$ \\
	knight moves & $\surd$ & $\times$ & $\surd$ \\
	overlaps & $\surd$ & $\times$ & $\surd$ \\
	direction changes & $\surd$ & $\times$ & $\surd$ \\
	intersections & $\surd$ & $\times$ & $\surd$ \\
	segment frequencies & $\times$ & $\surd$ & $\surd$ \\
	dot frequencies & $\times$ & $\surd$ & $\surd$ \\
	start point distribution & $\surd$ & $\surd$ & $\surd$ \\
	end point distribution & $\surd$ & $\surd$ & $\surd$ \\
	SysPal specific features &  NA & $\surd$ & $\surd$\\	
\bottomrule
\end{tabular}
\caption{Pattern characteristics reported in our earlier TinPal paper \cite{TinPal}, SysPal paper \cite{cho2017syspal} and the current submission (for both TinPal and SysPal).}\label{tab:cmp}
\end{table}

To improve the security of $3 \times 3$ patterns, Cho {\em et al.} proposed System-guided pattern lock (SysPal) that mandates users to include system-assigned random dot(s) at any position while creating their pattern \cite{cho2017syspal}. Note that unlike SysPal, TinPal does not mandate users to include any specific dot(s) in their pattern, it just informs users about the next available dot choices by highlighting them in real-time while the pattern is being drawn. The authors \cite{cho2017syspal} evaluated three different SysPal policies, 1-dot, 2-dot and 3-dot which required study participants to use one, two or three system-assigned dots in their pattern respectively. They conducted an online study involving 1,717 participants to measure security and usability of all three SysPal policies. Later they conducted a lab study with 46 participants to measure memorability of  patterns created using 1-dot and 2-dot SysPal policies. The results of their online study show that 1-dot and 2-dot SysPal policies improved the guessing resistance of $3 \times 3$ patterns without much impacting memorability. The authors of SysPal paper also reported few pattern characteristics such as line segment frequencies, dot frequencies, start point distribution, end point distribution and SysPal specific features such as position of mandated dots and distance between mandated dots (Table \ref{tab:cmp}). However, they did not report important characteristics such as stroke length, knight moves, overlaps, direction changes and intersections which are considered to be effective against guessing attacks \cite{Uellenbeck:guessing, Aviv:guessing, harshal:guessing} and shoulder-surfing attacks \cite{vonZezschwitz:shouldersurfing, Aviv:shouldersurfing, Sun:shouldersurfing, Song:shouldersurfing, Andriotis:guessing}. Further, some of the characteristics reported in the SysPal paper \cite{cho2017syspal} were not reported in our TinPal paper \cite{TinPal}.

\begin{figure}[h]
\centering
\begin{subfigure}[b]{0.25\textwidth}
  \centering
  \includegraphics[scale=0.40]{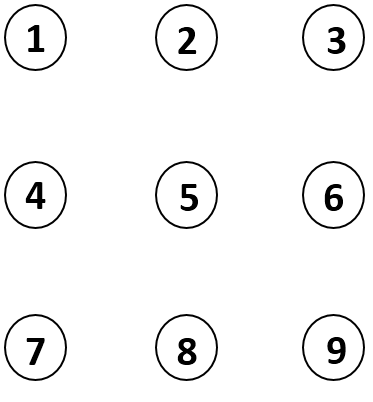}
  \captionsetup{font=scriptsize}
  \caption{$3\times 3$ labels}~\label{fig:3X3}
\end{subfigure}%
\begin{subfigure}[b]{0.25\textwidth}
  \centering
  \includegraphics[scale=0.40]{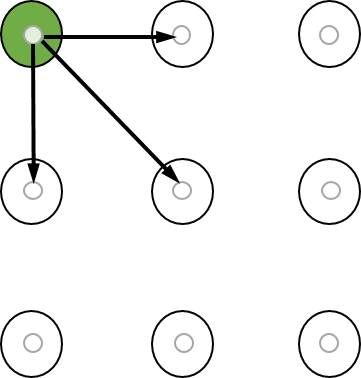}
  \captionsetup{font=scriptsize}
  \caption{Corner (Simple)}~\label{fig:corner_simple}
\end{subfigure}%
\begin{subfigure}[b]{0.25\textwidth}
  \centering
  \includegraphics[scale=0.40]{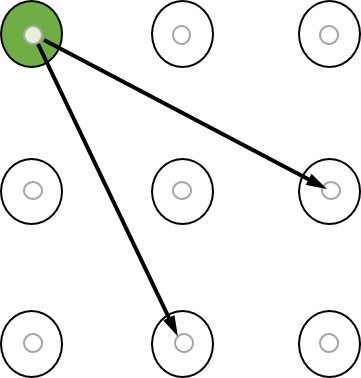}
  \captionsetup{font=scriptsize}
  \caption{Corner (Knight)}~ \label{fig:corner_knight}
\end{subfigure}%
\begin{subfigure}[b]{0.25\textwidth}
  \centering
  \includegraphics[scale=0.40]{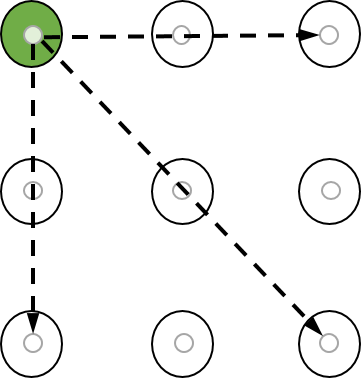}
  \captionsetup{font=scriptsize}
  \caption{Corner (Overlap)}~ \label{fig:corner_overlap}
\end{subfigure}

\begin{subfigure}[b]{0.25\textwidth}
  \centering
  \includegraphics[scale=0.40]{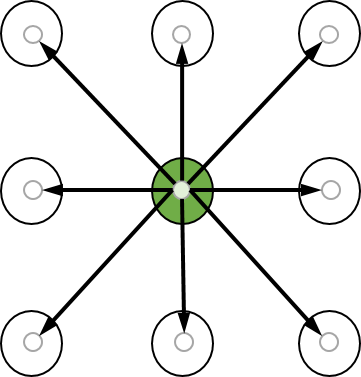}
\captionsetup{font=scriptsize}
  \caption{Center}~\label{fig:center}
\end{subfigure}%
\begin{subfigure}[b]{0.25\textwidth}
  \centering
  \includegraphics[scale=0.40]{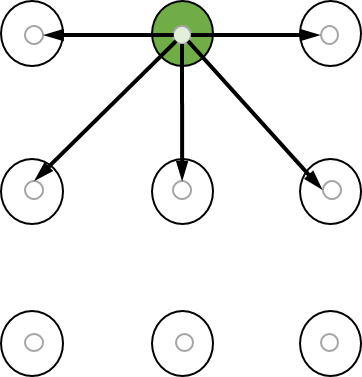}
\captionsetup{font=scriptsize}
  \caption{Side (Simple)}~\label{fig:side_simple}
\end{subfigure}%
\begin{subfigure}[b]{0.25\textwidth}
  \centering
  \includegraphics[scale=0.40]{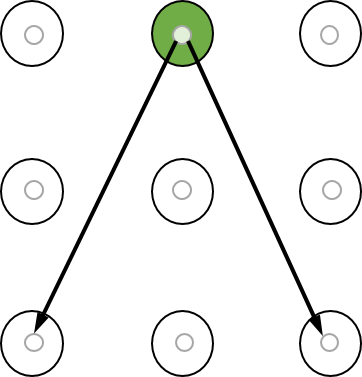}
\captionsetup{font=scriptsize}
  \caption{Side (Knight)}~\label{fig:side_knight}
\end{subfigure}%
\begin{subfigure}[b]{0.25\textwidth}
  \centering
  \includegraphics[scale=0.40]{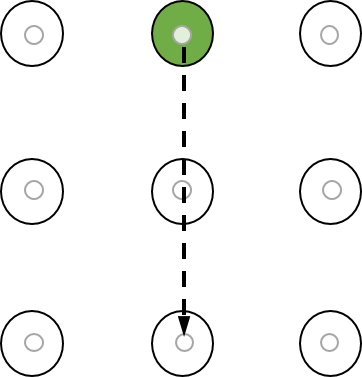}
\captionsetup{font=scriptsize}
  \caption{Side (Overlap)}~\label{fig:side_overlap}
\end{subfigure}
\caption{Connectivity rules for corner, center and side dots depicted using three types of moves {\em simple}, {\em knight} and {\em overlap}.}~\label{fig:reach}
\end{figure}

To sum up, previous studies \cite{TinPal, cho2017syspal} separately collected SysPal and TinPal patterns under different settings (online versus lab) and evaluated them using different criterias (as depicted in Table \ref{tab:cmp}). Therefore, it was important to collect TinPal and SysPal patterns under similar settings and evaluate them with the same criteria. In the current paper, we aim to close this gap. Specifically, our new contributions are as follows.
\begin{itemize}
\item We conduct a new user study with 147 participants and collect $3 \times 3$ patterns using three different SysPal policies, {\em i.e.}, 1-dot, 2-dot and 3-dot (section \ref{sec:userstudy}). We ensured that the setup used in the new study is identical to our earlier study \cite{TinPal} and the same participant is not enrolled in more than one group. Therefore, we have five different groups (Original, TinPal, 1-dot, 2-dot and 3-dot) and total 246 (99 + 147) participants with each group containing at least 48 participants. 
\item We perform detailed security analysis of patterns created across all five groups using various characteristics including pattern length, stroke length, knight moves, overlaps, direction changes, intersections, dot frequencies, segment frequencies, start point distribution and end point distribution (Table \ref{tab:cmp}). We also analyse the position of the mandated dots in the SysPal patterns. Further, we measure the guessing resistance of patterns in all five groups using Markov model based guessing algorithm. We train separate Markov models on two different datasets, one collected in our current study and one collected in \cite{harshal:guessing}. We refer the reader to section \ref{sec:securityresults} for more details.
\item We compare memorability and efficiency of patterns created in all five groups (section \ref{sec:usabilityresults}). We measure memorability in terms of the number of participants who successfully recalled the patterns and the number of login attempts during the recall phase. We measure efficiency using the pattern creation time, pattern redraw time and pattern recall time. 
\end{itemize}
\noindent
\textbf{Notations}. To make it easier when referring to a particular pattern, we label all dots arranged in $3\times 3$ grid in the row-major order, where the upper-left dot is labelled as 1 and the lower-right dot is labelled as 9 as shown in Figure \ref{fig:3X3}. A pattern is therefore represented as an ordered sequence of dots, e.g., $5213847$ (Figure \ref{fig:overlap2}). When referring to a line segment (connection) between two consecutive dots $i$ and $j$ in a pattern, we use the notation $i \rightarrow j$. For instance, the line segment between consecutive dots 3 and 8 in the pattern $5213847$ is represented as $3 \rightarrow 8$. The line segments used for creating $3\times 3$ patterns are not all similar. For instance, the line segments $1 \rightarrow 2$, $1 \rightarrow 3$, $1 \rightarrow 5$, $1 \rightarrow 6$ and $1 \rightarrow 9$ all have different physical lengths. To capture this physical length notion, we use the concept of stroke length \cite{Aviv:guessing, harshal:guessing} which is defined as the sum of Euclidean distances of all line segments within the pattern. In order to compute the Euclidean distance of a line segment, we label the upper-left dot in $3\times 3$ grid as (0,0) and the lower-right dot as (2,2). Thus, the Euclidean distance of the segment $1 \rightarrow 2$ is 1, that of $1 \rightarrow 5$ is $\sqrt{2}$, that of $1 \rightarrow 3$ is $2$, that of $1 \rightarrow 6$ is $\sqrt{5}$ and that of $1 \rightarrow 9$ is $2\sqrt{2}$.

\subsection{TinPal Motivation}
\begin{figure}[h]
\centering
\begin{subfigure}[b]{0.25\textwidth}
  \centering
  \includegraphics[scale=0.32]{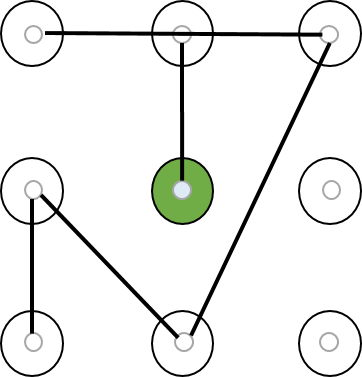}
  \captionsetup{font=scriptsize}
  \caption{Pattern $5213847$}   \label{fig:overlap2}
\end{subfigure}%
\begin{subfigure}[b]{0.25\textwidth}
  \centering
  \includegraphics[scale=0.32]{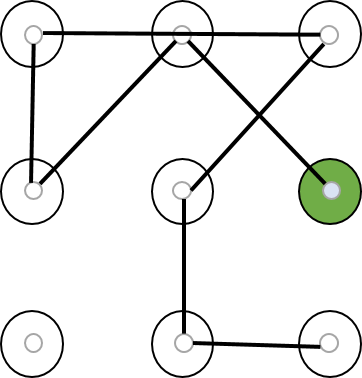}
  \captionsetup{font=scriptsize}
  \caption{Pattern $62413589$}   \label{fig:overlap1}
\end{subfigure}%
\begin{subfigure}[b]{0.25\textwidth}
  \centering
  \includegraphics[scale=0.32]{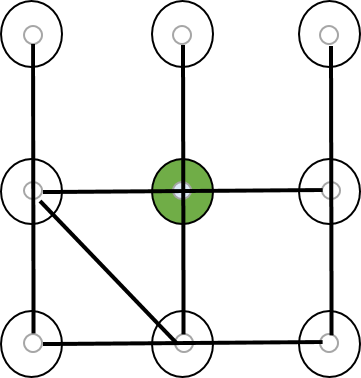}
  \captionsetup{font=scriptsize}
  \caption{Pattern $528463971$}  \label{fig:overlap_max}
\end{subfigure}%
\begin{subfigure}[b]{0.25\textwidth}
  \centering
  \includegraphics[scale=0.32]{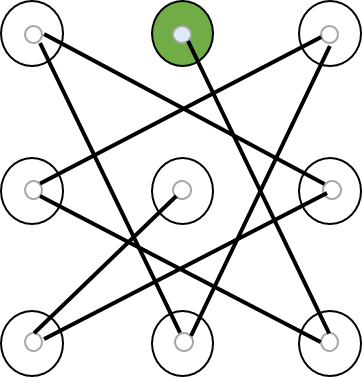}
  \captionsetup{font=scriptsize}
  \caption{Pattern $294381675$} \label{fig:knight_max}
\end{subfigure}%
\caption{Illustration of simple, knight and overlap moves.}~\label{fig:overlapex}
\end{figure}

The pattern drawing rules of a $3 \times 3$ grid are difficult to comprehend. Further, {\em all rules are never communicated to the user before or while drawing the pattern}. These rules are as follows \cite{Uellenbeck:guessing}:\\
(R1) At least 4 dots must be selected.\\
(R2) No dot can be used more than once.\\
(R3) Only straight lines are allowed.\\
(R4) One cannot jump over dots not visited before.\\
We observed that rules (R1), (R2) and (R3) are enforced by the original pattern lock interface. For instance, if the user creates a pattern with less than four dots, the existing pattern lock interface rejects the pattern and displays an error message that says ``Connect at least 4 dots. Try Again". Further, it allows two dots to be connected using straight lines only. However, {\em there is no mechanism in the original interface to inform users about rule (R4). As a result, users might not be aware of all feasible connection options}. We give an example. Consider different connection choices pertaining to dot 1 in $3 \times 3$ grid.
\begin{enumerate}
\item One can connect dot $i$ to dot $j$ that is one unit away in the horizontal direction or one unit away in the vertical direction or one unit away in both directions. We refer to such line segments as {\em simple moves}. The line segments $1 \rightarrow 2$,  $1 \rightarrow 4$ and $1 \rightarrow 5$ depicted in Figure \ref{fig:corner_simple} are examples of simple moves. Previous studies show that a majority of users employ only simple moves in their pattern \cite{Uellenbeck:guessing, Aviv:guessing, harshal:guessing, cho2017syspal}.
\item More complex connections are also possible from dot 1 which is not immediately obvious from the pattern drawing rules. One can connect dot $i$ to dot $j$ that is two units away in the horizontal (vertical) direction and one unit away in the vertical (horizontal) direction. Such connections are referred to as {\em knight moves}. For instance, one can connect dot 1 to dot 6 or dot 8 using slanted straight lines $1 \rightarrow 6$ and $1 \rightarrow 8$ as shown in Figure \ref{fig:corner_knight}.
\item  Rule (R4) on the other hand implies that dot $i$ can be connected to dot $j$ directly, if all dots along the (straight line) path are already connected. Such connections are referred to as {\em overlaps}. For instance, one can connect dot 1 to dot 3 if dot 2 is already connected. This connection can happen in two ways.
\begin{enumerate}
\item {\em Overlapping segment} : It occurs when the connection to dot 2 is immediately followed by dots 1 and 3. In this case, the line segment $2 \rightarrow 1$ is covered completely by the line segment $1 \rightarrow 3$. An example of such pattern is $5213847$  (Figure \ref{fig:overlap2}).
\item {\em Overlapping dot} : It occurs when the connection to dot 2 is followed by some other dot(s) followed by dots 1 and 3. An example of such pattern is $62413589$ (Figure \ref{fig:overlap1}).
\end{enumerate}
Similarly, one can also join $1 \rightarrow 7$  if dot $4$ is already connected or join $1 \rightarrow 9$ if dot $5$ is already connected (Figure \ref{fig:corner_overlap}). 
\end{enumerate}

Many research studies demonstrate that knight moves and overlaps play an important role in thwarting both guessing attacks \cite{Andriotis:guessing} and shoulder-surfing attacks \cite{vonZezschwitz:shouldersurfing, Sun:shouldersurfing, Song:shouldersurfing}. However, there is no mechanism in the original pattern lock interface to inform users about the feasibility of such moves. Consequently, users restrict themselves to simple moves such as $1 \rightarrow 2$, $1 \rightarrow 4$ or $1 \rightarrow 5$ to create their pattern. Spelling out these connection choices in text form is a tedious task. This problem is further aggravated since the connection choices vary depending on whether the dot is located at the corner, center or to the side in $3\times 3$ grid as shown in Figure \ref{fig:reach}. There are four corner dots $\{1, 3, 7, 9\}$, one center dot $\{5\}$ and four side dots $\{2, 4, 6, 8\}$. A corner dot can be connected directly to any of the five non-corner dots (Figure \ref{fig:corner_simple},\ref{fig:corner_knight}), center dot can be connected directly to any of the remaining eight dots (Figure \ref{fig:center}) while a side dot can be connected directly to seven dots (Figure \ref{fig:side_simple},\ref{fig:side_knight}). Because of rule (R4), a corner dot can be connected to any other corner dot if the side dot between them is already connected ((Figure \ref{fig:corner_overlap}), whereas a side dot can be connected to the remaining eighth dot if dot 5 is already connected (Figure \ref{fig:side_overlap}). 

\begin{table}[h]
  \centering
 \scriptsize
  \begin{tabular}{c r r}
\toprule
     \textit{\#Overlaps}
     & \textit{Count}
      & \textit{Percentage}\\
    \midrule
    0 & 139,880 & 35.95\% \\
    1 & 159,480 & 40.98\% \\
    2 & 69,896 & 17.96\% \\
    3 & 16,912 & 4.35\% \\
    4 & 2,688 & 0.69\% \\
    5 & 256 & 0.07\% \\
    \midrule
     \#Patterns & 389,112 & 100\% \\		
    \bottomrule
  \end{tabular}
  \caption{Theoretical distribution of overlaps in $3 \times 3$ patterns.}~\label{tab:overlapDist}
\end{table}

\begin{table}[h]
  \centering
 \scriptsize
  \begin{tabular}{c r r}
\toprule
     \textit{\#Knight moves}
     & \textit{Count}
      & \textit{Percentage}\\
    \midrule
    0	&  10,096	& 2.60\% \\
    1	&  52,120	& 13.40\% \\
    2	& 109,496	& 28.14\% \\
    3	& 117,592	& 30.22\% \\
    4	& 71,488	& 18.37\% \\
    5	& 23,704	& 6.09\% \\
    6	& 4,240	& 1.09\% \\
    7	& 376		& 0.10\% \\
    \midrule
     \#Patterns & 389,112 & 100\% \\		
    \bottomrule
  \end{tabular}
  \caption{Theoretical distribution of knight moves in $3 \times 3$ patterns.}~\label{tab:knightDist}
\end{table}

Tables \ref{tab:overlapDist} and \ref{tab:knightDist} show the distribution of overlaps and knight moves in all possible $3\times 3$ patterns (389,112) respectively. If overlaps are never used in patterns, then the search space diminishes to 139,880, {\em i.e.}, about 1/3rd of 389,112. On the other hand, if knight moves are never used in patterns, then the search space reduces to just 10,096, {\em i.e.}, about 1/40th of 389,112. The maximum number of overlaps that can occur is 5, and it is observed for instance in patterns that begin with the center dot, followed by all side dots, followed by all corner dots. One such pattern ($528463971$) is shown in Figure \ref{fig:overlap_max}. The maximum number of knight moves that can occur is 7, and it is observed in patterns that begin with a side dot, followed by alternating corner and side dots, and optionally followed by the center dot. One such pattern ($294381675$) is shown in Figure \ref{fig:knight_max}.

\subsection{Contribution}
In this work, we alter the original $3\times 3$ interface with a visual indicator mechanism to make users aware of different connection choices. Specifically, as users draw their pattern, the new $3\times 3$ interface highlights the next set of unconnected dots that can be reached from the currently connected dot, thus making users aware of all available connection options at each step during pattern creation as well as during recall. We refer to this highlighting interface as {\em TinPal}. We note that unlike SysPal \cite{cho2017syspal}, the proposed interface does not force or persuade users to connect any particular dot(s), instead it simply informs users about the set of next choices available from the currently connected dot. 

\begin{figure*}[h]
\centering
\begin{subfigure}[b]{0.20\textwidth}
  \centering
  \includegraphics[scale = 0.1]{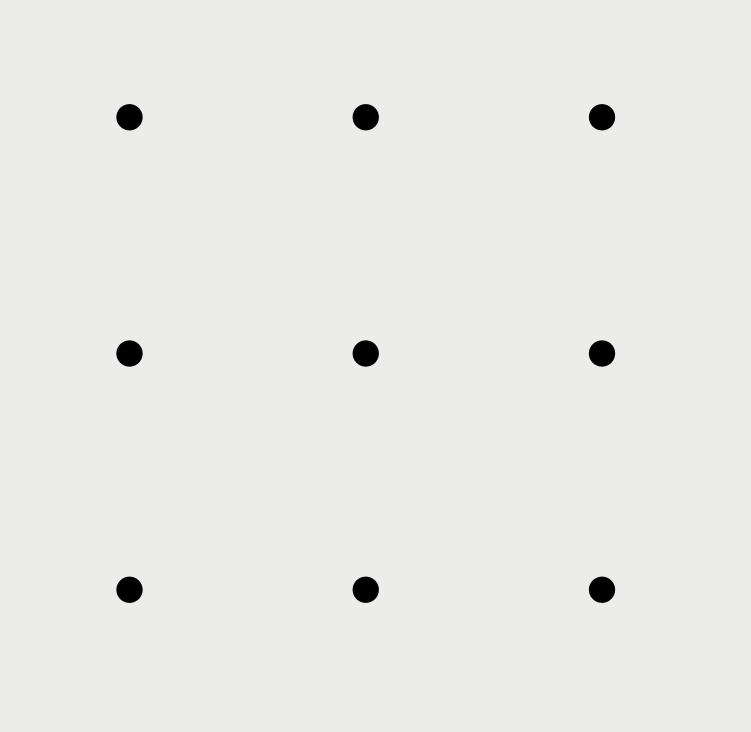}
  \captionsetup{font=scriptsize}
  \caption{$3 \times 3$ grid}~\label{fig:F0}
\end{subfigure}%
\begin{subfigure}[b]{0.20\textwidth}
  \centering
  \includegraphics[scale = 0.1]{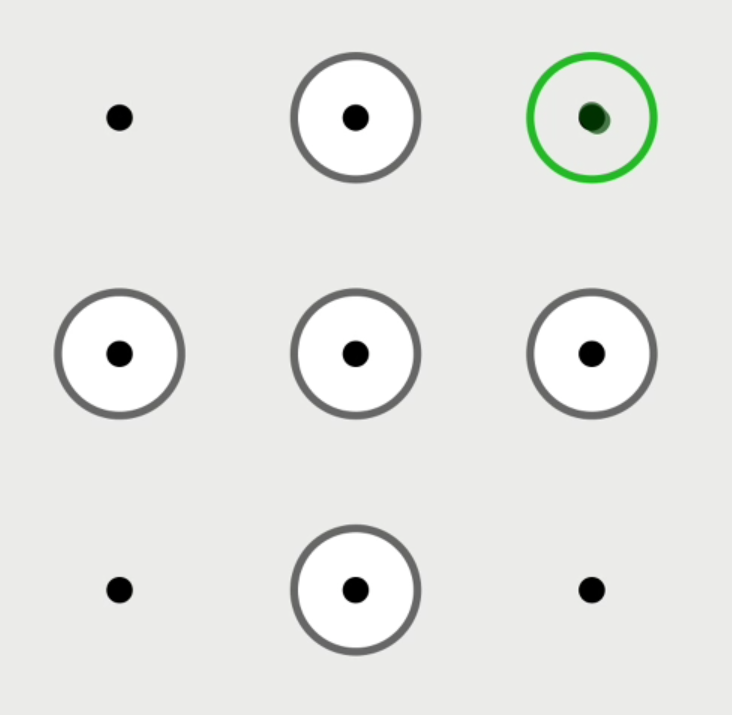}
  \captionsetup{font=scriptsize}
  \caption{Start dot 3}~\label{fig:F1}
\end{subfigure}%
\begin{subfigure}[b]{0.20\textwidth}
  \centering
  \includegraphics[scale = 0.1]{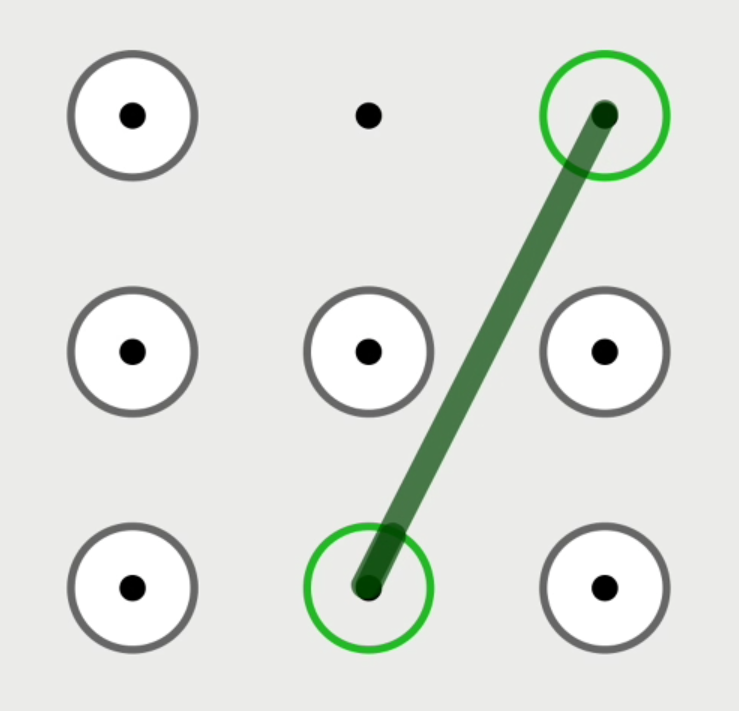}
  \captionsetup{font=scriptsize}
  \caption{Current dot 8}~\label{fig:F2}
\end{subfigure}%
\begin{subfigure}[b]{0.20\textwidth}
  \centering
  \includegraphics[scale = 0.1]{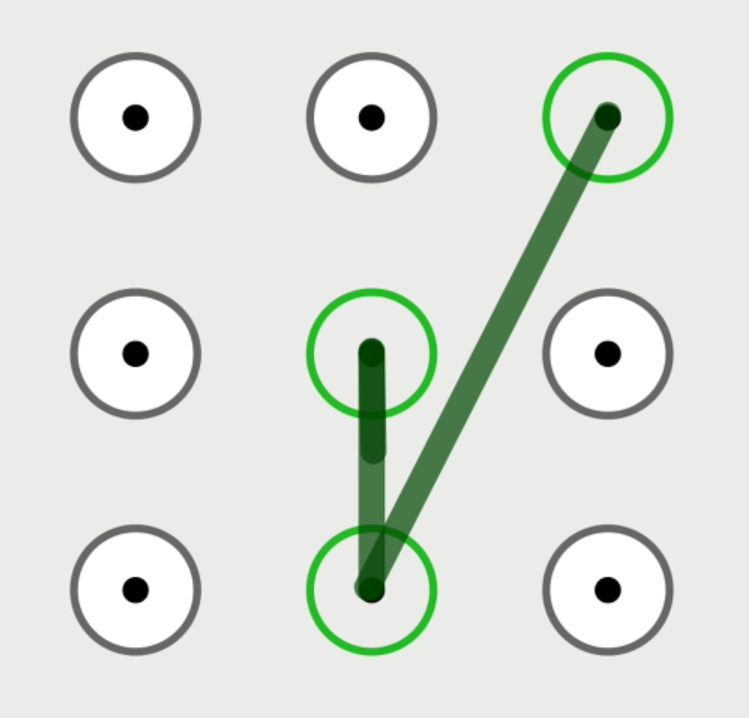}
  \captionsetup{font=scriptsize}
  \caption{Current dot 5}~\label{fig:F3}
\end{subfigure}%
\begin{subfigure}[b]{0.20\textwidth}
  \centering
  \includegraphics[scale = 0.1]{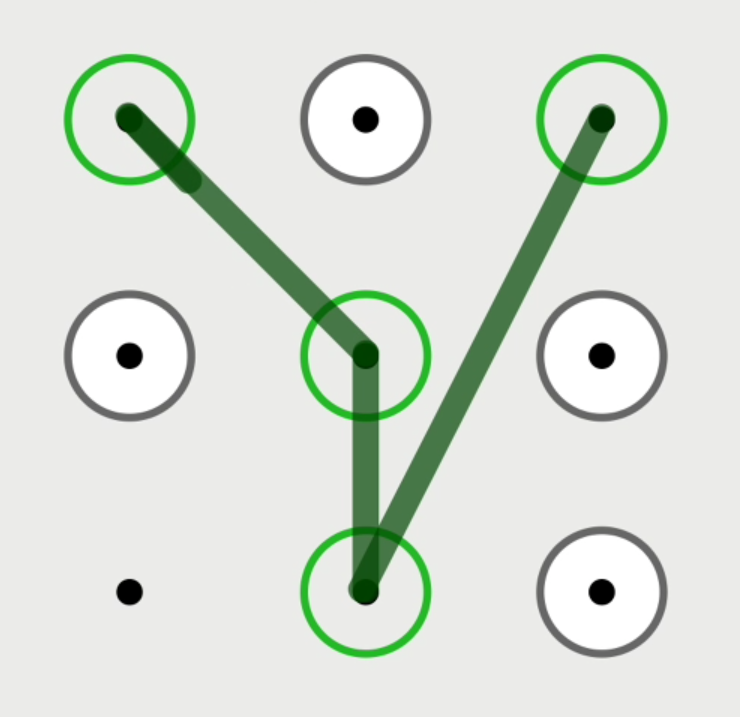}
  \captionsetup{font=scriptsize}
  \caption{Current dot 1}~\label{fig:F4}
\end{subfigure}%

\begin{subfigure}[b]{0.20\textwidth}
  \centering
  \includegraphics[scale = 0.1]{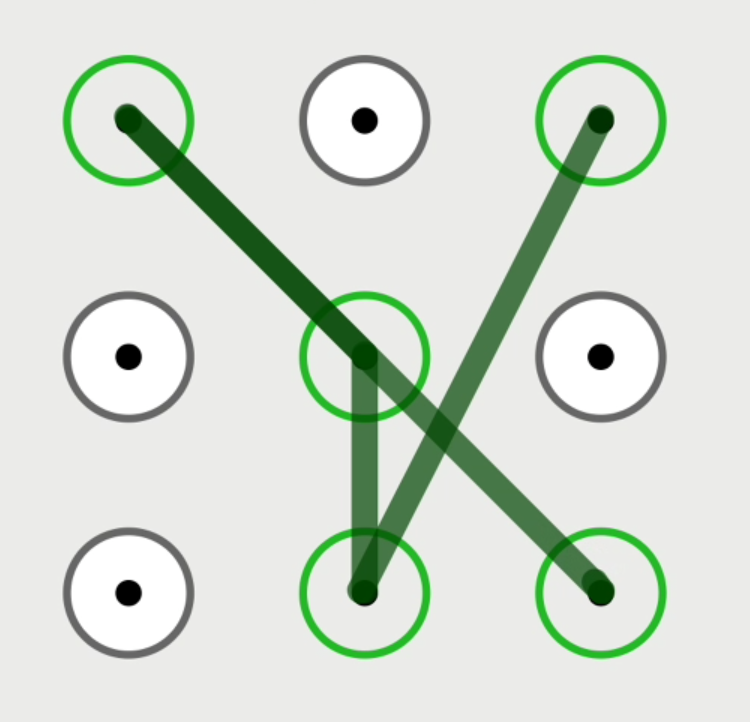}
  \captionsetup{font=scriptsize}
  \caption{Current dot 9}~\label{fig:F5}
\end{subfigure}%
\begin{subfigure}[b]{0.20\textwidth}
  \centering
  \includegraphics[scale = 0.1]{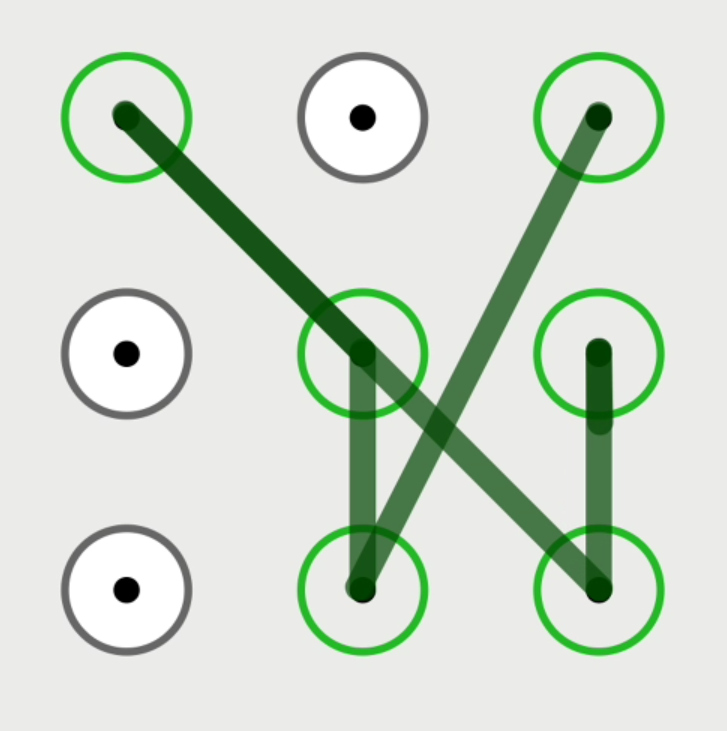}
  \captionsetup{font=scriptsize}
  \caption{Current dot 6}~\label{fig:F6}
\end{subfigure}%
\begin{subfigure}[b]{0.20\textwidth}
  \centering
  \includegraphics[scale = 0.1]{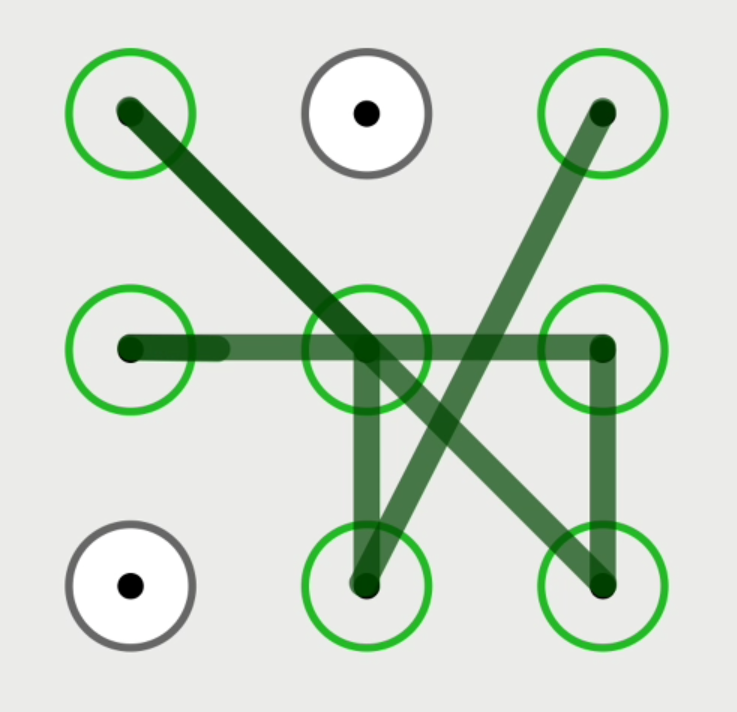}
  \captionsetup{font=scriptsize}
  \caption{Current dot 4}~\label{fig:F7}
\end{subfigure}%
\begin{subfigure}[b]{0.20\textwidth}
  \centering
  \includegraphics[scale = 0.1]{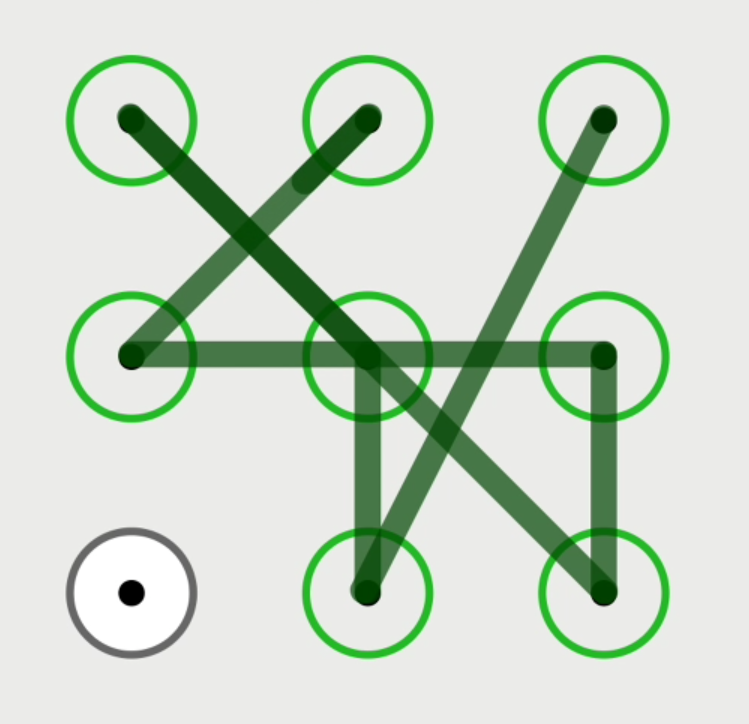}
  \captionsetup{font=scriptsize}
  \caption{Current dot 2}~\label{fig:F8}

\end{subfigure}%
\begin{subfigure}[b]{0.20\textwidth}
  \centering
  \includegraphics[scale = 0.1]{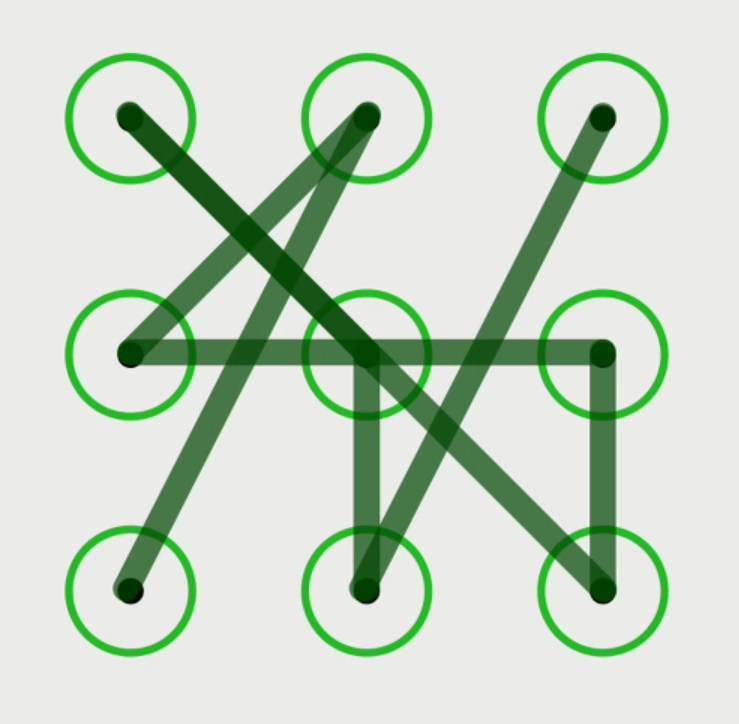}
  \captionsetup{font=scriptsize}
  \caption{Current dot 7}~\label{fig:F9}
\end{subfigure}%
\caption{A step-by-step illustration of pattern creation on TinPal. This interface highlights the next set of unconnected dots that can be visited from the currently connected dot. The connection choices are conveyed to users in real-time while the pattern is being drawn. The pattern created in the above example is 385196427.}~\label{fig:example}
\end{figure*}

The working of TinPal is illustrated in Figure \ref{fig:example}. It shows a step-by-step snapshot of TinPal while the pattern is being drawn by the user. 
\begin{enumerate}
\item Suppose that the user starts her pattern with dot 3. From dot 3, the user can visit any non-corner dot $\{2,4,5,6,8\}$. However, the user cannot visit dot 1 as dot 2 is still unconnected or dot 7 as dot 5 is unconnected or dot 9 as dot 6 is unconnected (R4). Hence, only non-corner dots $\{2,4,5,6,8\}$ are highlighted by TinPal as shown in Figure \ref{fig:F1}. 
\item Next, the user visits dot 8 (knight move). From dot 8, the user can visit any dot except dot 2 as dot 5 is still not connected (R4). Therefore, the set of unconnected dots $\{1,4,5,6,7,9\}$ are highlighted as shown in Figure \ref{fig:F2}. 
\item Subsequently, the user connects dot 5 (simple move). From dot 5, the user can visit any unconnected dot $\{1,2,4,6,7,9\}$ as highlighted in Figure \ref{fig:F3}.
\item The user chooses dot 1 (simple move). From there, the user can visit any unconnected non-corner dot $\{2,4,6\}$. Since dot 5 is already connected, the user can also visit dot 9 (R4). However, the user cannot visit dot 7 as dot 4 is not yet connected. Hence, the set $\{2,4,6,9\}$ is highlighted by TinPal as depicted in Figure \ref{fig:F4}.
\item Next choice is dot 9 (overlap). From there, the user can visit any unconnected non-corner dot $\{2,4,6\}$. Further, since dot 8 is already connected, the user can now visit dot 7 (R4). Hence, the set $\{2,4,6,7\}$ is highlighted as indicated in Figure \ref{fig:F5}.
\item Next, the user visits dot 6 (simple move). From dot 6, in addition to unconnected dots 2 and 7, the user can also visit dot 4 since dot 5 is already connected (R4). Therefore, the set $\{2,4,7\}$ is highlighted in Figure \ref{fig:F6}. 
\item Next choice is dot 4 (overlap). From there, the user can go to either dot 2 or dot 7 as highlighted in Figure \ref{fig:F7}.
\item The user connects dot 2 (simple move). Now, the only choice available is dot 7 which is highlighted in Figure \ref{fig:F8}.
\item Finally, the user connects dot 7 (knight move) and the pattern $385196427$ is recorded as shown in Figure \ref{fig:F9}.
\end{enumerate}

Our contributions are as follows:
\begin{itemize}
\item We enhance the original $3\times 3$ interface with a visual indicator mechanism to help users choose more diverse patterns. Specifically, the new interface, referred to as TinPal highlights the set of reachable dots from the currently connected dot, thus making pattern drawing rules more salient to users. The highlighting mechanism works in real-time while the pattern is being drawn.
\item We also describe an algorithm that takes the currently connected dot as an input and outputs the next set of unconnected dots that can be visited from the current dot in $3 \times 3$ grid. 
\item We evaluate the impact of our visual indicator mechanism with a comparative user study involving 246 (99+147) participants. Specifically, we measure the usability and security of $3 \times 3$ patterns created using the original interface, three SysPal policies (1-dot, 2-dot and 3-dot), and TinPal. Our results show that participants in the TinPal group used significantly large number of dots and complex features such overlaps, knight moves and direction changes to create their pattern as compared to the other groups.
\item We also estimate the guessing resistance of patterns created in all five groups using a Markov model based guessing algorithm. Within first 20 attempts, the guessing algorithm cracked 12\% patterns in the Original group, 10\% patterns in the 1-dot group, 8\% patterns in the 2-dot group, 20\% patterns in the 3-dot group, but none (0\%) in the TinPal group. Another Markov model trained on \cite{harshal:guessing} dataset cracked 12.24\% patterns in the Original group,  8\% patterns in the 1-dot group, 6.12\% patterns in the 2-dot group,  18.75\% patterns in the 3-dot group, but only 4\% patterns in the TinPal group.
\end{itemize}
The organization of this paper is as follows. First, we provide a brief overview of graphical passwords and review the work related to the security of the pattern lock scheme. Subsequently, we describe the design choices we made along with the working mechanism for the proposed interface, TinPal. Next, we describe the user study, and present security and usability results. Finally, we discuss the future work.

\section{Related Work}
Graphical passwords are considered to be promising alternative to textual passwords since many research studies \cite{shepard1967recognition,Paivio1968,yuille1983imagery} show that graphical information is easier to remember than textual information. Based on the difficulty of retrieving graphical information from the visual memory, graphical schemes are broadly divided into three categories \cite{Biddle:survey}: {\em recognition-based}, {\em cued recall-based} and {\em recall-based}.
A typical example of the recognition-based scheme is {\em PassFaces} \cite{PassFaces} in which the user selects a face from a set of nine image faces during registration and correctly recognizes the face among a set of decoy image faces during login. To attain the desired level of security, there are multiple iterations in {\em PassFaces}, with each iteration employing a different set of nine images. An example of the cued recall-based scheme is {\em Pass-points} \cite{PassPoints} in which the user selects a sequence of points on a system-assigned image. During login, this image acts as a memory cue that helps the user to recall the selected points in the correct order. An example of the recall-based scheme is {\em Pass-Go} \cite{PassGo} in which the user draws one or more strokes on a $n\times n$ grid. The $3\times 3$ pattern lock scheme is a special case of Pass-Go, where $n$ is set to 3 so that the grid could fit on small screen of smartphones. 

\subsection{Android Pattern Lock}
Androids' pattern lock scheme is the most widely deployed graphical based authentication system on smartphones, hence its security has been well studied in the literature. In 2010, Aviv {\em et al.} demonstrated that it is possible to reconstruct the user's entire pattern from the oily traces left on the screen \cite{Aviv:sidechannel}. This attack is popularly known as {\em smudge attack}. Later in 2013, Andriotis {\em et al.} showed that it is possible to recover the entire pattern even from the partial traces by exploiting users' biased choices \cite{Andriotis:sidechannel}. They surveyed 144 participants and found that more than 50\% of the participants started their pattern from the upper-left dot. Further analysis revealed that 18.75\% of the participants used the path $1 \rightarrow 2 \rightarrow 3$ in their pattern. These observations along with the partial physical traces reduced the search space drastically.

In 2013, Uellenbeck {\em et al.} found that users' pattern choices are biased and prone to guessing attacks \cite{Uellenbeck:guessing}. They collected approximately 2900 patterns from 584 participants on 5 different layouts. Their analysis revealed that most users create $3\times 3$ patterns with horizontal (e.g., $1 \rightarrow 2$) and vertical strokes (e.g., $1 \rightarrow 4$) which reduces the security of $3\times 3$ patterns to just 3-digit random PINs. 
They also found that patterns created on a circular interface (eight dots on the circumference and one dot in the center) were more secure than $3\times 3$ patterns. Similar results were reported by Tupsamudre {\em et al.} \cite{harshal:guessing}. They proposed a different circular interface, called Pass-O, with all nine dots on the circumference and evaluated it with a large-scale study (21,053 users, 123,190 patterns). However, both studies \cite{Uellenbeck:guessing, harshal:guessing} focused on security and lacked rigorous usability evaluation. In 2016, Aviv {\em et al.} \cite{Aviv:guessing} studied the security of $4 \times 4$ patterns and found that the majority of them are just extended versions of $3\times 3$ patterns, and hence insecure. 

Most graphical password schemes are susceptible to shoulder-surfing attacks \cite{Biddle:survey}. In 2015, Zezschwitz {\em et al.} performed a systematic evaluation of the shoulder-surfing susceptibility of the pattern lock scheme  \cite{vonZezschwitz:shouldersurfing}. They found that line visibility, pattern length, number of knight moves, number of overlaps and number of intersections (refer to section \ref{definitions} for definitions) play an important role in thwarting shoulder-surfing attacks. More recently, Aviv {\em et al.} showed that different viewing angles, hand positions and phone sizes can also affect the efficacy of shoulder-surfing attacks \cite{Aviv:shoulder}. In order to encourage users to create visually complex patterns, various strength meters have been proposed in the literature \cite{Andriotis:guessing, Sun:shouldersurfing, Song:shouldersurfing}. These strength meters typically nudge users to draw longer, and complex patterns containing overlaps, knight moves, direction changes and intersections. However, the impact of these strength meters can be limited if users are not aware of all possible connection choices while creating their pattern.

Recently, Ye {\em et al.} demonstrated a side-channel attack that recovered 95\% of the $3 \times 3$ patterns from a video recorded using a smartphone within just five attempts  \cite{Ye:2018:VAA:3232648.3230740}. Further, they could uncover 97.5\% of the complex patterns in just one attempt. However, we note that for such an attack to succeed, certain events are required to happen. First, the user unlocks her device in public. During the same time, the attacker who is at a distance of 2 meters from the user records the entire unlocking process without raising any suspicion. Further, the attack requires that both the user's fingertip and part of the target device surface are visible during recording. Subsequently, the attacker processes the recorded video offline and infers a set of candidate patterns using a computer vision algorithm. Finally, the attacker gets physical access to the target device for a short duration in order to try the inferred candidate patterns. In our work, we use a more practical threat model as described in \cite{Uellenbeck:guessing, Aviv:guessing, harshal:guessing, cho2017syspal}, {\em i.e.}, {\em we evaluate the security of $3 \times 3$ patterns against a dictionary based guessing attack where an attacker gets access to an Android device for a short duration, and attempts most probable (popular) patterns first to unlock it}. Interestingly, when \cite{Ye:2018:VAA:3232648.3230740} asked participants to infer complex patterns by watching the pattern unlocking videos, the success rate was less than 10\% in five attempts. The participants could adjust the video speed and watch video multiple times for a period of 10 minutes.

To improve the security of the pattern lock scheme, Siadati {\em et al.} proposed a persuasive interface that suggests a random starting point to the user \cite{siadati:persuasive} while Cho {\em et al.} proposed three SysPal policies that mandate users to include one, two or three randomly assigned dots at any position in the pattern \cite{cho2017syspal}. However, we note that the use of system-assigned random dots does not ensure that the resulting patterns will exhibit secure characteristics such as knight moves and overlaps simply because users may not be aware about the feasibility of such connections. In fact, Cho {\em et al.} found that in all three SysPal policies, the most frequently used segments were $1 \rightarrow 2$ and $2 \rightarrow 3$, and $i \rightarrow i+1$ was more frequently used than $i+1 \rightarrow i$ for all $i = 1, 2, 4, 5, 7, 8$, implying that most patterns were drawn from left to right. 

On the other hand, TinPal, the highlighting interface proposed in our earlier work \cite{TinPal}, does not mandate users to connect any specific dot(s), it just informs them about the set of reachable dots in each step during pattern creation as well as during recall. Our study results show that patterns drawn on TinPal were composed using significantly longer strokes and large number of knight moves, overlaps and direction changes as compared to those created using 1-dot, 2-dot and 3-dot SysPal policies. Moreover, the proportion of SysPal patterns cracked in 20 attempts were: 10\% (1-dot), 8\% (2-dot) and 20\% (3-dot). These numbers suggest (also stated in \cite{cho2017syspal}) that {\em mandating too many points could potentially reduce the overall password space}. Further, none of the TinPal patterns were cracked in 20 attempts. 

\section{Interface Design and Mechanics}
We employ two design principles, namely, {\em visibility} and {\em consistency} \cite{rogers2011interaction}, to enhance the original $3 \times 3$ interface. According to the visibility principle, the system should have proper mechanisms to convey to users what actions are possible. The visual indicator mechanism in TinPal makes the set of available choices visible to users. Specifically, it highlights the next set of unconnected dots that can be visited from the currently connected dot to help users choose diverse $3 \times 3$ patterns. This highlighting of dots happens in real-time while the pattern is being drawn.

The {\em consistency} principle on the other hand makes the interface intuitive to use. The highlighting of dots in TinPal happens not only during pattern creation, but also during recall. This eliminates the confusion since the behaviour of the interface is consistent during creation and recall. Further, the highlighting of dots could also serve as cue when the user is trying to retrieve the pattern from her memory. However, we note that the highlighting of nodes can potentially leak the information about dots used in the pattern. Therefore, we can also provide an option to turn off the highlighting mode once the pattern is created.

TinPal also retains the {\em feedback} mechanism of the original interface, thereby enforcing the minimum pattern length requirement on $3 \times 3$ grid \cite{Uellenbeck:guessing}. For instance, if the user connects less than 4 dots, it displays the {\em feedback} message, ``Connect at least 4 dots. Try again". Further, as only unconnected dots are highlighted by TinPal, the user is informed that a dot can be connected only once. 

\subsection{Mechanics}
\begin{figure*}[h]
\centering
\begin{subfigure}[b]{0.32\textwidth}
  \centering
  \includegraphics[scale=0.4]{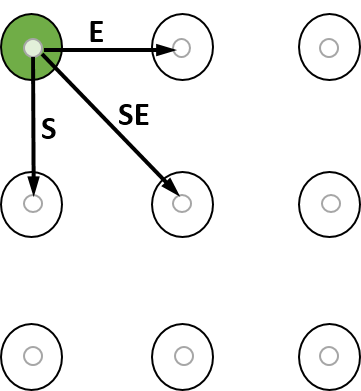}
  \captionsetup{font=scriptsize}
  \caption{Neighbours of dot 1} \label{fig:n1}
\end{subfigure}%
\begin{subfigure}[b]{0.32\textwidth}
  \centering
  \includegraphics[scale=0.4]{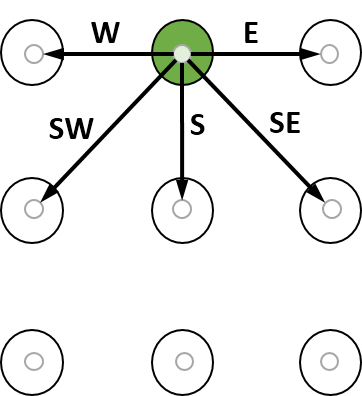}
  \captionsetup{font=scriptsize}
  \caption{Neighbours of dot 2}   \label{fig:n2}
\end{subfigure}%
\begin{subfigure}[b]{0.32\textwidth}
  \centering
  \includegraphics[scale=0.4]{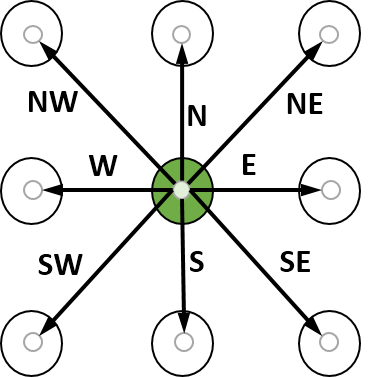}
  \captionsetup{font=scriptsize}
  \caption{Neighbours of dot 5}  \label{fig:n5}
\end{subfigure}
\caption{Neighbours of dot 1 (corner), dot 2 (side) and dot 5 (center).}~\label{fig:neighbour}
\end{figure*}

Now, we present an algorithm that takes the currently connected dot $d$ as an input and outputs the next set of unconnected dots $R$ that are reachable from the current dot $d$. This algorithm is invoked whenever the user connects a new dot on TinPal. The set of dots returned by this algorithm are highlighted on the interface. To simplify the algorithm, we define a neighbourhood relation on $3\times 3$ grid. We say that dot $j$ is a neighbour of dot $i$ on $3\times 3$ grid if $j$ is the nearest dot in any of the eight directions (north, north-east, east, south-east, south, south-west, west and north-west). Figure \ref{fig:neighbour} depicts the neighbourhood of dots 1, 2 and 5 in all possible directions. Dot 1 being a corner dot has 3 neighbours, {\em i.e.}, dots 2 (east), 4 (south) and 5 (south-east). Dot 2 being a side dot has 5 neighbours, {\em i.e.}, dots 1 (west), 3 (east), 4 (south-west), 5 (south), and 6 (south-east), whereas the center dot 5 has all other dots as its neighbours.

\begin{table}[h]
  \centering
\scriptsize
  \begin{tabular}{c | c c c c c c c c}
\toprule
     \textit{Dot} & \textit{N} $\uparrow$ & \textit{NE} $\nearrow$ & \textit{E}$\rightarrow$ & \textit{SE} $\searrow$ & \textit{S} $\downarrow$ & \textit{SW} $\swarrow$& \textit{W} $\leftarrow$ & \textit{NW} $\nwarrow$\\  
\midrule
1& $\perp$ & $\perp$ &2&5&4& $\perp$ & $\perp$ & $\perp$\\ 
2& $\perp$ & $\perp$ &3&6&5& 4&1& $\perp$ \\
3& $\perp$ & $\perp$ & $\perp$ & $\perp$ &6& 5 &2& $\perp$\\ 
4&1&2&5&8&7& $\perp$ & $\perp$ & $\perp$\\ 
5&2&3&6&9&8& 7&4&1\\
6&3& $\perp$ & $\perp$ & $\perp$ &9& 8&5&2\\
7&4&5&8& $\perp$ & $\perp$ & $\perp$ & $\perp$ & $\perp$\\ 
8&5&6&9& $\perp$ & $\perp$ & $\perp$&7&4\\
9&6& $\perp$ & $\perp$ & $\perp$ & $\perp$ & $\perp$&8& 5\\ 
    \bottomrule
  \end{tabular}
  \caption{Neighbourhood table for $3\times 3$ grid.}~\label{tab:neighbour}
\end{table}

\begin{figure*}[h]
\centering
\begin{subfigure}[b]{0.20\textwidth}
  \centering
  \includegraphics[scale = 0.1]{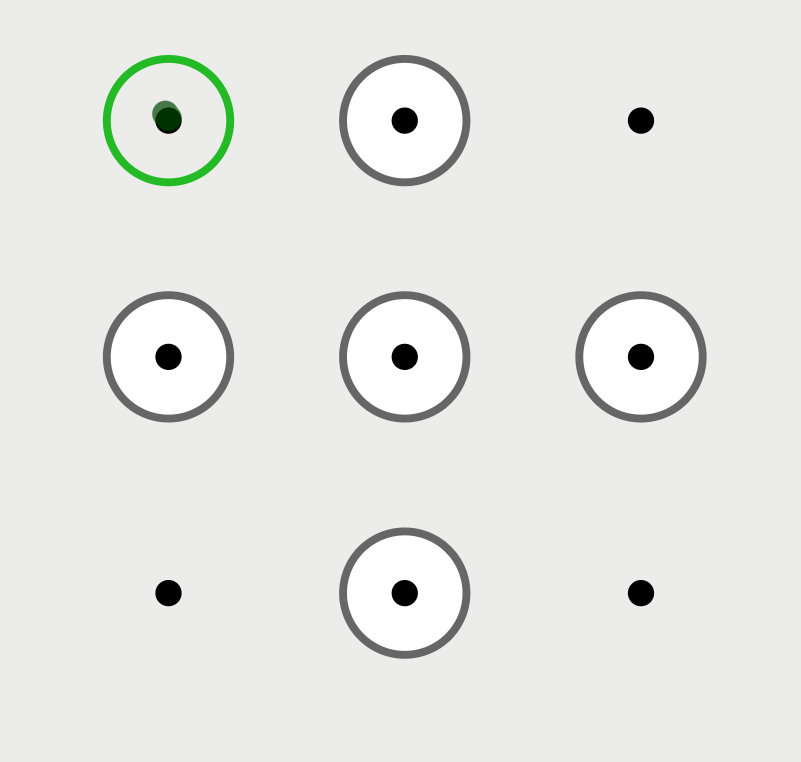}
  \captionsetup{font=scriptsize}
  \caption{Start dot 1}~\label{fig:algofig1}
\end{subfigure}%
\begin{subfigure}[b]{0.20\textwidth}
  \centering
  \includegraphics[scale = 0.1]{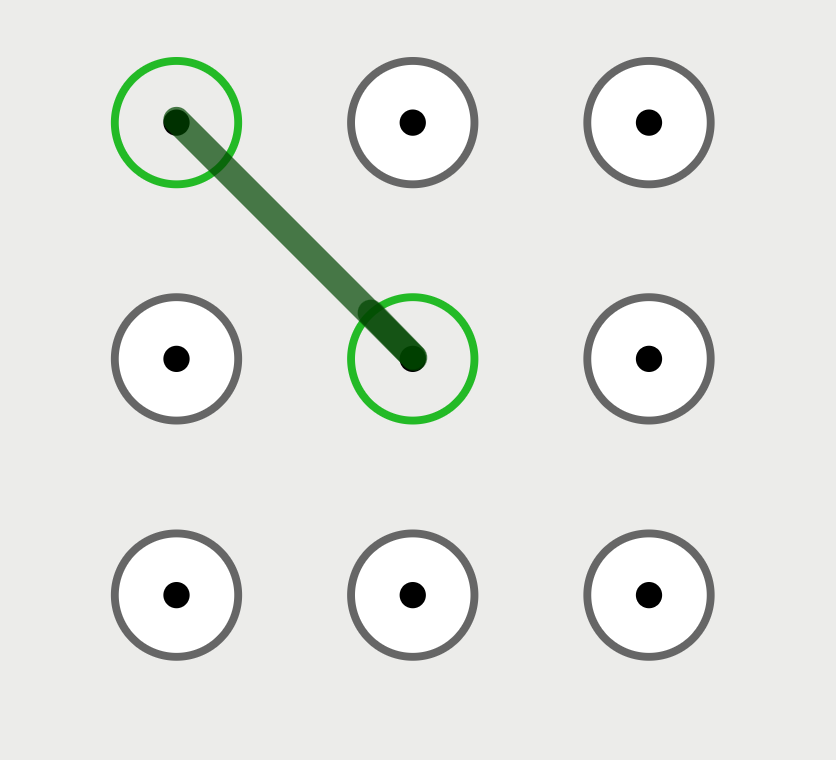}
  \captionsetup{font=scriptsize}
  \caption{Current dot 5}~\label{fig:algofig2}
\end{subfigure}%
\begin{subfigure}[b]{0.20\textwidth}
  \centering
  \includegraphics[scale = 0.1]{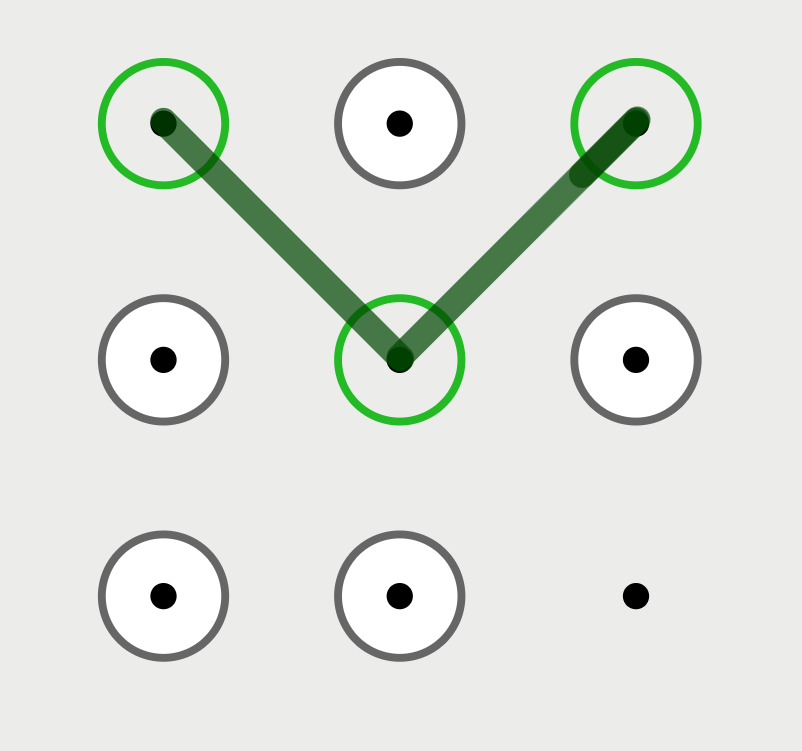}
  \captionsetup{font=scriptsize}
  \caption{Current dot 3}~\label{fig:algofig3}
\end{subfigure}%
\begin{subfigure}[b]{0.20\textwidth}
  \centering
  \includegraphics[scale = 0.1]{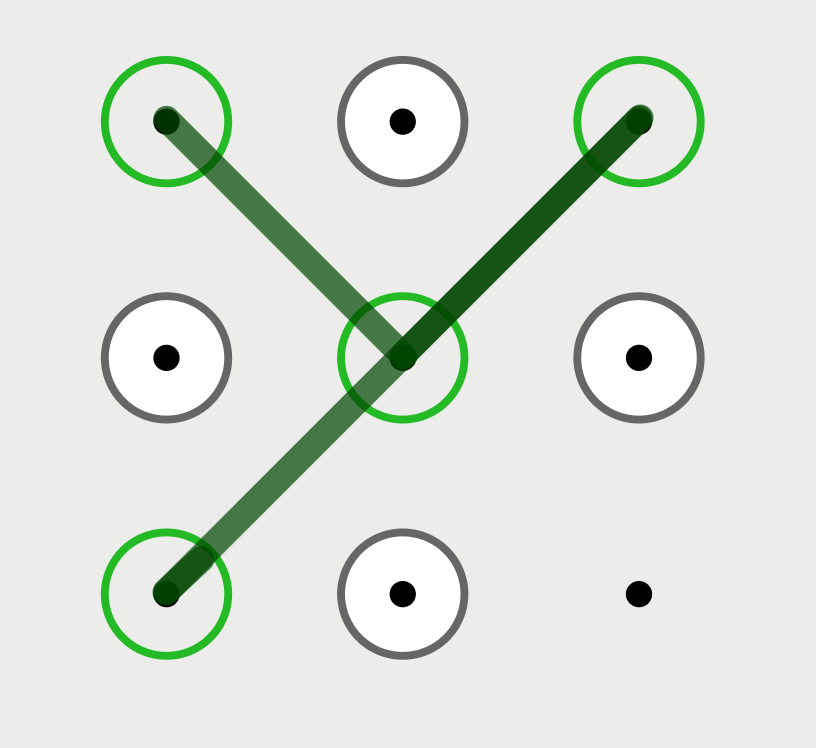}
  \captionsetup{font=scriptsize}
  \caption{Current dot 7}~\label{fig:algofig4}
\end{subfigure}%
\begin{subfigure}[b]{0.20\textwidth}
  \centering
  \includegraphics[scale = 0.1]{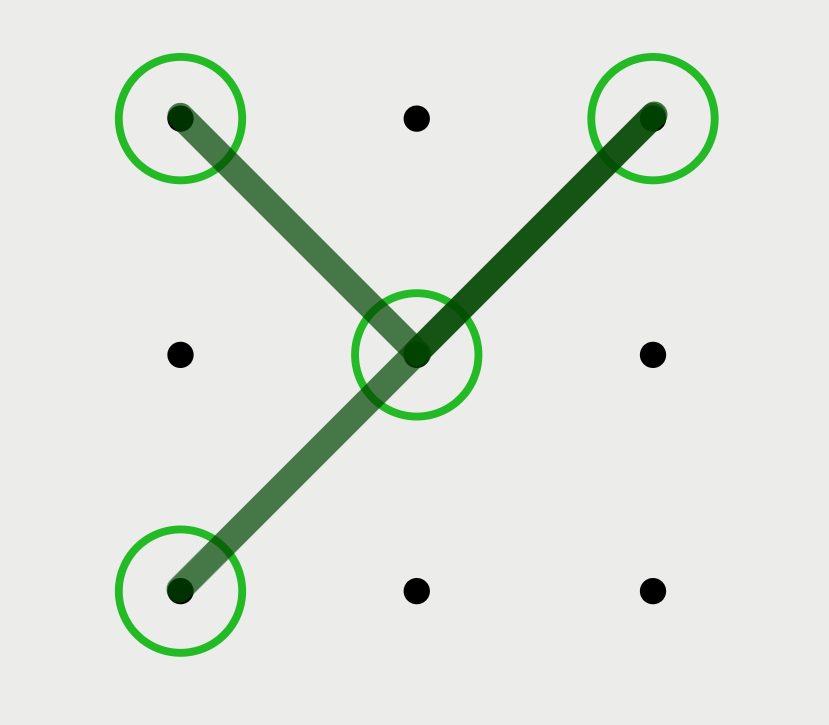}
  \captionsetup{font=scriptsize}
  \caption{Pattern $1537$}~\label{fig:algofig5}
\end{subfigure}%
\caption{A step-by-step illustration of pattern creation on TinPal. This interface highlights the next set of unconnected dots that can be reached from the currently connected dot as computed by our elimination algorithm. The connection choices are conveyed to users in real-time while the pattern is being drawn. The pattern created in the above example is 1537.}~\label{fig:algoexample}
\end{figure*}

Algorithm 1 starts by adding all unconnected dots to the set $R$ (line 4). Then it eliminates those dots from $R$ that are not reachable from the current dot $d$ (lines 6-12). For elimination, the algorithm employs Table \ref{tab:neighbour} which contains the neighbours of all 9 dots in $3\times 3$ grid. The table has nine rows corresponding to each dot and eight columns corresponding to each direction. If a dot $i$ does not have any neighbour in a particular direction $\delta$, then the corresponding entry in the neighbourhood table is marked with symbol $\perp$ (indicating empty). To eliminate the unreachable dots from the set $R$, the algorithm scans the entire $d^{th}$ row, {\em i.e.}, all 8 neighbours of the current dot $d$ in the neighbourhood table (lines 6-12). If the entry in the direction $\delta$ is marked with $\perp$ then the algorithm does not take any action. Otherwise, the algorithm checks if the neighbouring dot $e$ in the direction $\delta$ is already connected (line 9). If it is not, then the algorithm eliminates the neighbour of the neighbouring dot $e$ in the direction $\delta$ from $R$ (line 10). Finally, the set $R$ consisting of reachable dots from the current dot $d$ is returned for highlighting on $3 \times 3$ grid (line 13). We illustrate the working of this elimination algorithm with an example (Figure \ref{fig:algoexample}).

\begin{algorithm}[t]
\caption{Elimination Algorithm}~\label{Algo}
\begin{algorithmic}[1]
\Procedure{$ Elimination \ Algorithm$}{} \\
\textbf{Input:} Current dot $d$, set $U$ of unconnected dots and $9 \times 8$ neighbourhood table\\
\textbf{Output:} Set $R$ of unconnected dots that can be reached from the current dot $d$
\State $R \gets$ $U$
\State //scan the entire $d^{th}$ row (all 8 neighbours) in the neighbourhood table
\For{$\delta \in \{1,\ldots ,8\}$}
    \State $e \gets$ table[$d$][$\delta$] 
    \State //if neighbouring dot $e$ in direction $\delta$ is unconnected then eliminate neighbour of $e$ in direction $\delta$
    \If {$e \neq \perp$ $\&\&$ !isConnected($e$)}
    	\State $R \gets R \ \setminus$ \{table[$e$][$\delta$]\}
    \EndIf
\EndFor
\State return $R$
\EndProcedure
\end{algorithmic}
\end{algorithm} 

\begin{enumerate}
\item Suppose that the user starts her pattern by connecting dot 1. Subsequently, the elimination algorithm is invoked to determine the set of unconnected dots $R$ reachable from the current dot $1$. The algorithm begins by adding all unconnected dots $\{2,3,4,5,6,7,8,9\}$ to the set $R$ (line 4). The algorithm then scans the first row (corresponding to dot 1) of the neighbourhood table to identify and remove the unreachable dots in the set $R$ (lines 6-12). As its neighbour in the east (dot $2$) is still unconnected, the algorithm removes dot $3$ (which lies further to the east of dot $2$) from the set $R$ (line 10). This elimination occurs due to the overlap rule (R4) which says that dot 1 cannot be directly connected to dot 3 as dot 2 is still unconnected. Similarly, as the south neighbour dot $4$ is yet to be connected, the algorithm removes dot $7$ (which lies further to the south of dot $4$) from the set $R$. Also, as the south-east neighbour dot $5$ is unconnected, the algorithm removes dot $9$ (which lies further to the south-east of dot $5$) from the set $R$. Therefore, the set $R = \{2,4,5,6,8\}$ is determined to be reachable and returned (line 13) for highlighting on $3 \times 3$ grid (Figure \ref{fig:algofig1}).

\item Subsequently, the user chooses dot 5. The algorithm adds all unconnected dots $\{2,3,4,6,7,8,9\}$ to the set $R$ (line 4). The algorithm then scans the fifth row of the neighbourhood table to identify and remove the unreachable dots (from dot 5) from the set $R$ (lines 6-12). As all dots are reachable from dot 5, none of the dots is eliminated from the set $R$ and $\{2,3,4,6,7,8,9\}$ is returned (line 13) for highlighting purpose (Figure \ref{fig:algofig2}).

\item Next, the user chooses dot 3. The algorithm adds all unconnected dots $\{2,4,6,7,8,9\}$ to the set $R$ (line 4). The algorithm then scans the third row of the neighbourhood table to identify and remove the unreachable dots (from dot 3) from the set $R$ (lines 6-12). As its neighbour in the south, dot $6$ is still unconnected, the algorithm removes dot $9$ (which lies further to the south of dot $6$) from the set $R$ (line 10). This elimination occurs due to the overlap rule (R4) which says that dot 3 cannot be directly connected to dot 9 as dot 6 is still unconnected. As dot $5$ which lies to the south-west of dot $3$ is already connected, dot $9$ which lies further to the south-west of dot $5$ is retained in the set $R$. Therefore, the set $R = \{2,4,6,7,8\}$ is returned (line 13) for highlighting purpose (Figure \ref{fig:algofig3}).

\item The user chooses dot 7. The algorithm adds all unconnected dots $\{2,4,6,8,9\}$ to the set $R$ (line 4). The algorithm then scans the seventh row of the neighbourhood table to identify and remove the unreachable dots (from dot 7) from the set $R$ (lines 6-12). As its neighbour in the east, dot $8$ is still unconnected, the algorithm removes dot $9$ (which lies further to the east of dot $8$) from the set $R$ (line 10). This elimination occurs due to the overlap rule (R4) which says that dot 7 cannot be directly connected to dot 9 as dot 8 is still unconnected. Therefore, the set $R = \{2,4,6,8\}$ is returned (line 13) for highlighting on $3 \times 3$ grid (Figure \ref{fig:algofig4}).

\item Finally, the user releases her finger and the pattern $1537$ is recorded as shown in Figure \ref{fig:algofig5}.
\end{enumerate}

\section{User Study}\label{sec:userstudy}
To evaluate the impact of TinPal on users' pattern choices, we conducted two separate user studies. The first study was conducted in lab during August 2017. Participants were recruited using internal mailing lists within our organization. A total of 99 users responded to our e-mail and completed the first study. Out of 99 participants, 49 were randomly assigned to the Original group and 50 were assigned to the TinPal group. 
\begin{itemize}
\item \textbf{Original}. Participants in this group created their pattern on the original $3\times 3$ interface.
\item \textbf{TinPal}. Participants in this group created their pattern on TinPal which highlights the next set of available dots that can be visited from the currently connected dot.
\end{itemize}
\noindent
The second study was also conducted in lab during January 2018. The purpose of this study was to collect $3 \times 3$ patterns created using three different SysPal interfaces \cite{cho2017syspal}. We did not invite participants who already participated in the first study. A total of 147 users responded to our email and completed the second study. Out of 147 participants, 50 were randomly assigned to the 1-dot group, 49 to the 2-dot group and 48 to the 3-dot group.
\begin{itemize}
\item \textbf{1-dot}. Participants in this group were required to include one system-assigned dot in their $3 \times 3$ pattern.
\item \textbf{2-dot}. Participants in this group were required to include two system-assigned dots in their $3 \times 3$ pattern.
\item \textbf{3-dot}. Participants in this group were required to include three system-assigned dots in their $3 \times 3$ pattern.
\end{itemize}
Therefore, a total of 246 participants enrolled and completed our studies, with each group containing at least 48 participants. 

A research study \cite{Aviv:collection} found statistically significant difference between patterns collected on the mobile device of participants versus patterns collected using other collection methods. Therefore, we opted for pattern collection on the participant's own mobile device. To make our study available across all mobile devices, we created an HTML/Javascript web application using Java J2EE platform. Thus, participants could participate in the study using any standard web-browser on their mobile device without having to install any additional software. We simulated the look and feel of Android pattern lock as closely as possible. Participants were given a pen and a chocolate worth \$2 for completing the study. There is no IRB in our organization for approving studies involving human subjects. However, we took all necessary steps in order to be compliant with privacy regulations. The data was collected anonymously after obtaining consent from the participants. Further, the collected data was used for research purpose only.

\subsection{Participants}
The demographics of the participants are summarized in Table \ref{tab:demo}. Majority of participants were between 20 and 30 years of age, and right handed. The proportion of male participants in the experiment was slightly higher than that of the female participants. Also the proportion of participants with a background in Computer Science (CS)/Information Technology (IT)/Security exceeded those with no such background. Further, all participants had at least an undergraduate degree and belonged to the same nationality. We found no statistically significant difference in gender, handedness, age or background of participants across five groups ($p > 0.01$, corrected two-tailed Fischer's Exact Test).

During experiments, we also asked participants questions related to the mobile device they own and the screen locks they ever used (if any). These device and lock statistics are also presented in Table \ref{tab:demo}. A large fraction of participants used Android phones. Further, majority of them were familiar with the Android pattern lock scheme. We found no statistically significant difference in the mobile OS experience and pattern lock familiarity across five groups  ($p > 0.01$, corrected two-tailed Fischer's Exact Test).

\begin{table}[h]
  \centering
 \scriptsize
  \begin{tabular}{l r r r r r}
\toprule
   \textit{}
    & \textit{Original}
      & \textit{TinPal}
      & \textit{1-dot}
      & \textit{2-dot}
      & \textit{3-dot}\\
   \midrule
   \textbf{Gender} & & & & &\\
   Male & 51.02\% & 56\% & 64\% & 51.02\% & 58.33\%\\
   Female & 48.98\% &  44\% & 36\% & 48.98\% & 41.67\%\\
   \midrule
\textbf{Handedness} & & & & &\\
  Right & 95.92\% & 98\% & 96\% & 93.88\% & 87.50\%\\
 Left & 4.08\% & 2\% & 4\% & 6.12\% & 12.50\%\\
\midrule
  \textbf{Age Group} & & & & &\\
  20-25 & 67.35\% & 80\% & 70\% & 61.22\% & 70.83\%\\
  26-30 & 26.53\% & 16\% & 20\% & 24.49\% & 16.67\%\\
  31-35 & 6.12\% & 4\% & 10\% & 14.29\% &    12.50\%\\
 \midrule
\textbf{Background} & & & & &\\
  CS/IT/Security & 59.18\% & 54\% & 68\% &  63.26\% &	70.83\%\\
\  Others & 40.82\% & 46\% & 32\% & 36.74\% & 29.17\%\\
\midrule
\textbf{Mobile OS} & & & & &\\
  Android  & 95.92\%  &  92\% & 90\% & 93.88\% & 91.67\%\\
   iOS & 2.04\% &  6\% & 8\% & 4.08\% & 6.25\%\\
  Windows & 2.04\% & 2\% & 2\% & 2.04\% & 2.08\% \\
   \midrule
\textbf{Screen-lock} & & & & &\\
Android Pattern & 59.18\% & 58\% & 48\% & 65.31\% & 58.33\%\\
PIN & 30.61\% & 38\% & 38\% & 26.53\% & 33.33\%\\
Password & 24.29\% & 26\% & 18\% & 16.33\% & 10.42\%\\
Fingerprint & 38.78\% & 42\% & 50\%	& 57.14\% & 35.42\%\\
Slide-to-lock & 24.29\% & 26\% & 2\% & 4.08\% & 14.58\%\\
None & 10.20\% & 4\% & 6\% & 2.04\% & 10.42\%\\
\midrule
 \#Participants & 49 & 50 & 50 & 49 & 48\\
\bottomrule
  \end{tabular}
  \caption{Demographics of participants across five groups.}~\label{tab:demo}
\end{table}

\begin{figure*}[t]
\centering
\begin{subfigure}[b]{0.33\textwidth}
  \centering
  \includegraphics[scale = 0.089]{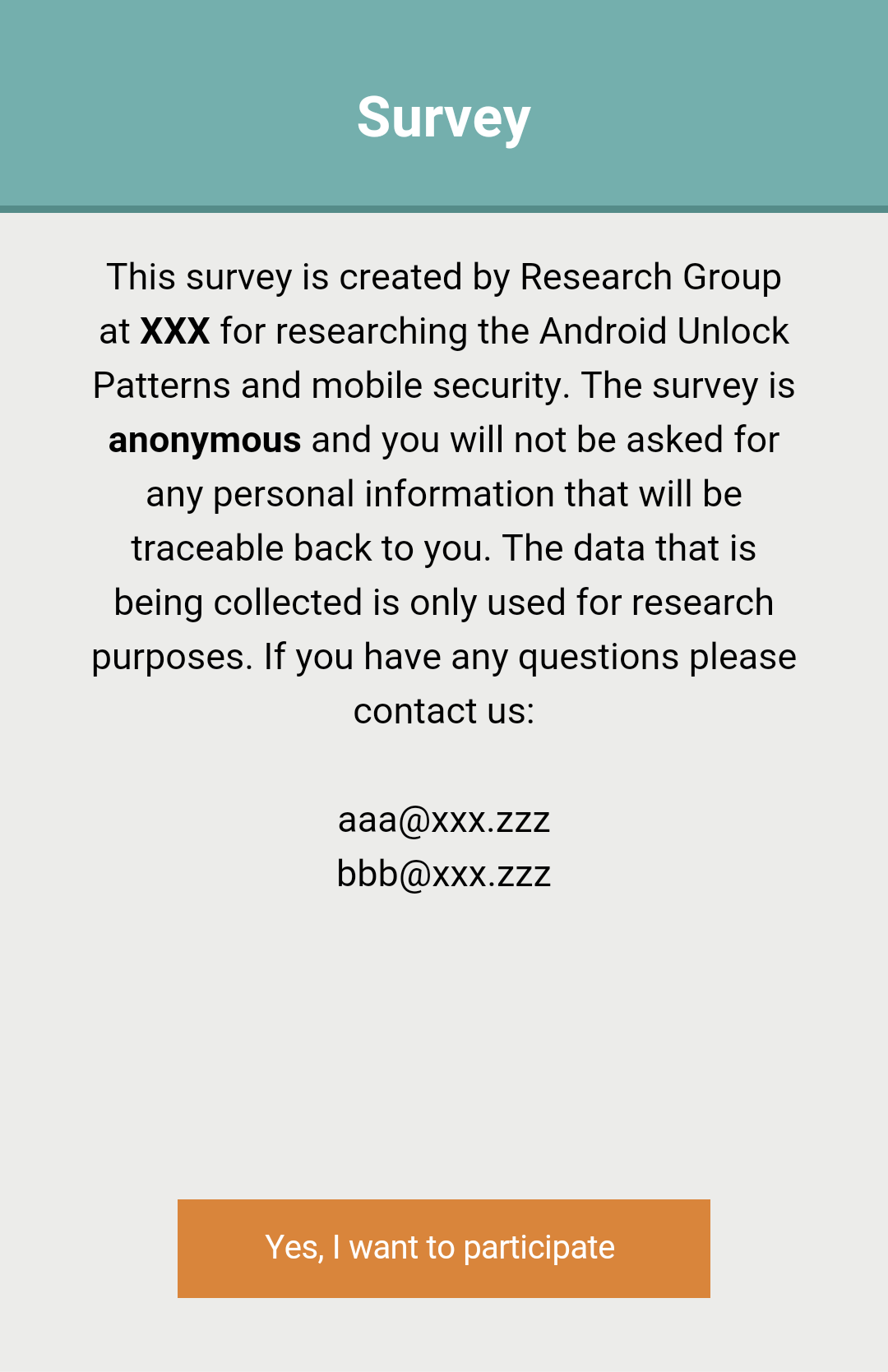}
  \captionsetup{font=scriptsize}
  \caption{Survey Information}~\label{fig:info}
\end{subfigure}%
\begin{subfigure}[b]{0.33\textwidth}
  \centering
  \includegraphics[scale = 0.089]{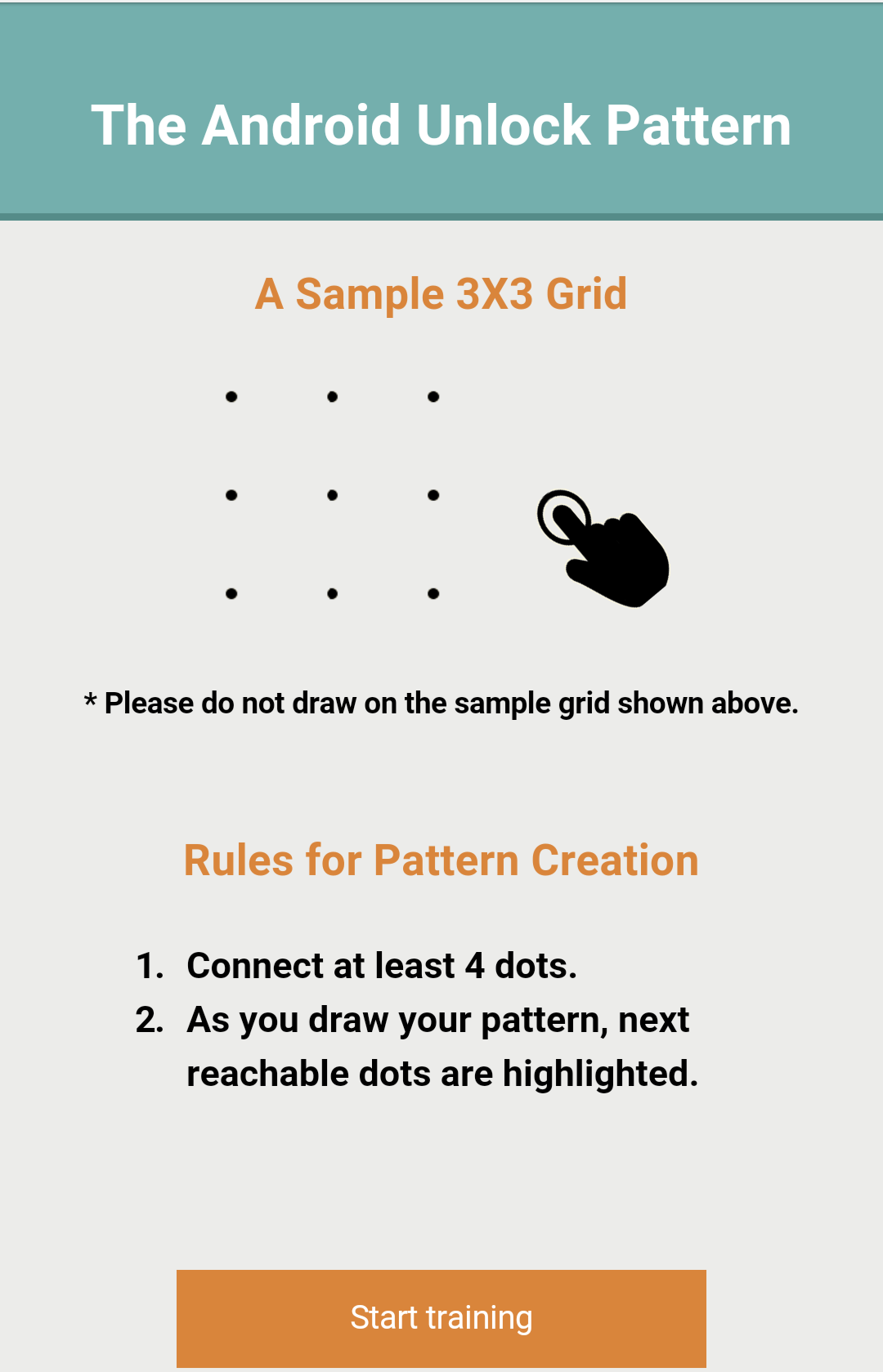}
  \captionsetup{font=scriptsize}
  \caption{Introduction (TinPal)}~\label{fig:policy}
\end{subfigure}%
\begin{subfigure}[b]{0.33\textwidth}
  \centering
  \includegraphics[scale = 0.089]{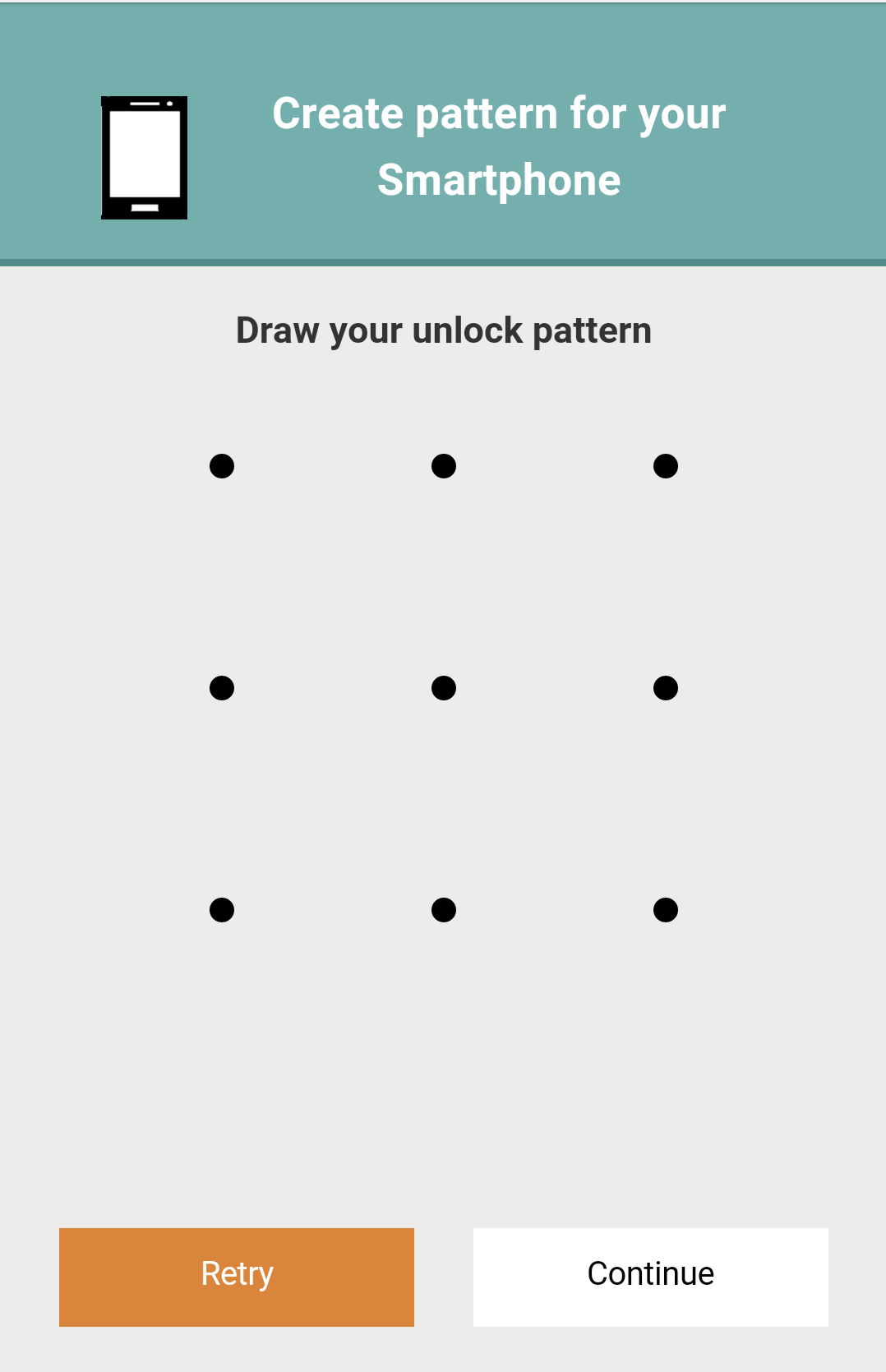}
  \captionsetup{font=scriptsize}
  \caption{Create Pattern}~\label{fig:create}
\end{subfigure}%

\begin{subfigure}[b]{0.33\textwidth}
  \centering
  \includegraphics[scale = 0.089]{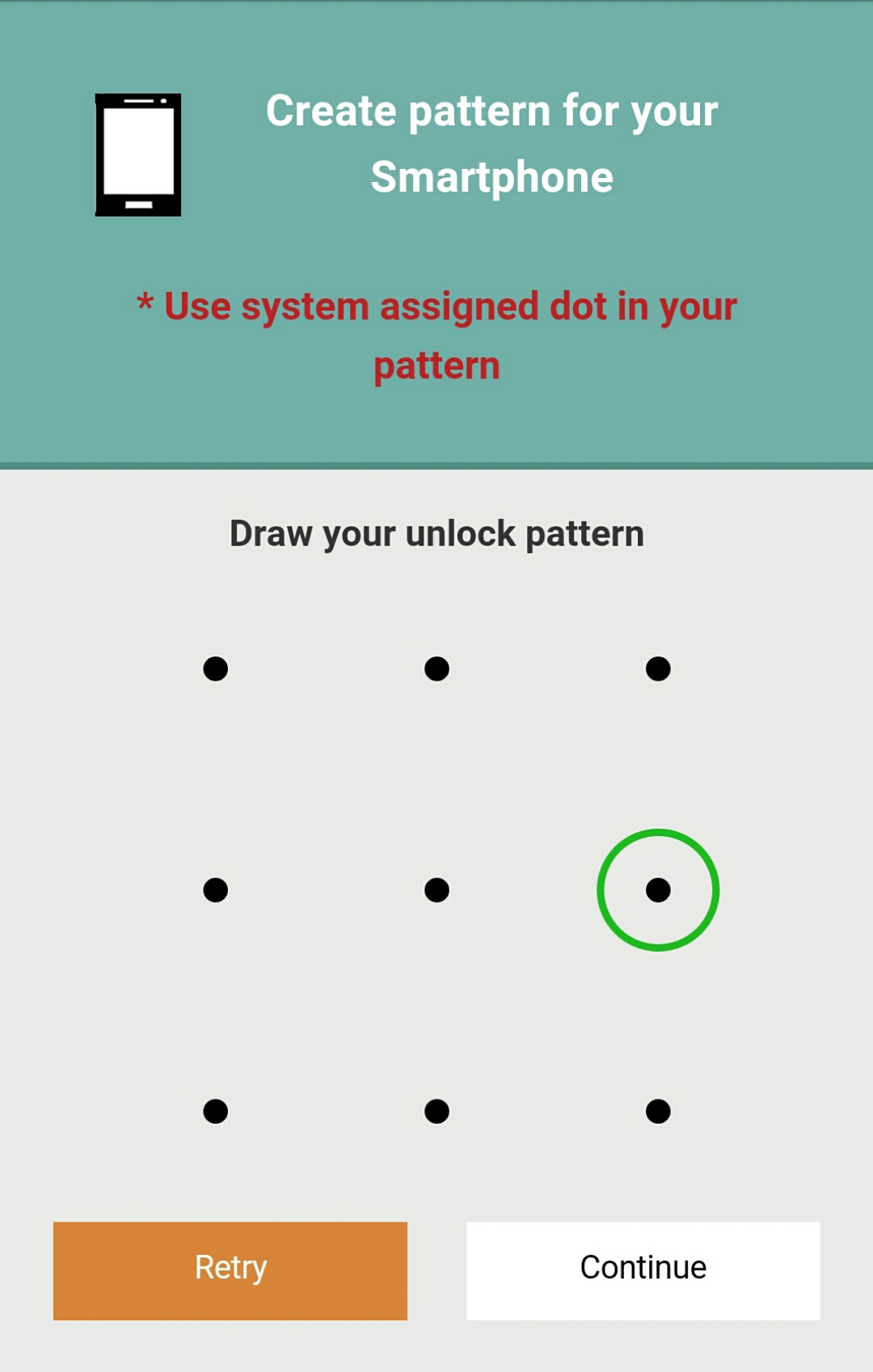}
  \captionsetup{font=scriptsize}
  \caption{Create Pattern (1-dot)}~\label{fig:one_dot}
\end{subfigure}%
\begin{subfigure}[b]{0.33\textwidth}
  \centering
  \includegraphics[scale = 0.089]{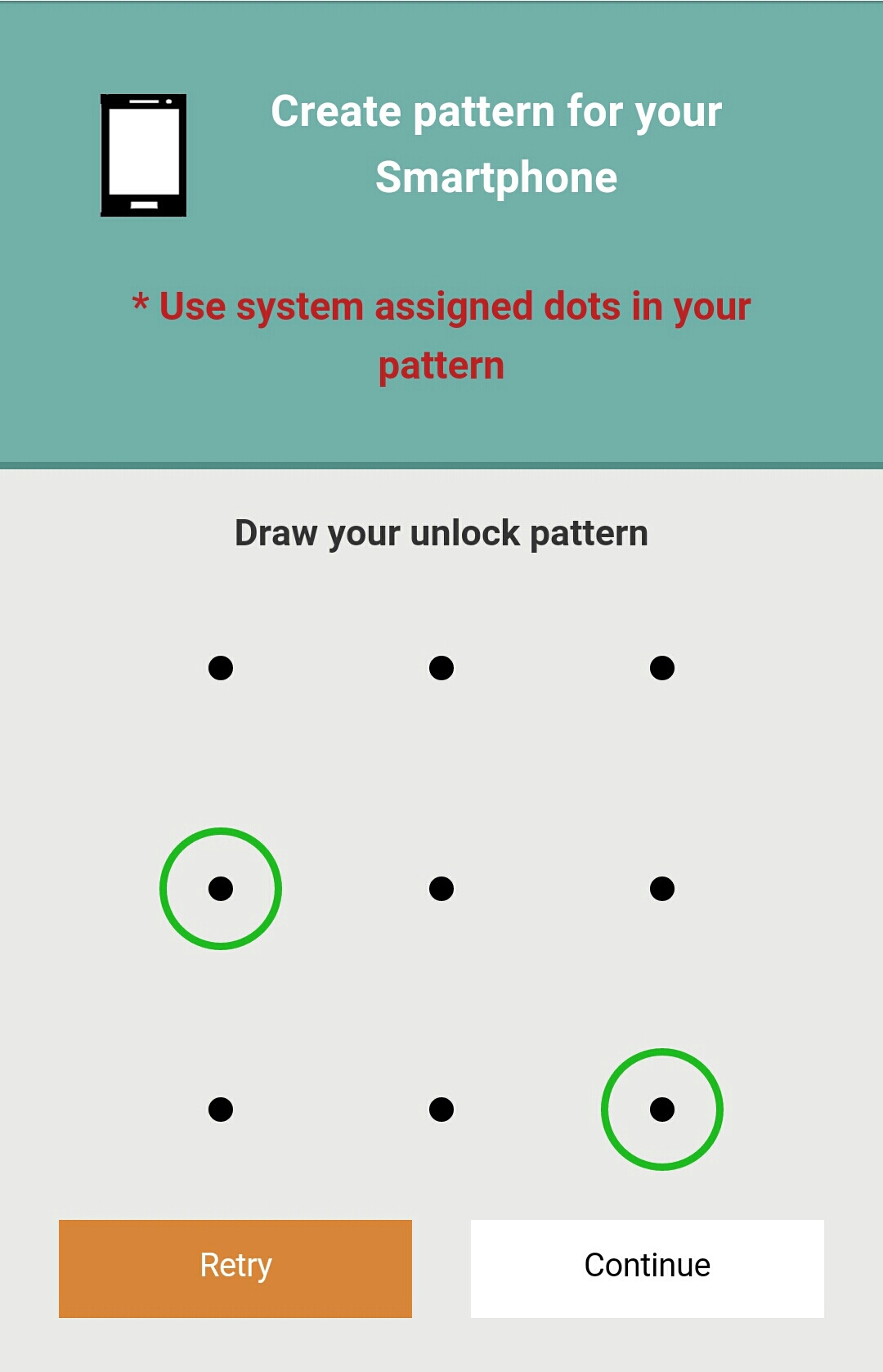}
  \captionsetup{font=scriptsize}
  \caption{Create Pattern (2-dot)}~\label{fig:two_dot}
\end{subfigure}%
\begin{subfigure}[b]{0.33\textwidth}
  \centering
  \includegraphics[scale = 0.089]{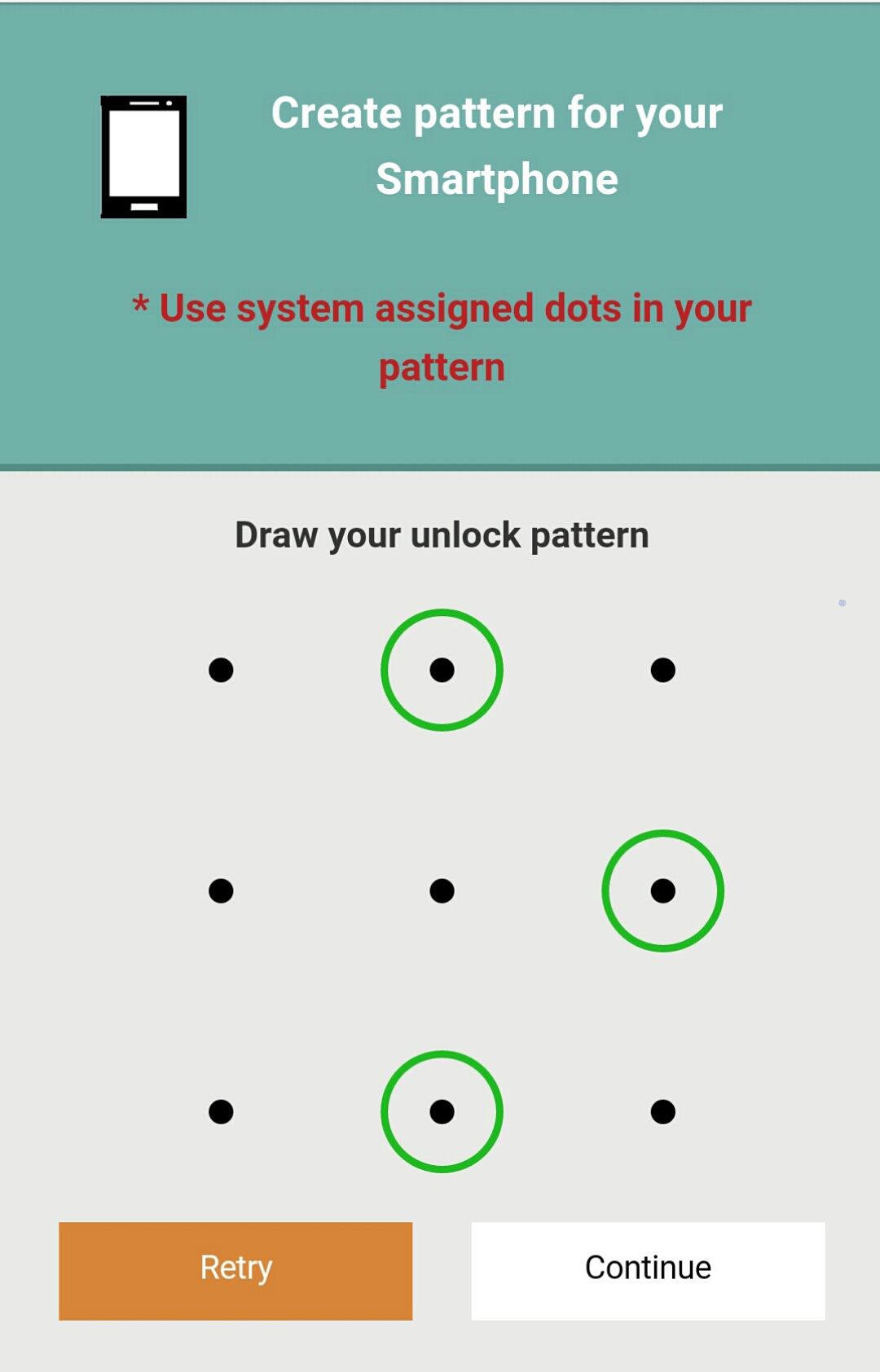}
  \captionsetup{font=scriptsize}
  \caption{Create Pattern (3-dot)}~\label{fig:three_dot}
\end{subfigure}%

\begin{subfigure}[b]{0.33\textwidth}
  \centering
  \includegraphics[scale = 0.089]{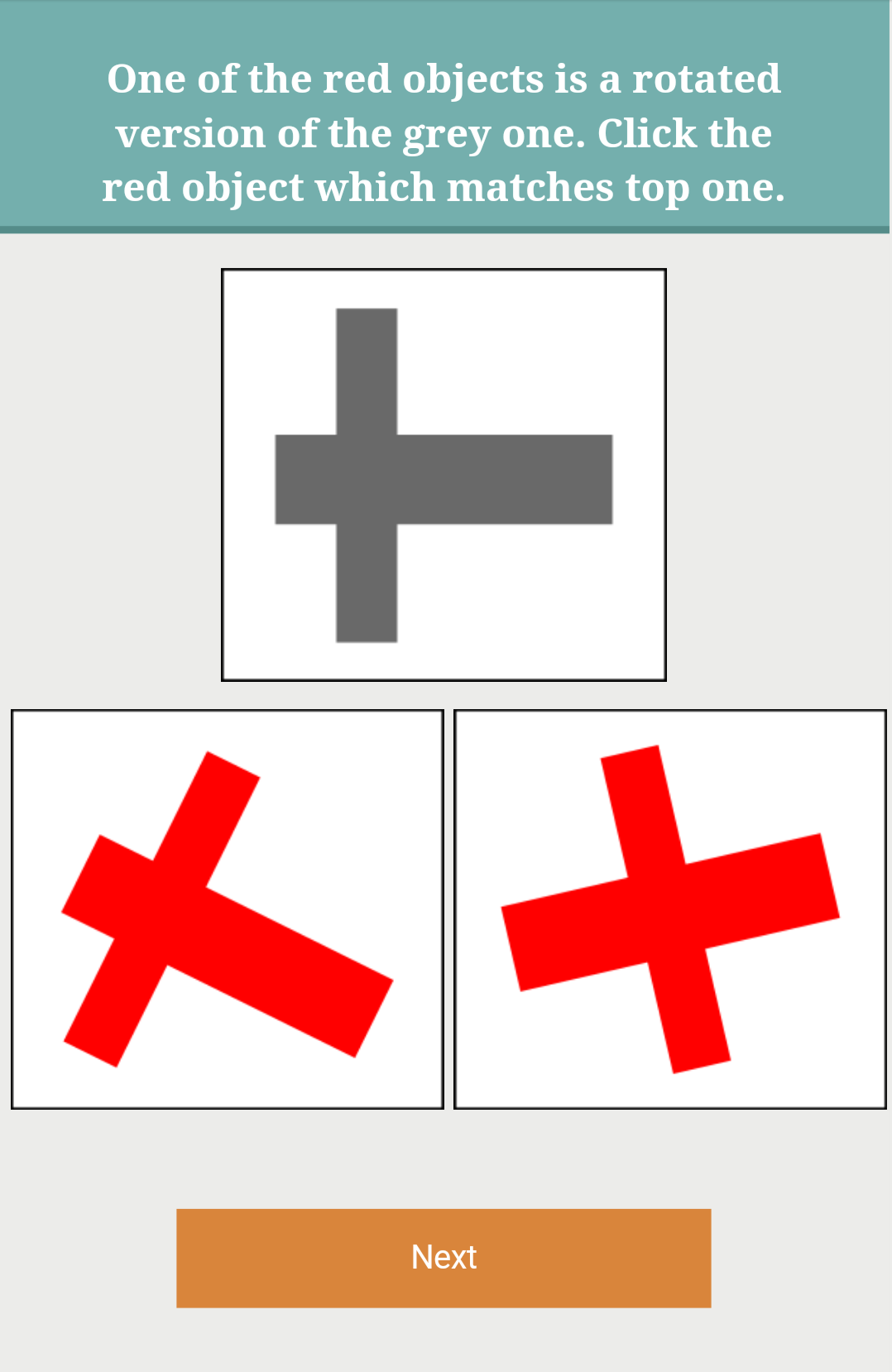}
  \captionsetup{font=scriptsize}
  \caption{Puzzle (Simple)}~\label{fig:easy_puzzle}
\end{subfigure}%
\begin{subfigure}[b]{0.33\textwidth}
  \centering
  \includegraphics[scale = 0.089]{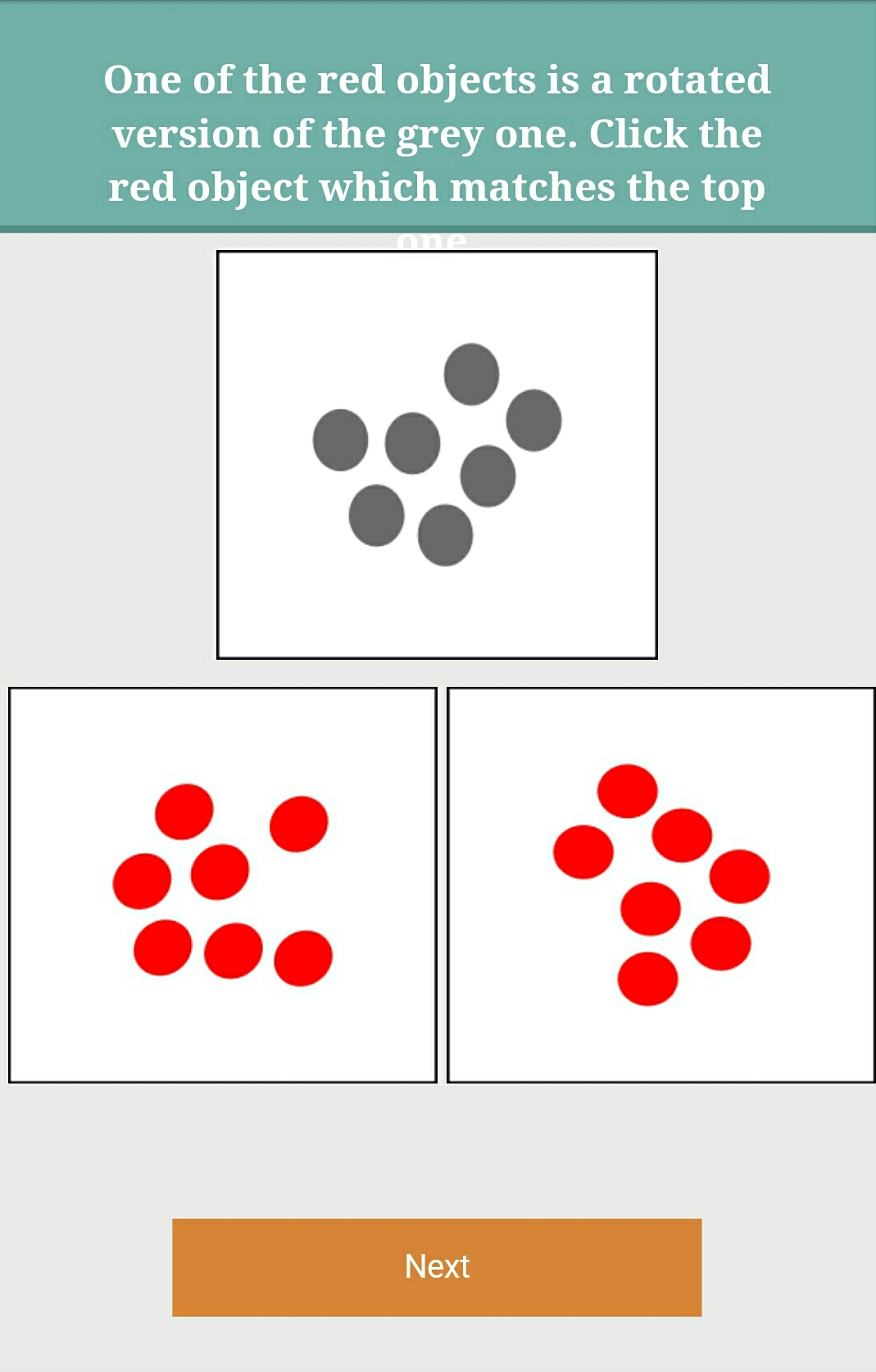}
  \captionsetup{font=scriptsize}
  \caption{Puzzle (Medium)}~\label{fig:medium_puzzle}
\end{subfigure}%
\begin{subfigure}[b]{0.33\textwidth}
  \centering
  \includegraphics[scale = 0.089]{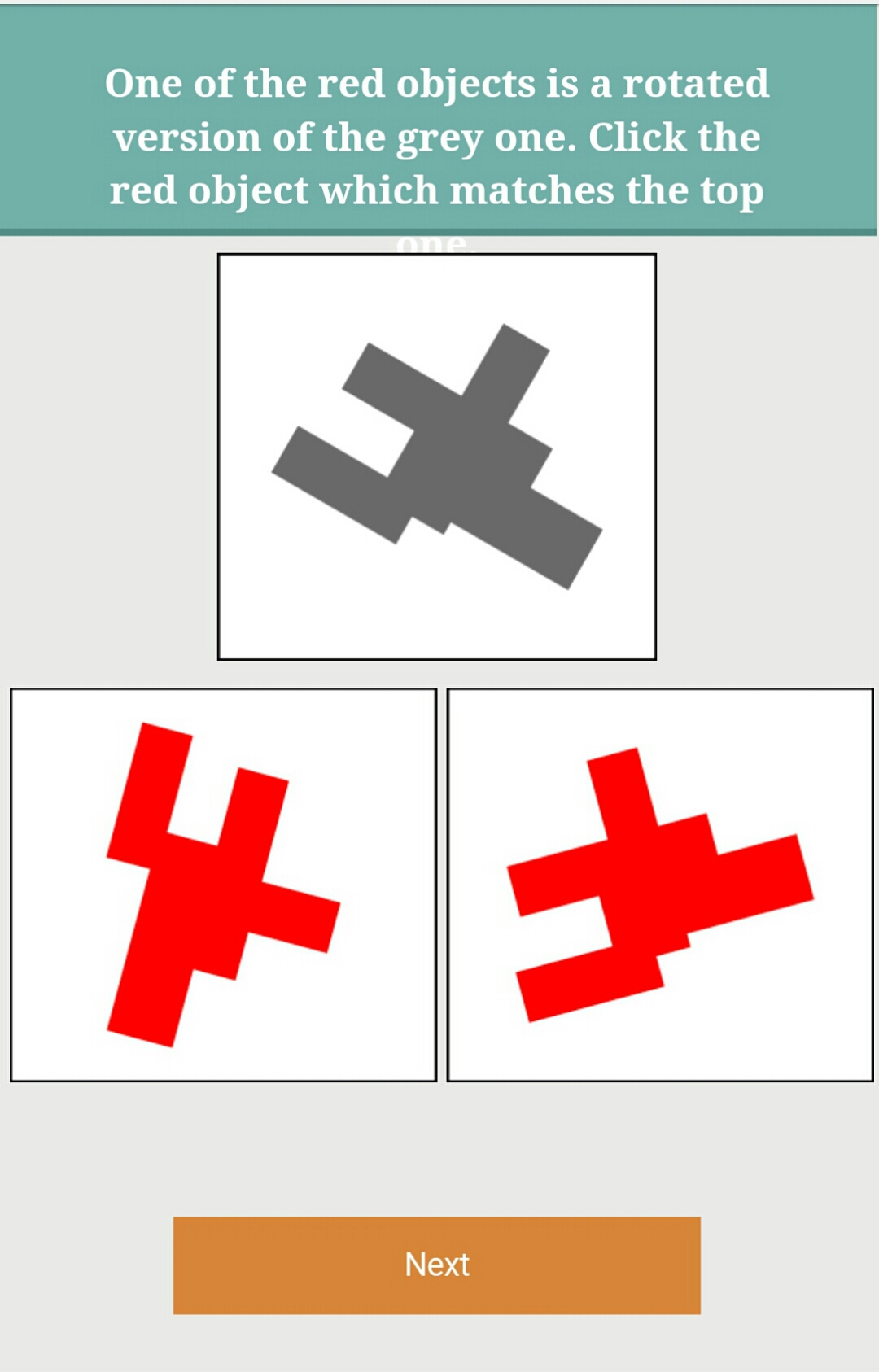}
  \captionsetup{font=scriptsize}
  \caption{Puzzle (Complex)}~\label{fig:complex_puzzle}
\end{subfigure}%
\caption{Selected screens of the user study.} ~\label{fig:study}
\vspace{-1cm}
\end{figure*} 

\subsection{Procedure}
The structure of our study is similar to the one described in \cite{Loge:study}. The study was conducted in lab in 7 stages: 1) Survey Information, 2) Brief introduction to Android Lock Patterns, 3) Training, 4) Pattern Creation, 5) Distraction Task, 6) Questions on demographics, device and screen-lock, and 7) Pattern Recall. The details of each stage are given below.

\begin{enumerate}
\item \textit{Survey Information.} Participants were requested to open the study link using a browser on their mobile device. On visiting the link, participants were shown the information page (Figure \ref{fig:info}) which contained a brief information about the study, its purpose and the data usage policy. This page was common to all groups. 

\item \textit{Introduction to Pattern Lock.} After agreeing to participate in the study, participants were shown a sample $3\times 3$ grid along with the pattern creation rules. Participants in the TinPal group were informed about the highlighting feature as depicted in Figure \ref{fig:policy} while participants in the control, 1-dot, 2-dot and 3-dot groups were informed about the four pattern creation rules given below \cite{Uellenbeck:guessing}. In addition, participants in all SysPal groups were also informed about the mandatory usage of the system-assigned random dots.\\
(R1) At least 4 dots must be selected.\\
(R2) No dot can be used more than once.\\
(R3) Only straight lines are allowed.\\
(R4) One cannot jump over dots not visited before.

\item \textit{Training.} To ensure that participants in each group become familiar with their assigned interface, we had a training page where the participants could explore the assigned interface by drawing as many patterns as they like. Further, we made sure that for a given participant, the system-assigned dot(s) do not change across different trials, {\em i.e.}, the system-assigned dot(s) remain the same even after pattern reset. For instance, once the system randomly selects dot 4 for a participant in the 1-dot group, the participant continues to see the same system-assigned dot even after she clicks on the reset button. Otherwise, the participant could click the reset button until she gets system-assigned dot(s) of her choice which would defeat the purpose of SysPal policies. The assigned dot(s) remained the same even in the next stage (Pattern Creation).

\item \textit{Pattern Creation.} After completing the training, participants were given the smartphone scenario \cite{Loge:study} in which they were asked to create a new pattern to protect their mobile device. Participants in all groups could create any pattern of their choice (Figure \ref{fig:create}), however participants in the 1-dot group, 2-dot group and 3-dot group were also asked to use system-assigned dots in their pattern (Figure \ref{fig:one_dot}, \ref{fig:two_dot}, \ref{fig:three_dot}). After submitting a valid pattern, participants in all groups were asked to re-confirm their pattern. If the confirmation failed, participants could try creating their pattern again, however in case of participants in 1-dot, 2-dot and 3-dot groups, the system-assigned dot(s) remain the same.

\item \textit{Distraction Task.} After creating the pattern, participants were given a distraction task involving graphical puzzles to solve. Specifically, participants were shown a target object at the top, and they had to identify which one of the bottom two objects matches the target as illustrated in Figures \ref{fig:easy_puzzle}, \ref{fig:medium_puzzle} and \ref{fig:complex_puzzle}. The purpose of these puzzles was to clear the visual working memory of participants. The dataset required for building the puzzles was downloaded from \cite{puzzle}. It consisted of 15 mental rotation puzzles. Based on the difficulty level, we manually categorized all 15 puzzles into three categories, namely, {\em easy}, {\em medium} and {\em complex}. Each category contained 5 mental rotation puzzles. Examples of easy, medium and complex puzzles are provided in Figures \ref{fig:easy_puzzle}, \ref{fig:medium_puzzle} and \ref{fig:complex_puzzle} respectively. Every participant had to attempt three graphical puzzles, one (picked randomly) from each category. 

\item \textit{Questions.} To further prolong the recall stage, we asked participants few questions pertaining to demographics (gender, age, handedness and educational background), their mobile device and the screen-lock they ever used (if any).

\item \textit{Pattern Recall.} Finally, participants were asked to recall their pattern within five attempts. The working of the TinPal interface was consistent during creation as well as recall, {\em i.e.}, the highlighting of dots happened not only during pattern creation, but also during recall. Participants in the SysPal groups were not shown the system-assigned dots.
\end{enumerate}

\subsection{User Data Collected}
To compare the usability and security of $3 \times 3$ patterns created on five different interfaces (Original, TinPal, 1-dot, 2-dot and 3-dot), we collected the following data points about every participant in each group:
\begin{itemize}
\item assigned interface
\item system-assigned random dot(s) in 1-dot, 2-dot and 3-dot groups
\item number of patterns tried in the training stage
\item time spent in the training stage
\item number of patterns tried in the creation stage
\item final pattern selected in the creation stage
\item time spent in drawing the final selected pattern
\item number of redraw attempts
\item time spent in redrawing the pattern 
\item time needed to solve the distraction task
\item survey responses
\item time needed to answer the survey
\item number of recall attempts
\item time spent in recalling the pattern
\end{itemize}

\subsection{Statistical Tests}
We compared categorical data such as gender, age-group, handedness, background, mobile OS, screen-lock in use, starting and ending points of patterns, and the number of unlocking attempts (maximum five) across all groups using two-tailed Fischer's Exact test (FET). Since pattern characteristics such as pattern length, stroke length, direction changes and intersections, and usability metrics such as pattern creation time, redraw time and recall time were not normally distributed (Shapiro-Wilk test rejects the null hypothesis that data is normal with a $p < 0.01$), we performed two-tailed Wilcoxon-Mann-Whitney (WMW) test \cite{marx2016edison} with a significance level of $\alpha = 0.01$. As there are five groups, 10 pairwise comparisons were performed. To account for multiple statistical comparisons, we applied Bonferroni correction, {\em i.e.}, we altered the value of $\alpha$ to $0.01/10 = 0.001$, and claimed statistical significance if $p < 0.001$. We claimed the result to be of possible significant interest if $p < 0.01$. We also compared the distribution of starting point and ending point choices of all groups using the {\em entropy} measure: 
\begin{align}
Entropy \ H = \sum\limits_{i=1}^9 p_i\cdot log_2(1/p_i)
\end{align}
where $p_i$ is the probability of choosing a dot $i$ as starting (or ending) point. 

\subsection{Limitations}
We performed two separate studies to collect TinPal and SysPal patterns, however we ensured that the setup used in both studies is identical and the same participants are not enrolled in more than one group. Similar data collection strategies have been reported in the previous studies \cite{Uellenbeck:guessing, Aviv:guessing}. The sample used in our comparative study is younger and more tech-savvy, and therefore, may have better memory than average which could influence the results. Further, as the study was conducted in a lab, the sample is small (246 participants), and with a large sample we could observe further patterns. However, the objective of our study was to determine whether informing users about connection options through the highlighting mechanism had any impact on their pattern choices which we found to be statistically significant with this small sample. 

Further, it is not easy to demonstrate if patterns created in the user study are realistic. Fahl {\em et al.} attempted to address the questions related to ecological validity in password studies and found that passwords created by participants in a role-play scenario resemble their real passwords \cite{Fahl:2013}. They also found that passwords created in the lab study were more representative of actual passwords than those created in the online study and priming subjects does not make any difference. Therefore, we conducted our study in lab and asked participants to create a pattern for the smartphone scenario \cite{Loge:study}. Moreover, recently researchers found statistically significant differences between patterns collected on the mobile device of participants and patterns collected using other collection methods \cite{Aviv:collection}. Therefore, we asked participants to create pattern on their own mobile device. 

\section{Security Results}\label{sec:securityresults}
To evaluate the impact of TinPal on users' pattern choices, we compared certain characteristics of patterns created across five groups. Specifically, we looked at the characteristics such as pattern length, stroke length, knight moves, overlaps, starting points, ending points, direction changes and intersections. These characteristics are considered to be effective in preventing guessing attacks \cite{Andriotis:sidechannel, Uellenbeck:guessing, Andriotis:guessing, Aviv:guessing, harshal:guessing} as well as shoulder-surfing attacks \cite{Sun:shouldersurfing, Song:shouldersurfing, vonZezschwitz:shouldersurfing}. We also measured the guessability of patterns across all groups using an attack technique based on $n$-gram Markov model \cite{Uellenbeck:guessing, Aviv:guessing, siadati:persuasive, harshal:guessing, cho2017syspal}. 

\subsection{Pattern Characteristics} \label{definitions}
\begin{figure*}[t]
\centering
\begin{subfigure}[b]{0.50\textwidth}
  \centering
  \includegraphics[scale = 0.40]{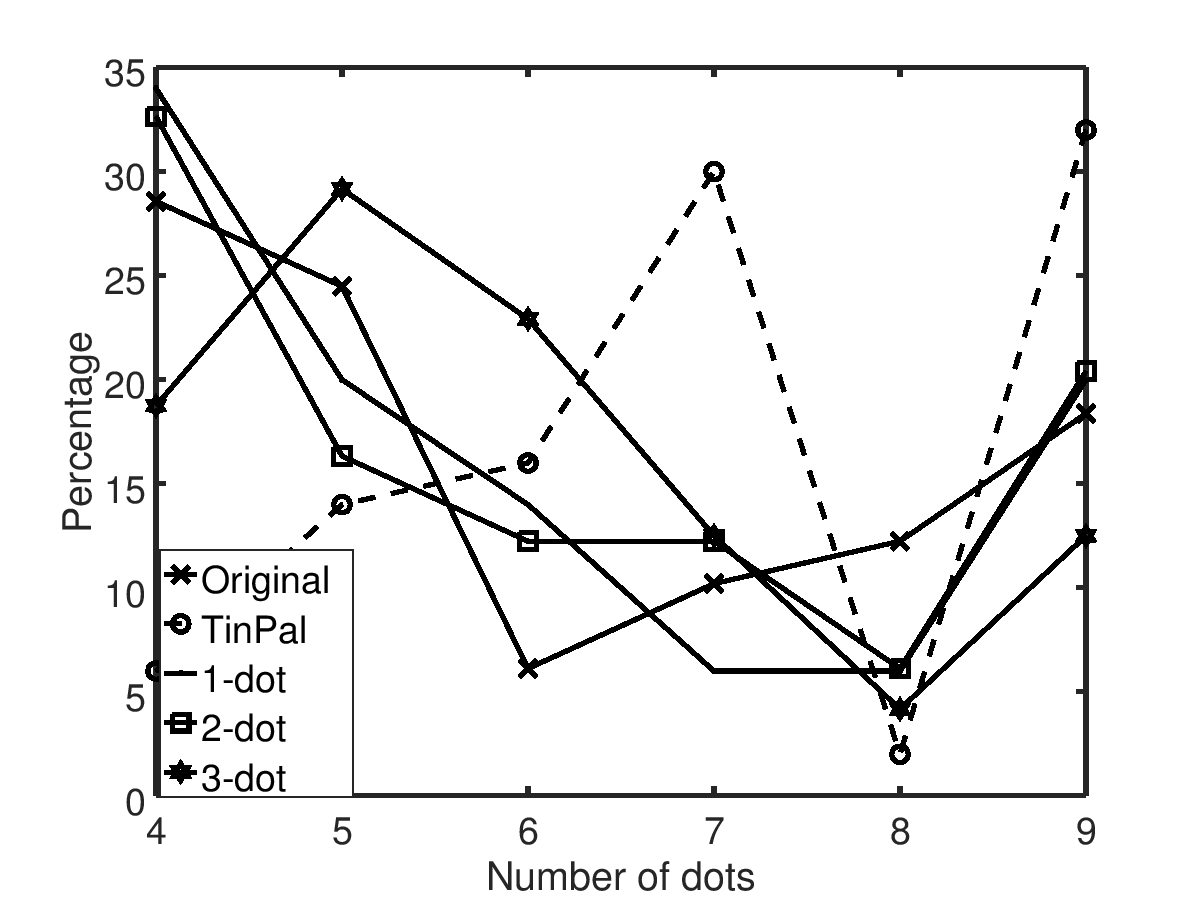}
  \captionsetup{font=scriptsize}
  \caption{Pattern length}~\label{fig:length}
\end{subfigure}%
\begin{subfigure}[b]{0.50\textwidth}
  \centering
  \includegraphics[scale = 0.40]{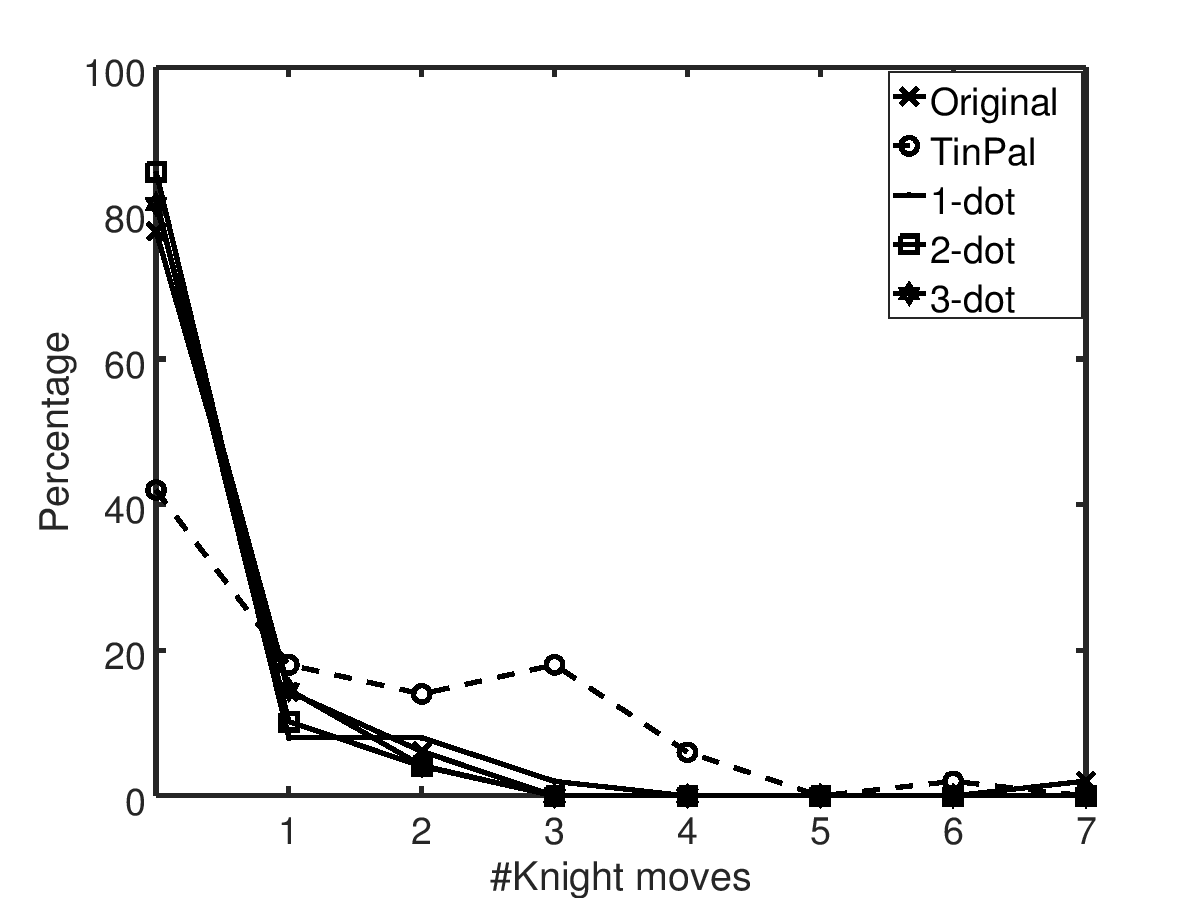}
  \captionsetup{font=scriptsize}
  \caption{Knight moves }~\label{fig:knight}
\end{subfigure}%

\begin{subfigure}[b]{0.50\textwidth}
  \centering
  \includegraphics[scale = 0.40]{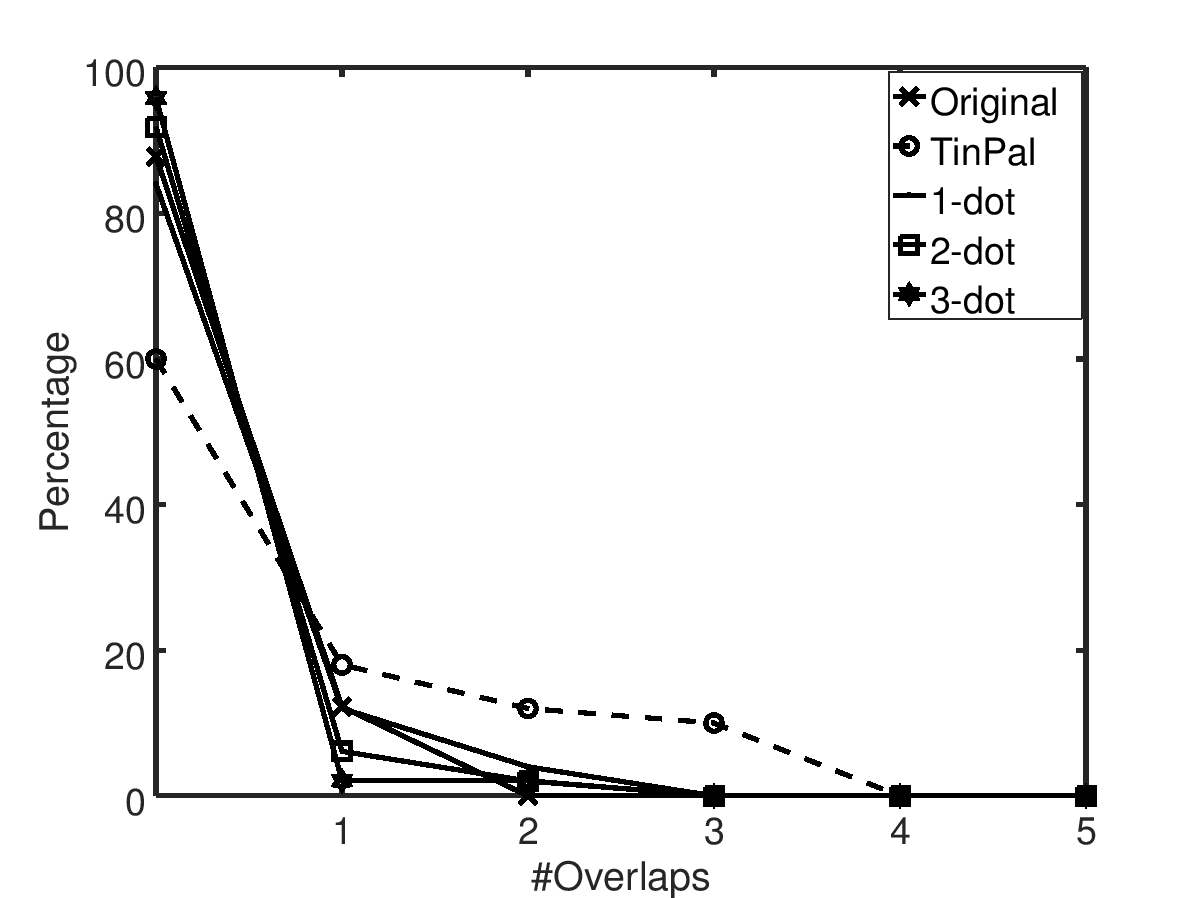}
  \captionsetup{font=scriptsize}
  \caption{Overlaps}~\label{fig:overlap}
\end{subfigure}%
\begin{subfigure}[b]{0.50\textwidth}
  \centering
  \includegraphics[scale = 0.40]{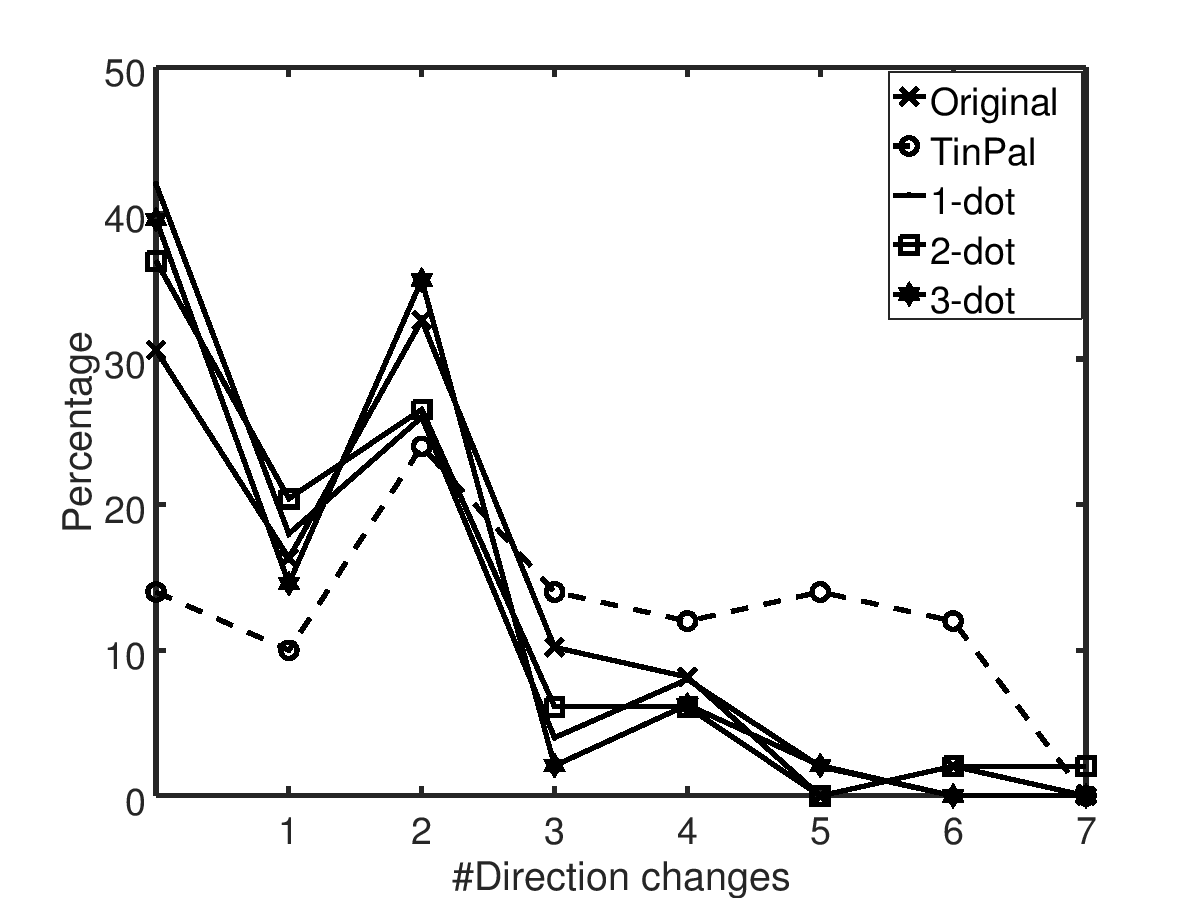}
  \captionsetup{font=scriptsize}
  \caption{Direction changes}~\label{fig:direction}
\end{subfigure}%
\caption{Distribution of length, knight moves, overlaps and direction changes across five groups.}
\end{figure*}

Table \ref{tab:stats} compares the characteristics of patterns created across all groups. The mean, the median and the standard deviation of each pattern characteristic is shown in the table. The distribution of TinPal patterns (as measured by mean, median and standard deviation) is relatively more similar to the distribution of theoretical patterns (389,112) compared to the other groups. We delve into each characteristic in more detail.
\begin{table}[h]
  \centering
 \scriptsize
  \begin{tabular}{l c c c c c c}
\toprule
   \textit{Characteristic}
    & \textit{Original}
      & \textit{TinPal}
    & \textit{1-dot} 
    & \textit{2-dot}
    & \textit{3-dot}  
    & \textit{Theory}\\
    \midrule
   \textbf{Pattern Length}  & & & & & &\\
   Mean & 6.08 & 7.04 & 5.90 & 6.04 & 5.92 & 7.97\\
   Standard deviation & 1.90 & 1.60 & 1.90 & 1.94 & 1.58 & 1.02\\
   Median & 5 & 7 & 5 & 6 & 6 & 8\\
    \midrule
\textbf{Stroke Length}  & & & & & &\\
   Mean & 6.05 & 8.39 & 5.80 & 5.81 & 5.50 & 11.03\\
   Standard deviation & 2.63 & 2.69 & 2.56 & 2.39 & 1.91 & 2.24\\
   Median & 5.41 & 8.27 & 5 & 5.83 & 5 & 11\\
 \midrule
\textbf{Knight Moves}  & & & & & &\\
   Mean & 0.41 & 1.36 & 0.30 & 0.18 & 0.23 & 2.71\\
   Standard deviation & 0.94 & 0.92 & 0.71 & 0.49 & 0.52 & 1.23\\
   Median & 0 & 1 & 0 & 0 & 0 & 3\\
 \midrule
\textbf{Overlaps} & & & & & &\\
   Mean & 0.12 & 0.72 & 0.20 & 0.10 & 0.06 & 0.93\\
   Standard deviation & 0.33 & 1.02 & 0.49 & 0.37 & 0.32 & 0.88\\
   Median & 0 & 0 & 0 & 0 & 0 & 1\\
 \midrule
\textbf{Direction Changes} & & & & & &\\
   Mean & 1.57 & 2.9 & 1.24 & 1.43 & 1.27 & 4.76\\
   Standard deviation & 1.40 & 1.91 & 1.36 & 1.59 & 1.30 & 1.31\\
   Median & 2 & 3 & 1 & 1 & 1 & 5\\
 \midrule
\textbf{Intersections}  & & & & & &\\
   Mean & 0.43 & 0.64 & 0.22 & 0.10 & 0.06 & 2.58\\
   Standard deviation & 1.99 & 1.25 & 0.65 & 0.31 & 0.32 & 2.13\\
   Median & 0 & 0 & 0 & 0 & 0 & 2\\
\midrule
\textbf{\#Bigrams (Segments)} & 44 & 59 & 44 & 37 & 36 & 72\\
\textbf{\#Patterns} & 49 & 50 & 50 & 49 & 48 & 389,112\\
\bottomrule
  \end{tabular}
  \caption{Comparison of pattern characteristics across five groups. The table also shows statistics pertaining to all possible 389,112 patterns.}~\label{tab:stats}
\end{table}
\\\\
\textbf{Pattern length.} It is the most basic feature and represents the number of dots connected in the pattern \cite{Uellenbeck:guessing, Andriotis:guessing, Sun:shouldersurfing, Song:shouldersurfing, Aviv:guessing, vonZezschwitz:shouldersurfing, harshal:guessing}. The theoretical distribution of pattern length is shown in Table \ref{tab:length}. Theoretically, more than 90\% of the patterns consist of at least 7 dots. Figure \ref{fig:length} depicts the pattern length distribution in all five groups. In the TinPal group, the number of participants who used at least 7 dots to connect their pattern was 64\%, whereas in the Original, 1-dot, 2-dot and 3-dot groups the numbers were 40.82\%, 32\%, 38.78\% and 29.17\% respectively (Figure \ref{fig:length}). Consequently, the average number of dots used for creating patterns in the TinPal group was relatively higher (7.04) than the other groups (Table \ref{tab:stats}). We found significant difference in the distribution of pattern length between the TinPal group and all SysPal groups except 2-dot group (all $p < 0.001$, corrected two-tailed WMW test). Further, we found significant interest in the pattern length distribution between the TinPal group and the remaining two groups, 2-dot and Original (all $p < 0.01$, corrected two-tailed WMW test). The results of all pairwise comparisons are shown in Table \ref{tab:statsDetailed} (Appendix A). We marked the entry with value $p< 0.001$ in the table using (**) and $p< 0.01$ using (*). The results suggest that TinPal not only made users aware of all connection choices, but also influenced users to connect more dots in their pattern. 

\begin{table}[h]
  \centering
 \scriptsize
  \begin{tabular}{c r r}
\toprule
     \textit{\#Pattern length}
     & \textit{Count}
      & \textit{Percentage}\\
    \midrule
    4	& 1,624	& 0.42\% \\
    5	& 7,152	& 1.84\% \\
    6	& 26,016	& 6.68\% \\
    7	& 72,912	& 18.74\% \\
    8	& 140,704	& 36.16\% \\
    9	& 140,704	& 36.16\% \\
    \midrule
     \#Patterns & 389,112 & 100\% \\		
    \bottomrule
  \end{tabular}
  \caption{Theoretical distribution of pattern length in $3 \times 3$ patterns.}~\label{tab:length}
\end{table}
~\label{fig:plot}

\noindent
\textbf{Stroke Length.} Not every line segment drawn on $3 \times 3$ grid has the same physical length. We measure the physical length of a line segment using Euclidean distance. To do so, we label the upper-left dot in $3\times 3$ grid as (0,0) and the lower-right dot as (2,2). Thus, the Euclidean distance of the segment $1 \rightarrow 2$ is 1, that of $1 \rightarrow 5$ is $\sqrt{2}$, that of $1 \rightarrow 3$ is $2$, that of $1 \rightarrow 6$ is $\sqrt{5}$ and that of $1 \rightarrow 9$ is $2\sqrt{2}$. The concept of stroke length captures the physical length of the pattern and is defined as the sum of Euclidean distances of all line segments within the pattern \cite{Aviv:guessing, harshal:guessing}. Patterns containing the same number of dots can have different stroke lengths. For instance, the stroke length of the pattern $12365$ is 4, whereas the stroke length of the pattern $16729$ is $4\sqrt{5} \sim 8.94$. Stroke length is considered as an important feature in resisting shoulder-surfing attacks \cite{Sun:shouldersurfing} as well as guessing attacks \cite{Aviv:guessing, harshal:guessing}. 

We found that the stroke length of patterns in the TinPal group (8.39) was significantly higher than any other group (Table \ref{tab:stats}). Normalizing the stroke length with respect to the pattern length reveals the mean length of the line segment used for creating patterns in all five groups. The normalized stroke length of patterns\footnote{We used the ratio of means since it is always less than or equal to the means of ratio by Jensens' inequality.} in the TinPal group was highest $\frac{8.39}{7.04} \sim 1.19$. The normalized stroke length in the Original, 1-dot, 2-dot and 3-dot group was $\frac{6.05}{6.08} \sim 0.99$, $\frac{5.80}{5.90} \sim 0.98$, $\frac{5.81}{6.04} \sim 0.96$ and $\frac{5.50}{5.92} \sim 0.93$ respectively. We found significant difference in the stroke length between the TinPal group and all other groups (all $p < 0.001$, corrected two-tailed WMW test).
\\
\\
\begin{figure*}[h]
\centering
\begin{subfigure}[b]{0.33\textwidth}
  \centering
  \includegraphics[scale = 0.30]{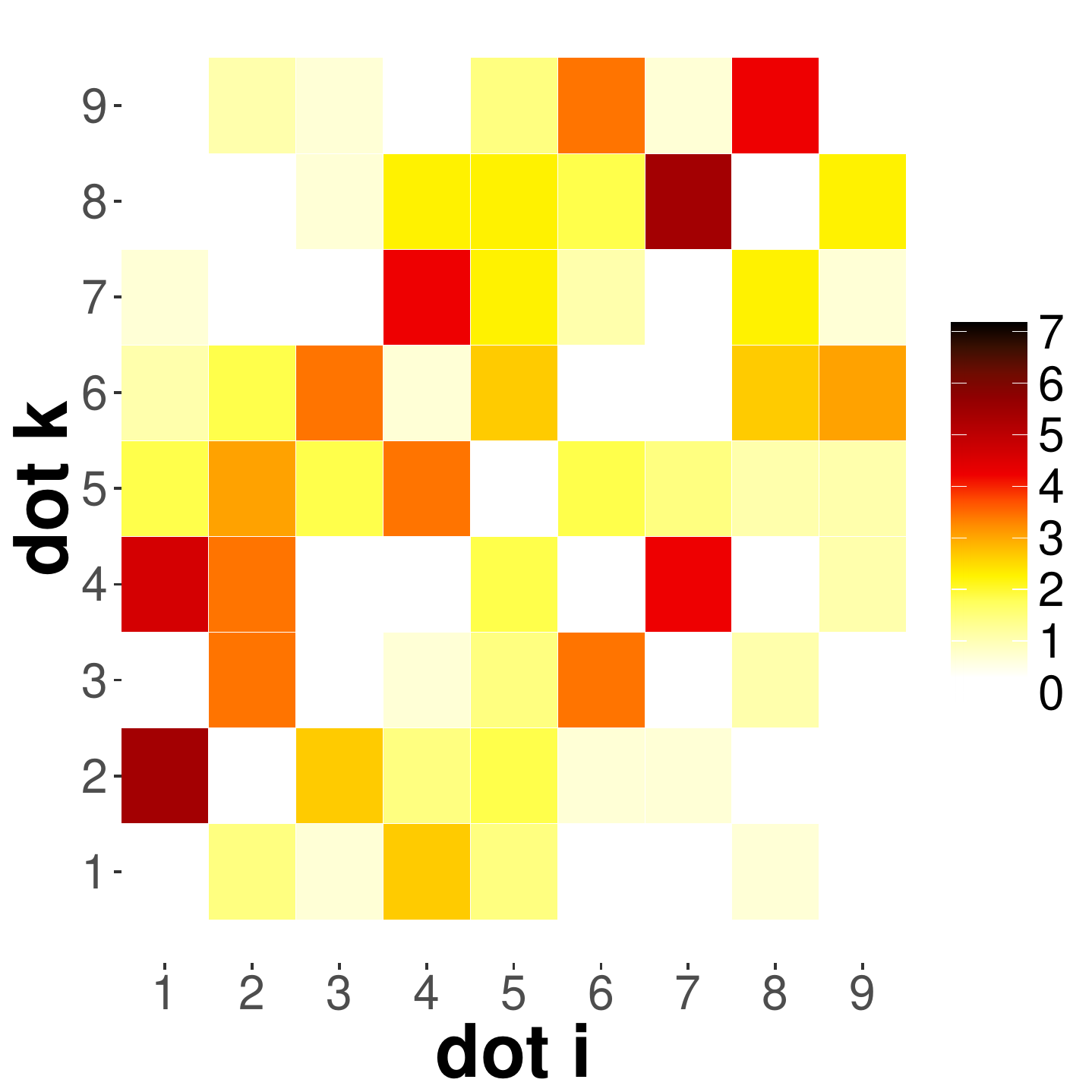}
  \captionsetup{font=scriptsize}
  \caption{Original}~\label{fig:existing_segments}
\end{subfigure}%
\begin{subfigure}[b]{0.33\textwidth}
  \centering
  \includegraphics[scale = 0.30]{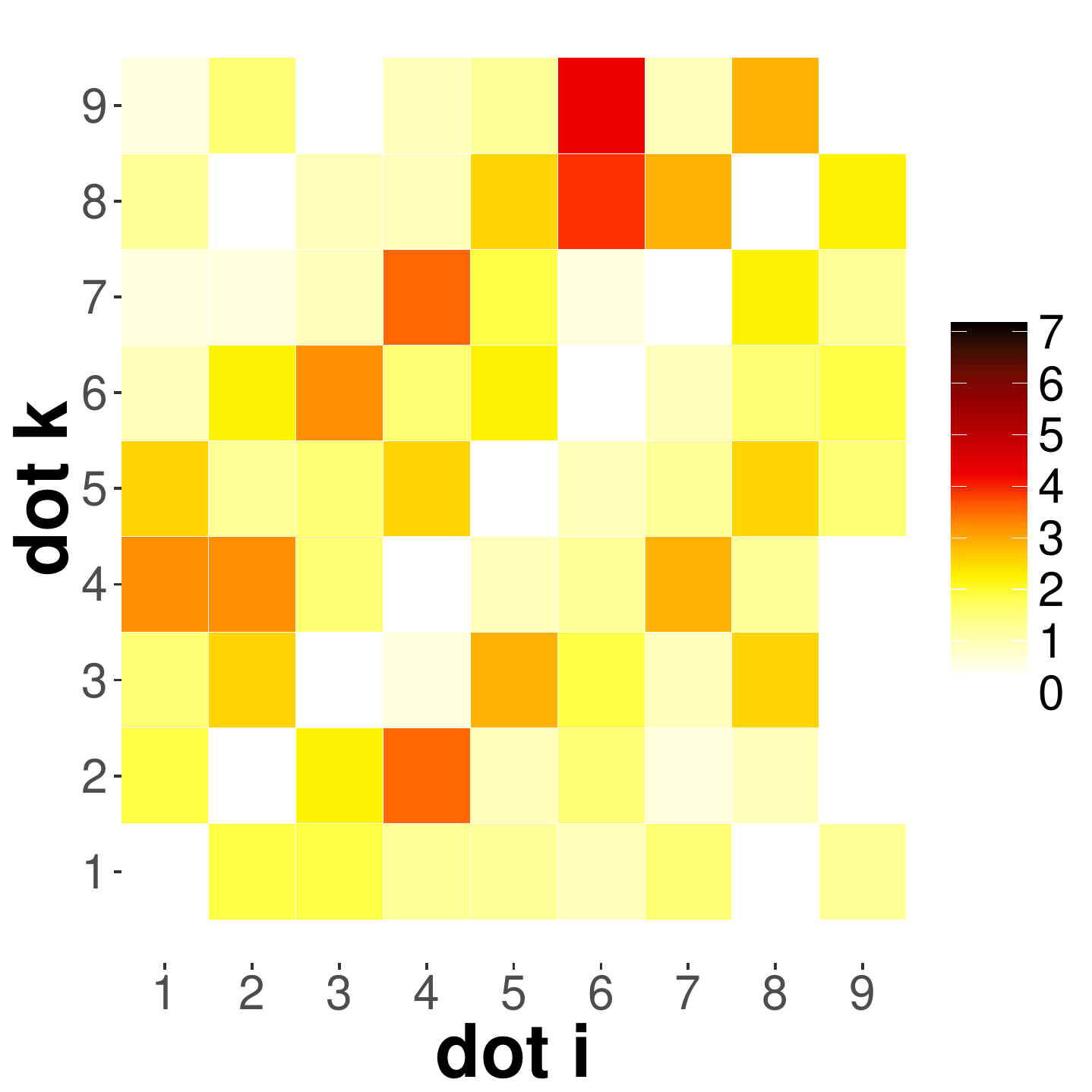}
  \captionsetup{font=scriptsize}
  \caption{TinPal}~\label{fig:enhanced_segments}
\end{subfigure}%
\begin{subfigure}[b]{0.33\textwidth}
  \centering
  \includegraphics[scale = 0.30]{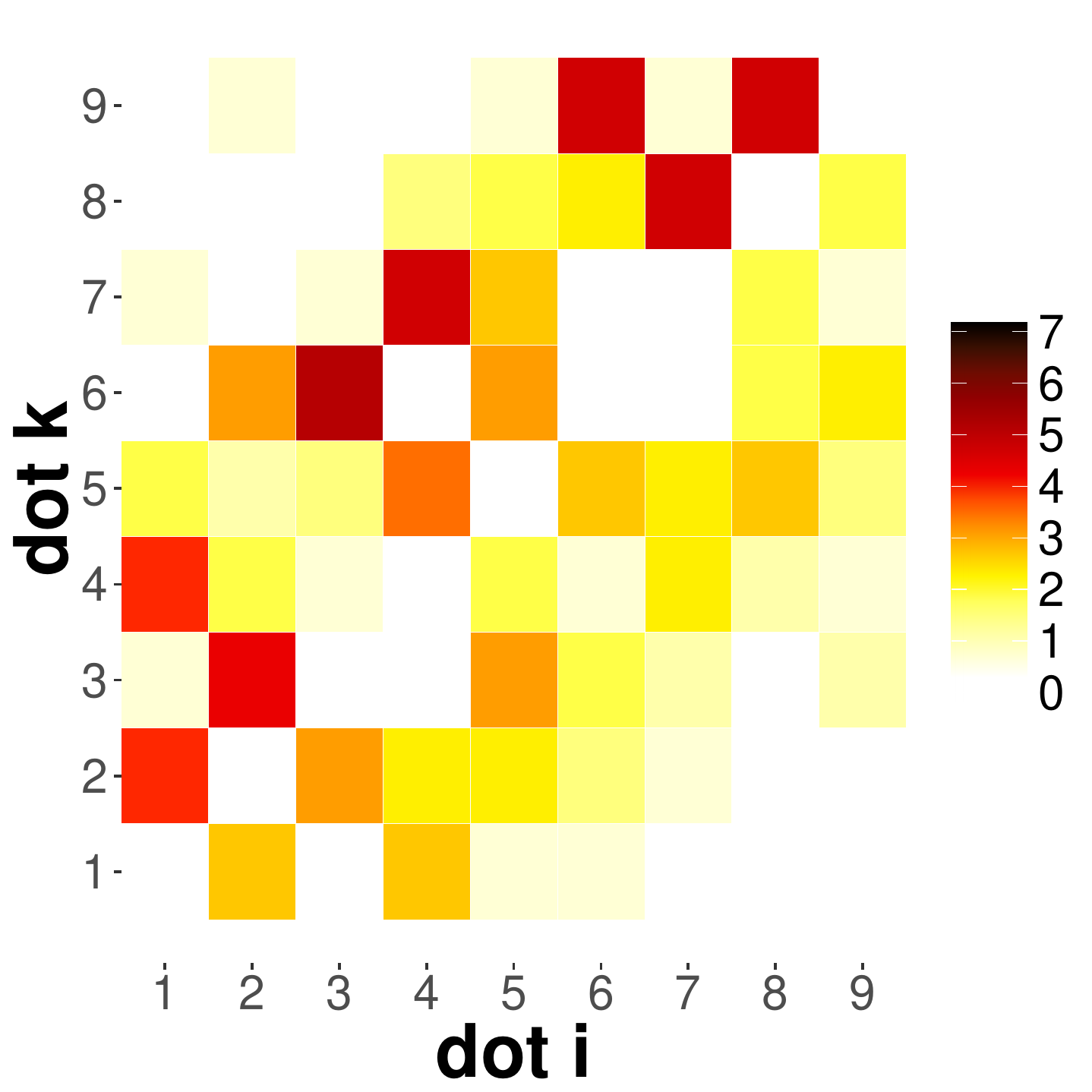}
  \captionsetup{font=scriptsize}
  \caption{1-dot}~\label{fig:syspal1_segments}
\end{subfigure}
\begin{subfigure}[b]{0.33\textwidth}
  \centering
  \includegraphics[scale = 0.30]{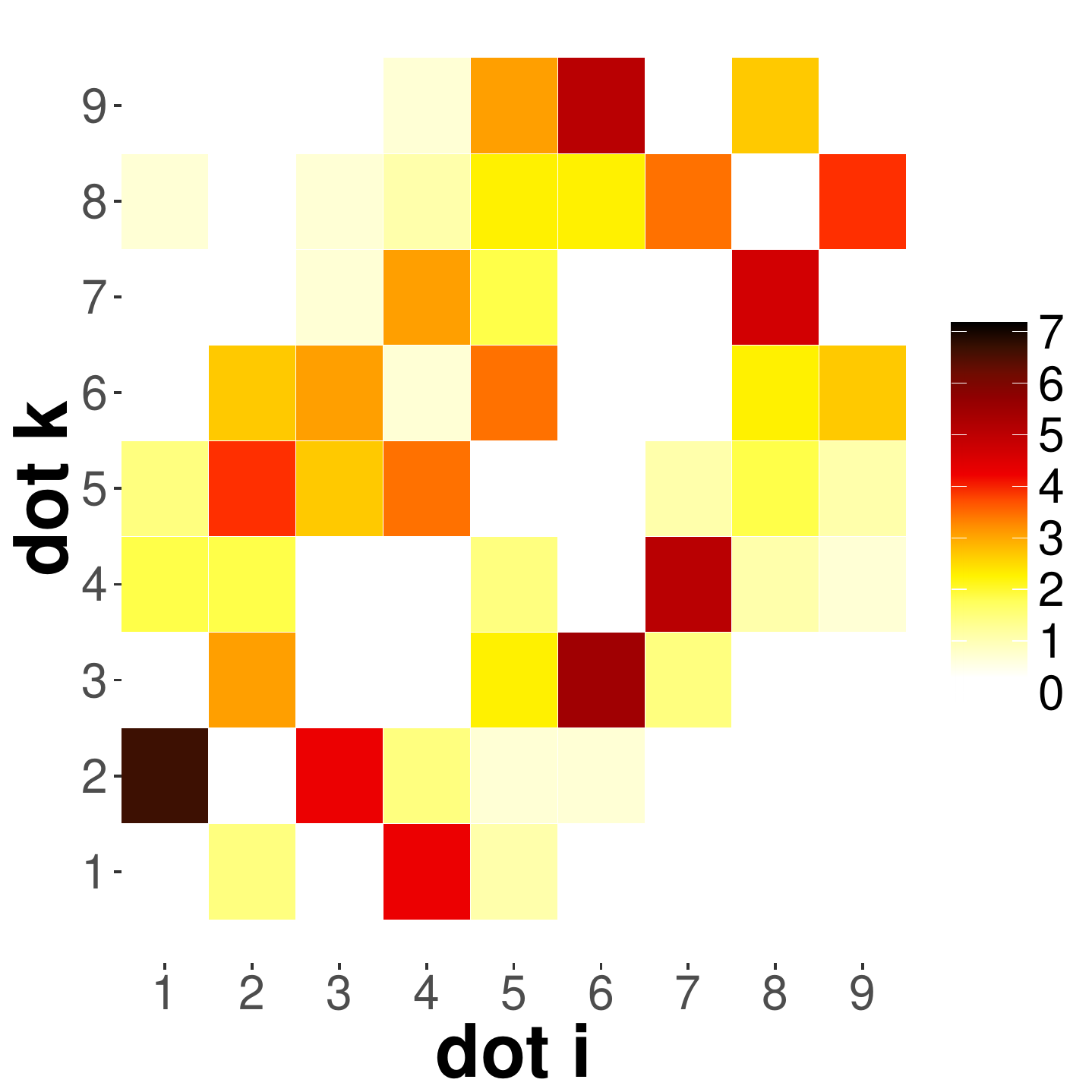}
  \captionsetup{font=scriptsize}
  \caption{2-dot}~\label{fig:syspal2_segments}
\end{subfigure}%
\begin{subfigure}[b]{0.33\textwidth}
  \centering
  \includegraphics[scale = 0.30]{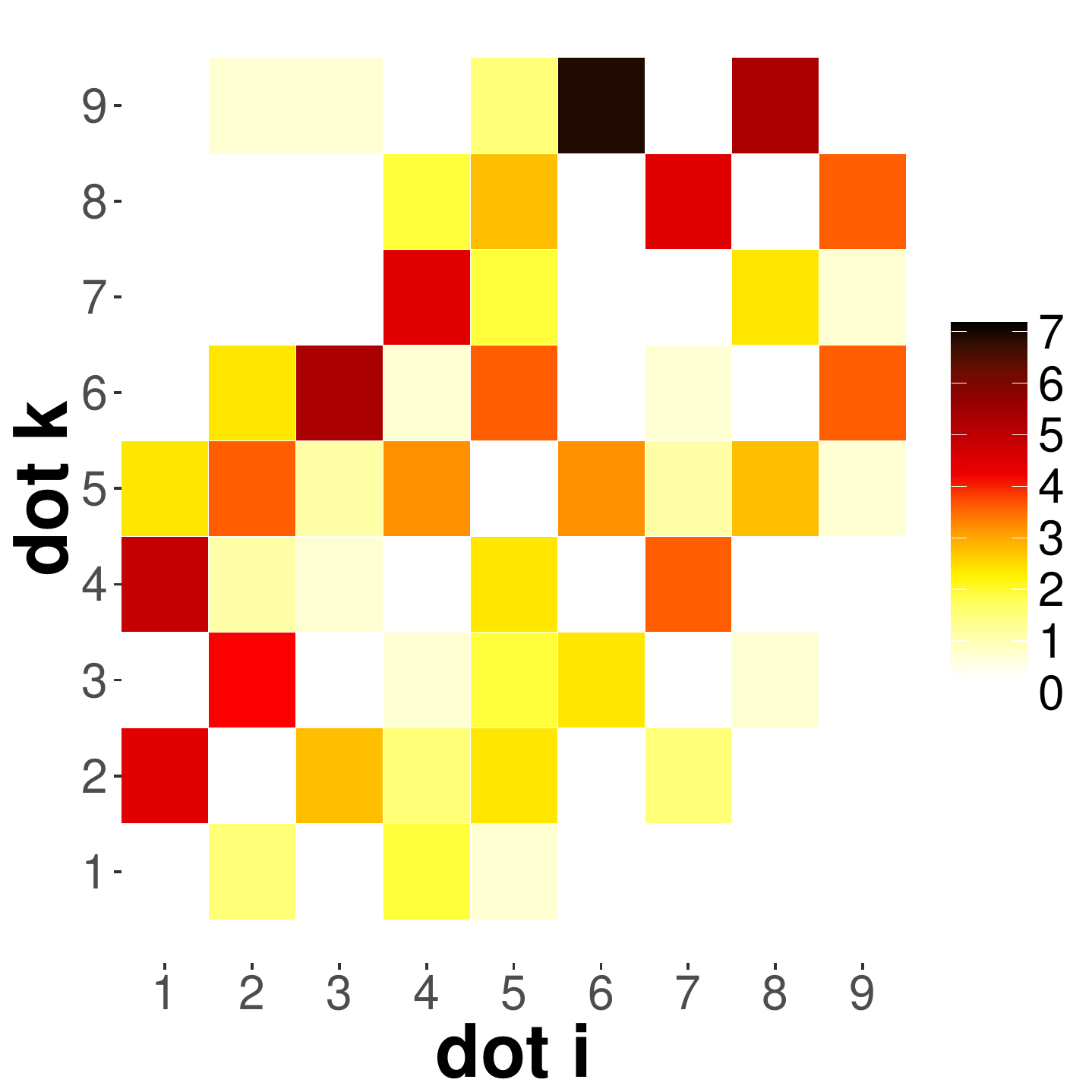}
  \captionsetup{font=scriptsize}
  \caption{3-dot}~\label{fig:syspal3_segments}
\end{subfigure}%
\caption{Usage of line segments ($i \rightarrow k$) across five groups. The more popular the line segment, the darker is the cell color.}~\label{fig:segmentsUsage}
\end{figure*}
\textbf{Frequency of Segments.} Next, we analysed the usage frequencies of each of the $9\cdot 8 = 72$ line segments across all five groups. Figure \ref{fig:segmentsUsage} illustrates the frequencies of all possible such segments. The more popular the line segment, the darker is the cell color. The number of unique line segments employed in the TinPal group was 59, whereas in the Original, 1-dot, 2-dot and 3-dot groups the numbers were 45, 44, 37 and 36 respectively. Therefore, participants in the TinPal group used more diverse line segments than the rest of the groups. Further, the number of unique line segments used decreased as the number of mandated points increased. Now, we report the usage of simple moves, knight moves and overlaps in more detail.

\begin{figure}[h]
\centering
\begin{subfigure}[b]{0.25\textwidth}
  \centering
  \includegraphics[scale=0.35]{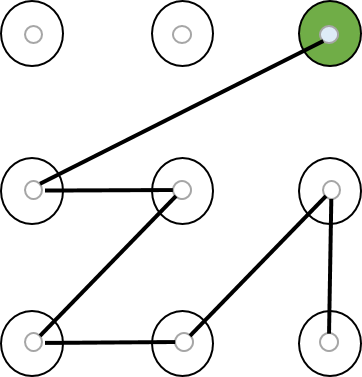}
  \captionsetup{font=scriptsize}
  \caption{Pattern with knight move} \label{fig:knighteg}
\end{subfigure}%
\begin{subfigure}[b]{0.25\textwidth}
  \centering
  \includegraphics[scale=0.35]{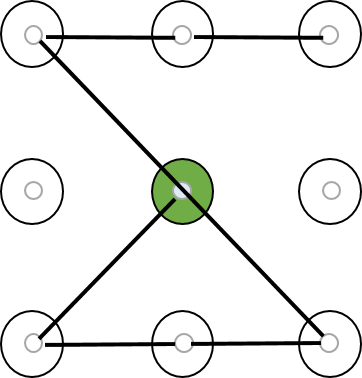}
  \captionsetup{font=scriptsize}
  \caption{With overlap move}   \label{fig:overlapeg}
\end{subfigure}%
\begin{subfigure}[b]{0.25\textwidth}
  \centering
  \includegraphics[scale=0.35]{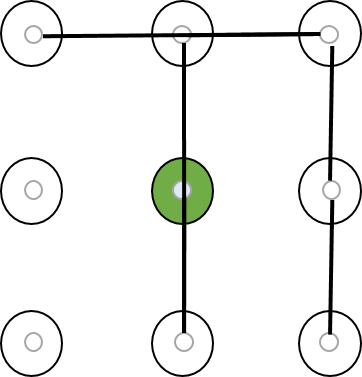}
  \captionsetup{font=scriptsize}
  \caption{With overlap move}   \label{fig:overlapeg2}
\end{subfigure}%
\begin{subfigure}[b]{0.25\textwidth}
  \centering
  \includegraphics[scale=0.35]{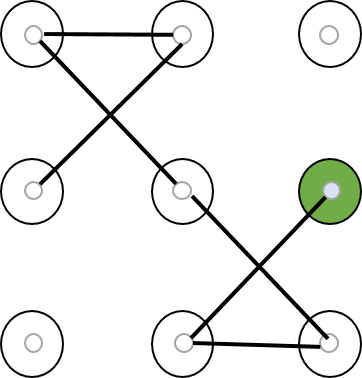}
  \captionsetup{font=scriptsize}
  \caption{With intersections}  \label{fig:intersecteg}
\end{subfigure}
\caption{Patterns with knight move, overlap move and intersections.}
\end{figure}

\begin{itemize}
\item{Simple Moves.} A simple move occurs when a dot is connected to another dot that is one unit away in either horizontal or vertical direction. In other words, line segments with the Euclidean distance of 1 or $\sqrt{2}$ are referred to as simple moves. Figure \ref{fig:segmentsUsage} shows that the most frequently used line segments in the Original, 1-dot, 2-dot and 3-dots groups were simple moves of the form $i \rightarrow i+1$ for all $i = 1, 2, 4, 5, 7, 8$ and $i \rightarrow i+3$ for all $i = 1, 2, 3, 4, 5, 6$, each having unit length. Few examples of popular simple moves are $1 \rightarrow 2$, $7 \rightarrow 8$, $3 \rightarrow 6$ and $6 \rightarrow 9$ .The distribution of simple moves in the TinPal group was relatively more uniform (less dark cells) as compared to the other groups.

\item{Knight Moves.} The concept of knight move is similar to the one defined in the game of chess. A knight move occurs when a dot is connected to another dot that is two units away in the horizontal (vertical) direction and one unit away in the vertical (horizontal) direction. In other words, line segments with the Euclidean distance of $\sqrt{5}$ are referred to as knight moves. For instance, the segment $3\rightarrow 4$ in the pattern $3457869$ is a knight move (Figure \ref{fig:knighteg}). The number of knight moves is considered to be an important feature in thwarting both shoulder-surfing attacks \cite{vonZezschwitz:shouldersurfing} and guessing attacks \cite{Andriotis:guessing}. Figure \ref{fig:segmentsUsage} shows that knight moves were used rarely in the Original and SysPal groups. The distribution of knight moves across all five groups is shown in Figure \ref{fig:knight}. In the TinPal group, the number of patterns containing at least one knight move was 58\%, whereas the numbers in the Original, 1-dot, 2-dot and 3-dot groups were 22.45\%, 18\%, 14.29\% and 18.75\% respectively. 

\item{Overlaps.} A line segment between two dots in a pattern is referred to as an overlap, if the dot between them is already connected. In other words, the line segments with the Euclidean distance of either 2 or $2\sqrt{2}$ are referred to as overlaps. For instance, the line segment $9 \rightarrow 1$ in the pattern $5789123$ constitutes an overlap (Figure \ref{fig:overlapeg}). Its length is $2\sqrt{2}$. Another instance of an overlap is the connection $8 \rightarrow 2$ in the pattern $5821369$ (Figure \ref{fig:overlapeg2}). Its length is 2. The number of overlap moves is considered to be an important feature in resisting both shoulder-surfing \cite{vonZezschwitz:shouldersurfing, Sun:shouldersurfing, Song:shouldersurfing} and guessing attacks \cite{Andriotis:guessing}. The distribution of overlaps across all five groups is shown in Figure \ref{fig:overlap}. In the TinPal group, the number of patterns containing at least one overlap move was 40\%, whereas the number of such patterns in the Original, 1-dot, 2-dot and 3-dot groups were 12.24\%, 16\%, 9.16\% and 4.17\% respectively. 
\end{itemize}
Therefore, TinPal interface influenced users to include more knight moves and overlaps in their pattern as compared to SysPal policies.

\begin{figure*}[h]
\centering
\begin{subfigure}[b]{0.20\textwidth}
  \centering
  \includegraphics[scale = 0.40]{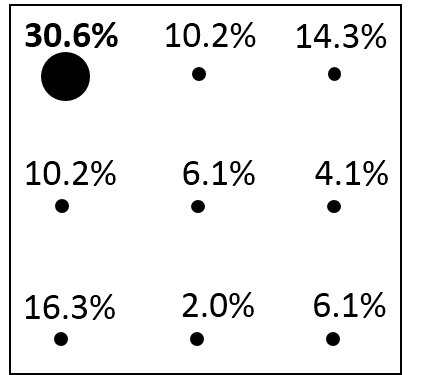}
  \captionsetup{font=scriptsize}
  \caption{Original}~\label{fig:existing_start}
\end{subfigure}%
\begin{subfigure}[b]{0.20\textwidth}
  \centering
  \includegraphics[scale = 0.40]{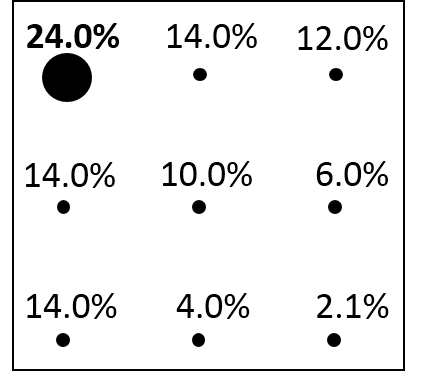}
  \captionsetup{font=scriptsize}
  \caption{TinPal}~\label{fig:enhanced_start}
\end{subfigure}%
\begin{subfigure}[b]{0.20\textwidth}
  \centering
  \includegraphics[scale = 0.40]{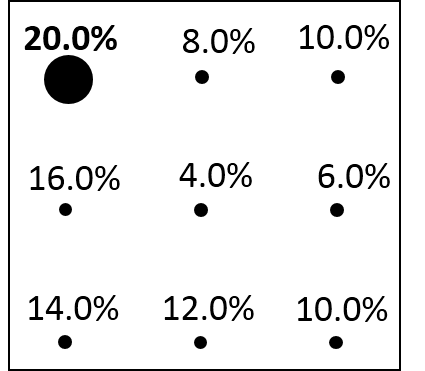}
  \captionsetup{font=scriptsize}
  \caption{1-dot}~\label{fig:syspal1_start}
\end{subfigure}%
\begin{subfigure}[b]{0.20\textwidth}
  \centering
  \includegraphics[scale = 0.40]{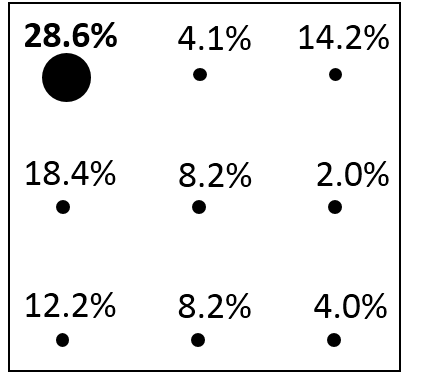}
  \captionsetup{font=scriptsize}
  \caption{2-dot}~\label{fig:syspal2_start}
\end{subfigure}%
\begin{subfigure}[b]{0.20\textwidth}
  \centering
  \includegraphics[scale = 0.40]{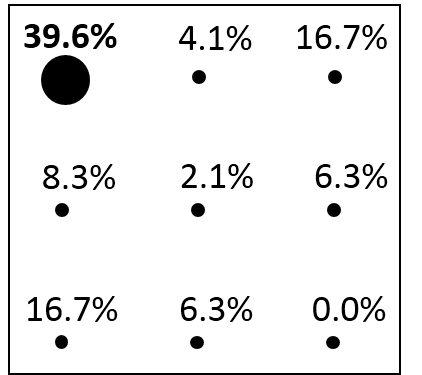}
  \captionsetup{font=scriptsize}
  \caption{3-dot}~\label{fig:syspal3_start}
\end{subfigure}%
\caption{Distribution of starting points across five groups.}~\label{fig:start}
\end{figure*}

\begin{figure*}[h]
\centering
\begin{subfigure}[b]{0.20\textwidth}
  \centering
  \includegraphics[scale = 0.40]{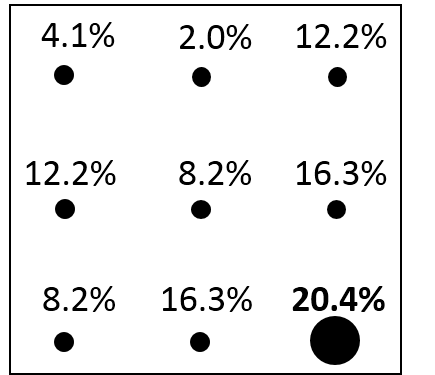}
  \captionsetup{font=scriptsize}
  \caption{Original}~\label{fig:existing_end}
\end{subfigure}%
\begin{subfigure}[b]{0.20\textwidth}
  \centering
  \includegraphics[scale = 0.40]{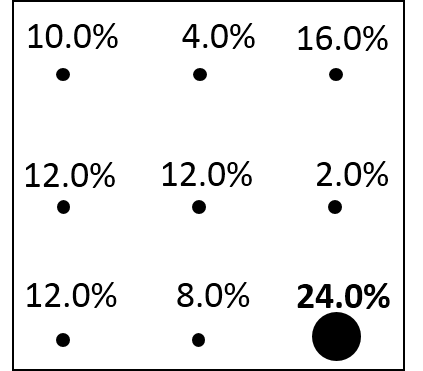}
  \captionsetup{font=scriptsize}
  \caption{TinPal}~\label{fig:enhanced_end}
\end{subfigure}%
\begin{subfigure}[b]{0.20\textwidth}
  \centering
  \includegraphics[scale = 0.40]{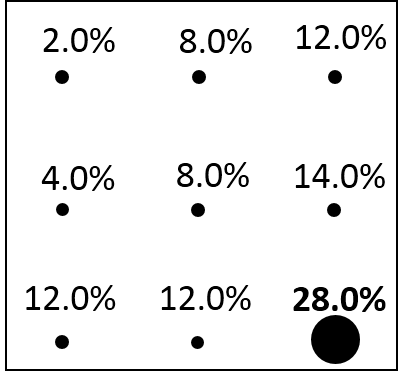}
  \captionsetup{font=scriptsize}
  \caption{1-dot}~\label{fig:syspal1_end}
\end{subfigure}%
\begin{subfigure}[b]{0.20\textwidth}
  \centering
  \includegraphics[scale = 0.40]{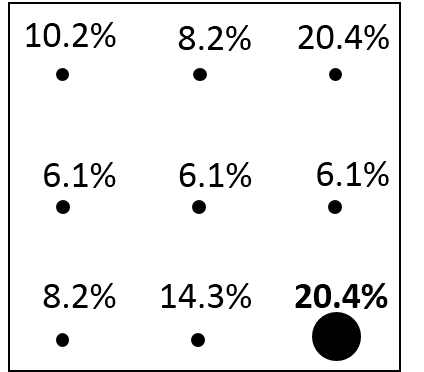}
  \captionsetup{font=scriptsize}
  \caption{2-dot}~\label{fig:syspal2_end}
\end{subfigure}%
\begin{subfigure}[b]{0.20\textwidth}
  \centering
  \includegraphics[scale = 0.40]{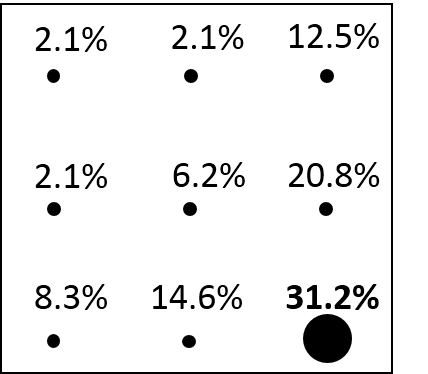}
  \captionsetup{font=scriptsize}
  \caption{3-dot}~\label{fig:syspal3_end}
\end{subfigure}%
\caption{Distribution of ending points across five groups.}~\label{fig:end}
\end{figure*}

\noindent
\\
\textbf{Direction Changes.} Another important pattern characteristics that adds to the complexity of a pattern is direction change. A direction change occurs when two consecutive line segments in a given pattern have different Euclidean distances \cite{harshal:guessing}. For instance, two consecutive line segments $3 \rightarrow 4$ and $4 \rightarrow 5$ in the pattern $3457869$ constitute a direction change since these segments have different Euclidean distances ($\sqrt{5}$ and 1 respectively). Simple patterns such as $321456987$ (`S' shape) composed of unit distance line segments do not comprise any direction change \cite{harshal:guessing}. The theoretical distribution of direction changes is shown in Table \ref{tab:direction}. Theoretically, about 95\% of the patterns contain more than two direction changes. The distribution of direction changes across all five groups is shown in Figure \ref{fig:direction}. In the TinPal group, the number of patterns containing more than two direction changes was 52\%, while the number of such patterns in the Original, 1-dot, 2-dot and 3-dot groups were 20.41\%, 14\%, 16.33\% and 10.42\% respectively. The difference in the number of direction changes between the TinPal group and all other groups was significant (all $p < 0.001$, corrected two-tailed WMW test). 
\begin{table}[h]
  \centering
 \scriptsize
  \begin{tabular}{c r r}
\toprule
     \textit{\#Direction Changes}
     & \textit{Count}
      & \textit{Percentage}\\
    \midrule
	0 & 664 & 0.17\%\\
	1 & 3,232 & 0.83\%\\
	2 & 15,800 & 4.06\%\\
	3 & 45,224 & 11.62\%\\
	4 & 89,096 &  22.90\%\\
	5 & 114,632 & 29.46\%\\
	6 & 89,640 & 23.04\%\\
	7 & 30,824 & 7.92\%\\
   \midrule
       \#Patterns & 389,112 & 100\% \\		
    \bottomrule
  \end{tabular}
  \caption{Theoretical distribution of direction changes in $3 \times 3$ patterns.}~\label{tab:direction}
\end{table}
\\\\
\textbf{Intersections.} An intersection occurs when two non-consecutive line segments in a pattern cross each other. For instance, the line segments $6 \rightarrow 8$ and $9 \rightarrow 5$ in the pattern $6895124$ intersect each other (Figure \ref{fig:intersecteg}). Here, another intersection occurs between the line segments $5 \rightarrow 1$ and $2\rightarrow 4$. The number of intersections in a pattern is considered as an important feature in countering shoulder-surfing attacks \cite{Sun:shouldersurfing, Song:shouldersurfing, vonZezschwitz:shouldersurfing}. The theoretical distribution of intersections is shown in Table \ref{tab:intersection}. After analysing the datasets, we found that the highlighting mechanism had no significant effect on the number of intersections across all five groups (all $p > 0.001$, corrected two-tailed WMW test). However, we found significant interest in the number of intersections between the TinPal group and two SysPal groups, 2-dot and 3-dot (all $p < 0.01$, corrected two-tailed WMW test).
\begin{table}[h]
  \centering
 \scriptsize
  \begin{tabular}{c r r}
\toprule
     \textit{\#Intersections}
     & \textit{Count}
      & \textit{Percentage}\\
    \midrule
0 & 58,771 & 15.10\% \\
1 & 85,536 &  21.98\% \\
2 & 73,432 & 18.87\% \\
3 & 61,775 &15.88\% \\
4 & 43,237 &	11.11\% \\
5 & 26,462 &	6.80\% \\
6 & 17,676 &	4.54\% \\
7 & 10,484 &	2.69\% \\
8 & 6,431  &	1.65\% \\
9 & 2,829  &	0.73\% \\
10 & 1,475 &	0.38\% \\
11 & 533   &	0.14\% \\
12 & 386   &	0.10\% \\
13 & 49  & 0.01\% \\
14 & 36 & 0.01\% \\
   \midrule
       \#Patterns & 389,112 & 100\% \\		
    \bottomrule
  \end{tabular}
  \caption{Theoretical distribution of intersections in $3 \times 3$ patterns.}~\label{tab:intersection}
\end{table}
\\\\
\noindent
\textbf{Frequency of Dots.} Previous studies \cite{Andriotis:sidechannel, Uellenbeck:guessing, Andriotis:guessing, Aviv:guessing, harshal:guessing} demonstrate that majority of users start their pattern with the upper-left dot and end their pattern with the lower-right dot. Unsurprisingly, we found that the upper-left dot was the most popular starting point (Figure \ref{fig:start}) and the lower-right dot was the most popular ending point (Figure \ref{fig:end}) in all five groups. 

Figure \ref{fig:start} shows that the number of patterns starting with the upper-left dot was much less in the 1-dot group (just 20\%) as compared to the other groups. This is because, 54\% (27/50) of the participants in the 1-dot group started their pattern with the system-assigned random dot. On the contrary, the number of patterns starting with the upper-left dot in the 3-dot group was 39.58\% (19/48) much higher than the other groups. This is because, in 89.47\% (17/19) of these cases, the upper-left dot was one of the three random dots assigned by the system. As a result, the entropy of the starting point distribution in the 1-dot group was highest (3.04 bits) and the entropy in the 3-dot group was lowest (2.50 bits). However, we did not find any significant difference in the usage of the upper-left dot as starting point between any pair of groups ($p > 0.01$, corrected two-tailed FET).

Figure \ref{fig:end} shows that the number of $3 \times 3$ patterns ending with the lower-right dot was highest in the 3-dot group, 31.25\% (15/48). Further analysis revealed that in 22.92\% (11/48) of the cases, the lower-right dot was one of the three random dots assigned by the system. As a result, the entropy of the end point distribution in the 3-dot group was just 2.67 bits, whereas in the other groups it was between 2.89 to 3.00 bits. However, we did not find any significant difference in the usage of the lower-right dot as an ending point between any pair of groups ($p > 0.01$, corrected two-tailed FET).

\begin{figure*}[h]
\centering
\begin{subfigure}[b]{0.20\textwidth}
  \centering
  \includegraphics[scale = 0.40]{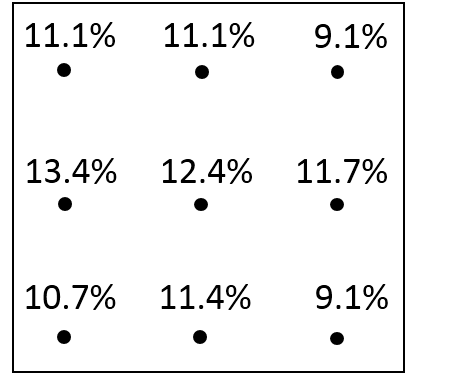}
  \captionsetup{font=scriptsize}
  \caption{Original}~\label{fig:existing_end}
\end{subfigure}%
\begin{subfigure}[b]{0.20\textwidth}
  \centering
  \includegraphics[scale = 0.40]{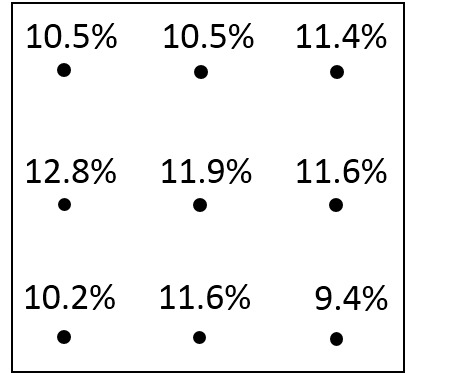}
  \captionsetup{font=scriptsize}
  \caption{TinPal}~\label{fig:enhanced_end}
\end{subfigure}%
\begin{subfigure}[b]{0.20\textwidth}
  \centering
  \includegraphics[scale = 0.40]{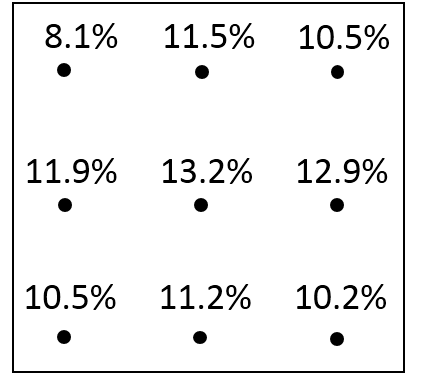}
  \captionsetup{font=scriptsize}
  \caption{1-dot}~\label{fig:syspal1_end}
\end{subfigure}%
\begin{subfigure}[b]{0.20\textwidth}
  \centering
  \includegraphics[scale = 0.40]{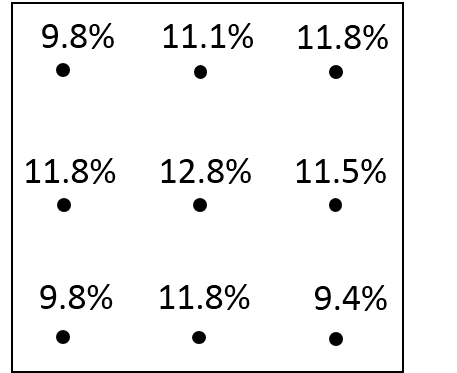}
  \captionsetup{font=scriptsize}
  \caption{2-dot}~\label{fig:syspal2_end}
\end{subfigure}%
\begin{subfigure}[b]{0.20\textwidth}
  \centering
  \includegraphics[scale = 0.40]{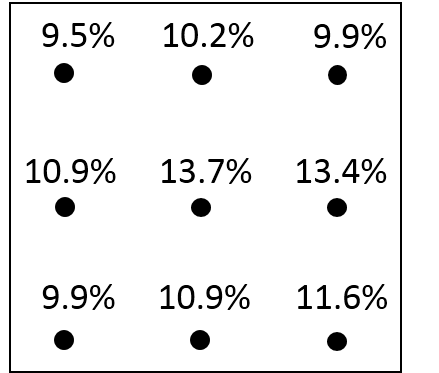}
  \captionsetup{font=scriptsize}
  \caption{3-dot}~\label{fig:syspal3_end}
\end{subfigure}%
\caption{Distribution of all 9 dots across five groups.}~\label{fig:dot}
\end{figure*}

We also analysed the usage frequencies of each of the nine dots in $3 \times 3$ grid across all five groups (Figure \ref{fig:dot}). The most frequently used dot in all SysPal groups was dot 5 (the center dot), whereas the most frequently used dot in the TinPal and Original groups was dot 4. Overall, frequencies of all nine dots across five groups were evenly distributed, and entropy in all five groups was nearly 3.16 bits. Note that maximum possible entropy is $log_2(9) \sim 3.17$ bits.
\\\\

\noindent
\textbf{Position of Mandated Dots.} SysPal policies mandate users to include system-assigned random dot(s) in their $3 \times 3$ pattern. We found that across all SysPal policies there was a tendency to use system-assigned dots at the beginning of the pattern. Specifically, 54\% of the participants in the 1-dot group used the system-assigned dot in the first position of their pattern, whereas 67.35\% of the participants in the 2-dot group and 79.16\% of the participants in the 3-dot group used one of the system-assigned dots in the first position of their pattern. Similar behaviour was also reported in \cite{cho2017syspal}. 

Next, we determined how often the system-assigned dots in the 2-dot and 3-dot patterns were used adjacent to each other. We use the definition of adjacency as given in \cite{cho2017syspal}. If there is a direct connection (line segment) between two system-assigned dots then we consider those two dots to be adjacently located. For instance, if dot 1 and dot 2 are system-assigned, and the pattern is $1236$, we consider those two dots to be located adjacently. However, if the pattern is $1452$, then we do not consider dots 1 and 2 as located adjacently. There are ${9\choose 2} = 36$ possible ways of selecting two dots $(i, j)$ in $3 \times 3$ grid. Of these 36 pairs, 20 pairs (55.56\%) can be connected directly using simple move, 8 pairs (22.22\%) can be connected directly using knight move and the remaining 8 pairs (22.22\%) can be connected using overlap (Table \ref{tab:randomPairs}). 

\begin{table}[h]
  \centering
 \scriptsize
  \begin{tabular}{l l l}
\toprule
        \textit{Segment Type}
      & \textit{$(i,j)$}
      & \textit{Count}\\
   \midrule
	Simple & (1,2), (1,4), (1,5) & 20 (55.56\%)\\
	           & (2,3), (2,4), (2,5), (2,6) &\\
		& (3,5), (3,6) & \\
		& (4,5), (4,7), (4,8) & \\
		& (5,6), (5,7), (5,8), (5,9) &\\
		& (6,8), (6,9) & \\
		& (7,8) & \\
		& (8,9) & \\
  \midrule
	Knight & (1,6), (1,8) & 8 (22.22\%)\\
		& (2,7), (2,9) & \\
		& (3,4), (3,8) & \\
		& (4,9) & \\
		& (6,7) & \\
  \midrule
	Overlap & (1,3), (1,7), (1,9) & 8 (22.22\%)\\
		   & (2,8) & \\
		   & (3,7), (3,9) & \\
                        & (4,6) & \\
		  & (7,9) & \\ 	
   \bottomrule 
\end{tabular}
\caption{Classification of line segments between every two dots $(i, j)$ in $3 \times 3$ grid.}~\label{tab:randomPairs}
\end{table}

In the 2-dot group, 53.06\% (26/49) of the participants were assigned two random dots $(i, j)$ which could be connected directly using simple move, 24.49\% (12/49) were assigned two dots which could be connected directly using knight move and the remaining 22.45\% (11/49) were assigned two dots which could be connected directly using overlap. We found that 73.08\% (19/26) of the participants in the first category placed system-assigned dots adjacently, {\em i.e.}, they used simple move. None of the participants (23/49) placed system-assigned dots adjacently in the other two categories, {\em i.e.}, they did not use knight move or overlap. In the 3-dot group, 77.08\% (37/48) of the participants used at least two system-assigned dots adjacently, whereas 27.08\% (13/48) of the participants used all three system-assigned dots adjacently. Again, we found that in 94.59\% (35/37) of the cases where system-assigned dots were used adjacently, they were connected using simple moves. Only in the remaining 5.41\% (2/37) of the cases, knight moves were used. Therefore, whenever system-assigned random dots were neighbours (Table \ref{tab:neighbour}), {\em i.e.}, they belong to the same row ({\em e.g.}, dots $1$ and $2$) or same column ({\em e.g.}, dots $1$ and $4$) or to the adjacent rows and columns ({\em e.g.}, dots $1$ and $5$), they were more likely to be connected with simple move.

\subsection{Pattern Guessability}
\textbf{Markov Model.} We use Markov model based attack technique to estimate the guessing resistance of patterns created in all five groups. We closely follow the attack methodology as described in \cite{Uellenbeck:guessing}. Markov models exploit the fact that subsequent choices in a human-generated sequence are mostly dependent on previous choices. For instance, in English language letter {\em h} is more likely to follow letter {\em t} than letter {\em z}. In case of $3\times 3$ patterns, dot 2 is more likely to follow dot 1 than dot 6. Based on these observations, $n$-gram Markov model predicts the next choice in a sequence using past $n-1$ choices. The probability of a $l$ length sequence $s_1s_2\ldots s_l$ can therefore be modelled as:
\begin{align}
P(s_1\ldots s_l) = P(s_1\ldots s_{n-1})\cdot \prod_{i=n}^{l} P(s_i|s_{i-n+1}\ldots s_{i-1})
\end{align}~\label{eq:markov}

\noindent
\textbf{Our dataset.} 
To build an $n$-gram Markov model, we need to decide the value of parameter $n$. If we choose the value of $n$ to be 2 (bigrams), then we require $9\cdot 8 = 72$ data points, and if we choose the value of $n$ to be 3 (trigrams), then we require $9 \cdot 8 \cdot 7 = 504$ data points. Since the control group dataset has 49 patterns and corresponding 298 data points that are insufficient to learn trigram probabilities, we resort to bigrams. We use Laplace smoothing to account for unseen bigrams.

We perform 10-fold cross-validation on each pattern set, {\em i.e.}, we split patterns collected in each group into 10 approximately equal-sized subsets. One of the subset is used as a test set and the remaining 9 subsets are combined into the training set. We use the training set to learn bigrams probabilities which in turn are used to estimate the probabilities of all possible 389,112 patterns by equation $(2)$. Subsequently, we sort all patterns in the decreasing order of probability and simulate guessing attack on the test set. We perform this validation 10 times where every subset was used in turn as a test set. We repeat this entire split-learn-simulate process 10 times and report the average results. 

The guessability results are summarized in Table \ref{tab:guess}. The guessing algorithm could not crack any pattern (0\%) in the TinPal group, whereas it cracked 12\%, 10\%, 8\% and 20\% patterns in the Original, 1-dot, 2-dot and 3-dot groups respectively. These numbers suggest that patterns created on the TinPal interface were more resistant to guessing attack than patterns created using any other group.
\\
\\
\textbf{ASIACCS'17 dataset.} We also had access to 69,797 $3 \times 3$ patterns collected by Tupsamudre et al. \cite{harshal:guessing}. Since their dataset is huge, we learn trigram probabilities and use equation $(2)$ to estimate the probabilities of all possible patterns (389,112). We then sort all possible patterns and simulate guessing attack on our datasets. The cracking results are given in Table \ref{tab:guess}. 

Within first 20 attempts, the guessing algorithm cracked 4\% of the patterns in TinPal group, whereas it cracked  12.24\%, 8\%, 6.12\% and 18.75\% of the patterns in the Original, 1-dot, 2-dot and 3-dot groups respectively. Again, overall patterns created using TinPal interface were more resistant to guessing attack than patterns created using the original interface and all SysPal policies. Further, as reported in \cite{cho2017syspal}, patterns created on 1-dot and 2-dot interfaces were more secure than the original interface whereas patterns created on the 3-dot interface were most vulnerable. Thus, mandating too many dots could reduce the search space for the attacker.

\begin{table}[h]
  \centering
 \scriptsize
  \begin{tabular}{l c c c c c}
\toprule
     \textit{Dataset}
     & \textit{Original}
      & \textit{TinPal}
     & \textit{1-dot}
     & \textit{2-dot}
     & \textit{3-dot}     
\\
    \midrule
    Current & 12\% & \textbf{0\%} & 10\% &  8\% & 20\%\\
    ASIACCS'17 &  12.24\% & \textbf{4\%} & 8\% & 6.12\% & 18.75\%\\
    \bottomrule
  \end{tabular}
  \caption{Guessability Results of all groups.}~\label{tab:guess}
\end{table}

\section{Usability Results}\label{sec:usabilityresults}
Now, we present the results pertaining to usability. Specifically, we compared memorability and efficiency of patterns created on different interfaces. \\
\textbf{Memorability.} 
We measured memorability using the following two metrics:
\begin{itemize}
\item number of users who successfully recalled their pattern
\item number of login attempts required to recall pattern
\end{itemize}

Before advancing to the pattern recall stage, participants in each group spent about 4 minutes to solve the distraction task and answer questions related to demographics, their devices and screen locks. The results pertaining to pattern recall are shown in Table \ref{tab:login}. More than 96.00\% of the participants in all five groups successfully recalled their pattern within just three attempts. We found no significant difference in the login attempts between any pair of groups (all $p\sim 1$, corrected two-tailed FET) which suggests that patterns created using the TinPal interface were not only longer and complex, but also easy to remember.
\begin{table}[h]
  \centering
 \scriptsize
  \begin{tabular}{l c c c c c}
\toprule
     \textit{}
         & \textit{Original}
      & \textit{TinPal}
     & \textit{1-dot}
     & \textit{2-dot}
     & \textit{3-dot}     
\\
    \midrule
    Attempt 1 & 35 (71.43\%) & 39 (78\%) & 39 (78\%) & 38 (77.55\%) & 36 (75\%)\\
    Attempt 2 & 11 (22.45\%) & 8 (16\%) & 7 (14\%) & 6 (12.24\%) & 8 (16.67\%)\\
    Attempt 3 & 2 (4.08\%) & 2 (4\%) & 2 (4\%) & 4 (8.16\%) & 3 (6.25\%)\\
    \midrule
    Successful & 48 (97.96\%) & 49 (98\%) & 48 (96\%) & 48 (97.96\%) & 47 (97.92\%)\\		
    Unsuccessful & 1 (2.04\%) & 1 (2\%) & 2 (4\%) & 1 (2.04\%) & 1 (2.08\%)\\	 
    \midrule
   Total & 49 & 50 & 50 & 49 & 48\\ 
   \bottomrule
  \end{tabular}
  \caption{Login attempts of participants across five groups.}~\label{tab:login}
\end{table}
\noindent
\\\\
\textbf{Efficiency. } We measure efficiency using the following three metrics:
\begin{itemize}
\item time required to create pattern during creation stage
\item time required to recall pattern during recall stage
\end{itemize}

\begin{figure*}[h]
  \centering
  \includegraphics[scale = 0.15]{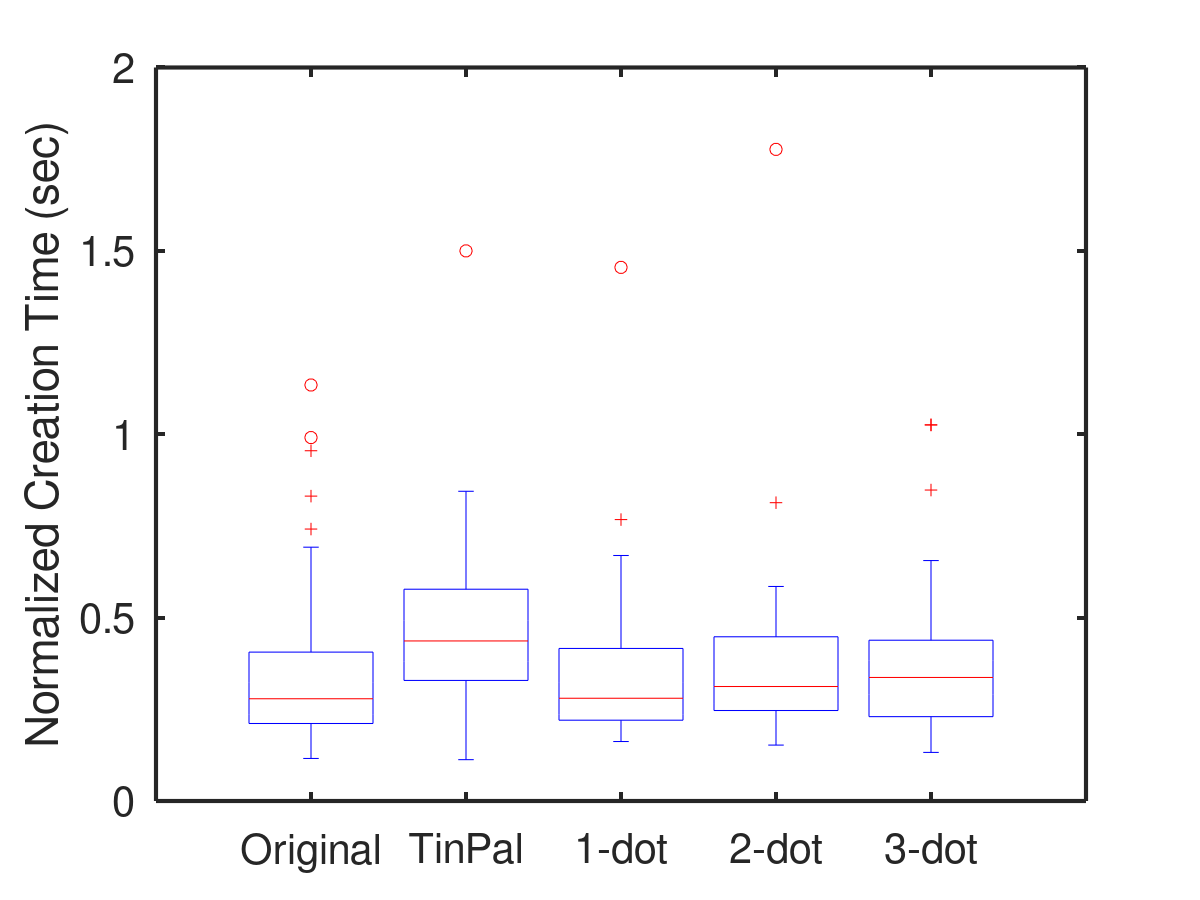}
  \captionsetup{font=scriptsize}
  \caption{Comparison of normalized pattern creation time across all five groups.}~\label{fig:box_creation}
\end{figure*}

Table \ref{tab:recall} shows the median time required to create, redraw and recall pattern across five groups. Before advancing to the pattern creation stage, participants in all groups tried multiple patterns in the training stage. This stage was exploratory, and provided participants with an opportunity to become familiar with the assigned interface. The time required to create a pattern in the TinPal group (median 3.91s) was maximum among all five groups (Table \ref{tab:recall}). However, we note that the stroke length of patterns created in the TinPal group (median 8.27) was also higher than the other groups (Table \ref{tab:stats}). Therefore, we normalize the pattern creation time {\em i.e.,} for each participant, we divide the time required to create the pattern with the stroke length of that pattern. The distribution of the normalized pattern creation time across all five groups is depicted in Figure \ref{fig:box_creation}. After normalization, we found no find significant difference in the pattern creation time between any pair of groups (all $p > 0.001$, corrected two-tailed WMW test). However, we found significant interest in the normalized creation time between the TinPal group and other groups ($p < 0.01$, corrected two-tailed WMW test)) which suggests that participants who used TinPal paid attention to the highlighted dots. All pairwise p-values are depicted in Table \ref{tab:usabilityStatsDetailed} (Appendix A).

\begin{figure*}[h]
  \centering
  \includegraphics[scale = 0.15]{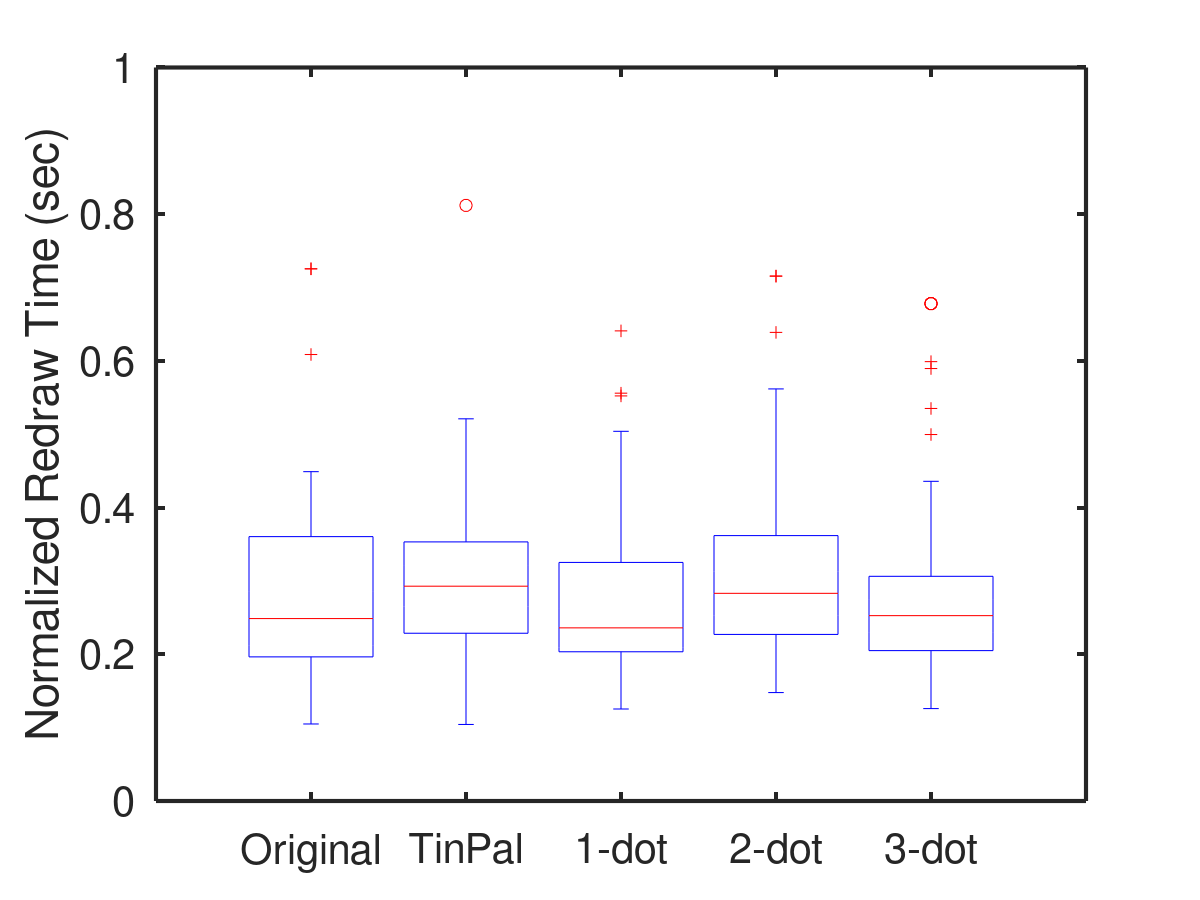}
  \captionsetup{font=scriptsize}
  \caption{Comparison of normalized pattern redraw time across all five groups.}~\label{fig:box_redraw}
\end{figure*}

\begin{figure*}[h]
  \centering
  \includegraphics[scale = 0.15]{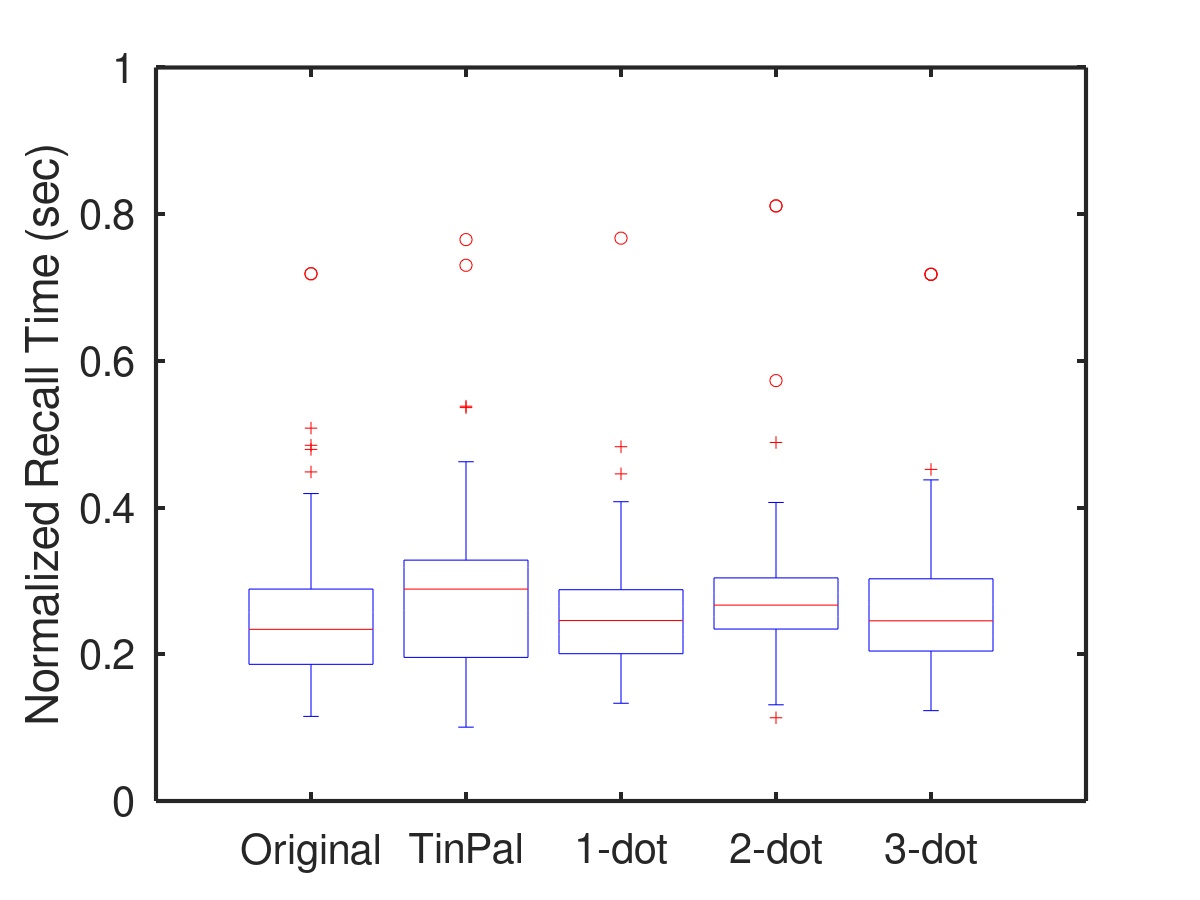}
  \captionsetup{font=scriptsize}
  \caption{Comparison of normalized pattern recall time across all five groups.}~\label{fig:box_recall}
\end{figure*}

Similarly, the median time required to recall pattern in the TinPal group (2.11s) was relatively higher than the other groups (Table \ref{tab:recall}). However, after normalizing the recall time with respect to the stroke length, the recall time of patterns in the TinPal group (0.29s) was similar to the recall time of patterns in the other groups. We did not find any significant difference or significant interest in the normalized recall time between the TinPal group and other groups (all $p > 0.01$, corrected two-tailed WMW test). The pattern redraw time of all five groups after normalizing was also similar. The distributions of the normalized redraw time and normalized recall time across all five groups are depicted in Figures \ref{fig:box_redraw} and \ref{fig:box_recall} respectively.
\\\\
\begin{table}[t]
  \centering
 \scriptsize
  \begin{tabular}{l c c c c c}
\toprule
     \textit{Median}
         & \textit{Original}
      & \textit{TinPal}
     & \textit{1-dot}
     & \textit{2-dot}
     & \textit{3-dot}     
\\
    \midrule

Creation time & 1.64s & 3.91s & 1.67s & 1.63s & 1.34s\\
Normalized time & 0.28s & 0.43s & 0.28s & 0.31s & 0.34s\\
 \midrule
Redraw time & 1.27s & 2.34s & 1.56s & 1.23s & 1.16s\\
Normalized time & 0.25s & 0.29s & 0.24s & 0.28s & 0.25s\\
\midrule
Recall time & 1.25s & 2.11s & 1.18s & 1.37s & 1.01s \\ 
Normalized time & 0.23s & 0.29s & 0.25s & 0.27s & 0.25s  \\
  \bottomrule
  \end{tabular}
  \caption{Median time required to create, redraw and recall pattern across five groups.}~\label{tab:recall}
\end{table}

\noindent
\textbf{Acceptability.} We asked participants in the TinPal group an open-ended question: ``{\em Which pattern lock interface do you prefer, the existing one that you have used before (on your phone) or the new one that you saw in our experiment}"? Of the 29 (58\%) experimental group participants who reported using Android pattern screen-lock before (Table \ref{tab:demo}), 27 (93.10\%) said that they would prefer the new one used in the experiment while 2 (6.90\%) said that they would prefer the original one. Remarks made by participants are as follows.
\\
 {\em `The new interface is better because of feedback feature.'} 
\\
{\em `The new one gives more idea about the variety of patterns we can make.'}
\\
{\em `I prefer the existing interface, need simple patterns only.'}

\section{Discussion and Future Work}
In this work, we proposed a new $3\times 3$ interface TinPal that informs users about different available connection options during pattern creation as well as recall. We also gave an efficient algorithm to determine the set of reachable dots from the currently connected dot in $3 \times 3$ grid. We evaluated the efficacy of TinPal and SysPal interfaces with a user study involving 246 participants. The results of our comparative study indicate that TinPal influenced users' pattern choices without much affecting the usability. Participants who used TinPal created significantly longer patterns containing visually complex features such as knight moves, overlaps and direction changes than those who used SysPal policies or the original interface. Guessability results also show that patterns created on TinPal are more resilient to guessing attacks than patterns created in any other group. These results are encouraging as TinPal just informed users about all available options and did not force them to choose a particular option.

SysPal policies mandate users to use one, two or three randomly chosen dots in their pattern. However, our study results show that these policies do not ensure that the resulting patterns will have important features such as knight moves ({\em e.g.}, $1 \rightarrow 6$) or overlaps ({\em e.g.}, $1 \rightarrow 3$) as users may not be aware of such connection options. Therefore, a large fraction of the search space remains unutilized. Comparatively, patterns drawn on TinPal used significantly longer strokes and large number of knight moves, overlaps and direction changes. Further, we argue that it is possible for users to circumvent SysPal policies by resetting the pattern until the desired dots are chosen by the system. TinPal, on the other hand, does not force or persuade users to include any particular dot(s) in their pattern. It just informs them about the next set of available options from the currently connected dot.

Among three SysPal groups, we found that patterns created in the 2-dot group were more secure than those created in the 1-dot and 3-dot groups. These findings are consistent with the previous study \cite{cho2017syspal}. However, contrary to the earlier findings \cite{cho2017syspal}, our study results show that patterns created using 3-dot policy were no better than those created using the original interface. In fact, guessability results suggest that pattern created using the original interface were more secure than those created using 3-dot policy (Table \ref{tab:guess}). Participants in the 3-dot policy tend to create shorter patterns mostly using simple moves (Table \ref{tab:stats}). The number of unique line segments employed in the Original group was 45 and in the 3-dot group the number was only 36. Similar was the case with the 2-dot group where the number of unique line segments was just 37. TinPal patterns were created using 59 unique line segments, the highest among all groups. Further, we found that whenever system-assigned dots were neighbours in $3 \times 3$ grid (Table \ref{tab:neighbour}) {\em i.e.}, they belong to the same row or same column or to the adjacent rows and adjacent columns then users were most likely to connect them directly with simple move. For example, if the system assigns two neighbouring dots 2 and 3 in the first row, then the user is more likely to connect them directly than if the system assigns two dots 2 and 9 that belong to different rows and columns. Note that, dots 2 and 9 could be connected directly using knight move but the user may not be aware of such connection. 

Our data indicates that TinPal had no effect on the starting point choices of the users, and they remain biased. For example, the upper-left dot is still the most popular starting choice for creating $3 \times 3$ patterns. One way to reduce this bias is to suggest a random starting point to the user as done in \cite{siadati:persuasive}. After the user starts from the suggested point, TinPal will spring into action and highlight the next set of reachable dots from the connected dot. Unlike \cite{siadati:persuasive}, 1-dot SysPal policy allows users to place system-assigned dot at any position in their pattern. However, our study results show that more than 50\% of the users in the 1-dot group started their pattern with the system-assigned random dot. 

There is a possibility of combining TinPal and SysPal, however it would be more interesting to study the combined effect of TinPal and different pattern strength meters proposed in the literature \cite{Andriotis:guessing, Sun:shouldersurfing, Song:shouldersurfing}. With the original interface, users might not be aware of all potential choices for creating their pattern which can reduce the impact of pattern strength meters. The combination of TinPal and a pattern strength meter could nudge users to create more secure patterns. As the user draws her pattern, TinPal will make her aware of all available choices that could be reached from the current dot and at the same time the strength meter will indicate which of the available choices are secure and which are not. 

\bibliographystyle{ACM-Reference-Format}
\bibliography{chi}
\newpage
\appendix{Appendix A}

\begin{table}[h]
  \centering
 \scriptsize
  \begin{tabular}{l l l l l}
\toprule
        \textit{Characteristic}
      & \textit{GroupA}
      & \textit{GroupB}
      & \textit{p-value}
      & \textit{significant}\\
   \midrule
   Pattern Length & TinPal & 1-dot    & 0.00096 & **\\
                          & TinPal & 2-dot    & 0.00489 & *\\
		     & TinPal & 3-dot    & 0.00057 & ** \\
		     & TinPal & Original & 0.00687 & *\\
                          & 1-dot & 2-dot & 0.73 & no\\
		     & 1-dot & 3-dot & 0.63 & no\\
		     & 1-dot & Original & 0.72 & no\\
		     & 2-dot & 3-dot & 0.98 & no\\
	                &  2-dot & Original & 0.97 & no\\
		     & 3-dot & Original& 0.98 & no\\ 	
\midrule
Stroke Length  & TinPal & 1-dot    & 0.0000022 & **\\
                          & TinPal & 2-dot    & 0.0000013 & **\\
		     & TinPal & 3-dot    & 0.000000025 & **\\
		     & TinPal & Original & 0.0000074 & **\\
                          & 1-dot & 2-dot &  0.90 & no\\
		     & 1-dot & 3-dot & 0.96 & no\\
		     & 1-dot & Original & 0.57 & no\\
		     & 2-dot & 3-dot & 0.75 & no\\
	                &  2-dot & Original  & 0.76 & no\\
		     & 3-dot & Original & 0.43 & no\\ 	
\midrule
Direction Changes  & TinPal & 1-dot    & 0.000006326 & **\\
                          & TinPal & 2-dot    & 0.00005988 & **\\
		     & TinPal & 3-dot    & 0.00001266 & **\\
		     & TinPal & Original  & 0.00040858 & **\\
                          & 1-dot & 2-dot & 0.57 & no\\
		     & 1-dot & 3-dot & 0.73 & no\\
		     & 1-dot & Original  & 0.17 & no\\
		     & 2-dot & 3-dot & 0.83 & no\\
	                &  2-dot & Original  & 0.41 & no\\
		     & 3-dot & Original  & 0.29 & no\\ 	
\midrule
Intersections  & TinPal & 1-dot    & 0.018432 & no\\
                          & TinPal & 2-dot    &0.003484 & *\\
		     & TinPal & 3-dot    & 0.001697 & *\\
		     & TinPal & Original  &  0.041358 & no\\
                          & 1-dot & 2-dot & 0.59 & no\\
		     & 1-dot & 3-dot & 0.45 & no\\
		     & 1-dot & Original  & 0.69 & no\\
		     & 2-dot & 3-dot & 0.74 & no\\
	                &  2-dot & Original  & 0.43 & no\\
		     & 3-dot & Original  & 0.20 & no\\ 	
\bottomrule
  \end{tabular}
  \caption{Pairwise comparisons of pattern characteristics across five groups. ** indicates the difference between two groups is statistically significant ($p < 0.001$), whereas * indicates there is significant interest ($p < 0.01$).}~\label{tab:statsDetailed}
\end{table}

\begin{table}[h]
  \centering
 \scriptsize
  \begin{tabular}{l l l l l}
\toprule
        \textit{Characteristic}
      & \textit{GroupA}
      & \textit{GroupB}
      & \textit{p-value}
      & \textit{significant}\\
   \midrule
Creation Time & TinPal & 1-dot    & 0.003 & *\\
                          & TinPal & 2-dot    & 0.008 & *\\
		     & TinPal & 3-dot    & 0.004 & * \\
		     & TinPal & Original  & 0.002 & *\\
                          & 1-dot & 2-dot & 0.41 & no\\
		     & 1-dot & 3-dot & 0.78 & no\\
		     & 1-dot & Original  & 0.74 & no\\
		     & 2-dot & 3-dot & 0.67 & no\\
	                &  2-dot & Original  & 0.30 & no\\
		     & 3-dot & Original  & 0.58 & no\\ 	
\midrule
Redraw Time & TinPal & 1-dot    & 0.13 & no\\
                          & TinPal & 2-dot    & 0.85 & no\\
		     & TinPal & 3-dot    & 0.06 & no\\
		     & TinPal & Original  & 0.10 & no\\
                          & 1-dot & 2-dot &  0.17 & no\\
		     & 1-dot & 3-dot & 0.83 & no\\
		     & 1-dot & Original & 0.77 & no\\
		     & 2-dot & 3-dot & 0.09 & no\\
	                &  2-dot & Original  & 0.11 & no\\
		     & 3-dot & Original  & 0.93 & no\\ 	
\midrule
Recall Time  & TinPal & 1-dot    & 0.11 & no\\
                          & TinPal & 2-dot    & 0.62 & no\\
		     & TinPal & 3-dot    & 0.13 & no\\
		     & TinPal & Original  & 0.04 & no\\
                          & 1-dot & 2-dot & 0.10 & no\\
		     & 1-dot & 3-dot & 0.96 & no\\
		     & 1-dot & Original  & 0.36 & no\\
		     & 2-dot & 3-dot & 0.13 & no\\
	                &  2-dot & Original  & 0.02 & no\\
		     & 3-dot & Original  & 0.38 & no\\ 	
\bottomrule
  \end{tabular}
  \caption{Pairwise comparisons of normalized pattern creation time, redraw time and recall time across five groups. ** indicates the difference between two groups is statistically significant ($p < 0.001$), whereas * indicates there is significant interest ($p < 0.01$).}~\label{tab:usabilityStatsDetailed}
\end{table}

\end{document}